\documentclass[]{elsarticle}

\usepackage[a4paper, margin=2.5cm]{geometry}

\usepackage{hyperref}
\usepackage{numprint}
\npthousandsep{,}\npthousandthpartsep{}\npdecimalsign{.}

\journal{Safety Science}

\usepackage{color}
\usepackage{comment}
\usepackage{enumitem}
\usepackage{tabularx}
\usepackage{array}
\usepackage{color, colortbl}
\usepackage{multirow}
\usepackage{longtable}
\usepackage{pdflscape}
\usepackage{amssymb}

\definecolor{Gray}{gray}{0.9}

\definecolor{mygreen}{RGB}{0, 150, 51}

\newcommand{\quotes}[1]{``{#1}''}

\usepackage{fancyhdr}
\begin{document}

\newcolumntype{R}[1]{>{\raggedright\let\newline\\\arraybackslash\hspace{0pt}}m{#1}}
\newcolumntype{C}[1]{>{\centering\let\newline\\\arraybackslash\hspace{0pt}}m{#1}}
\newcolumntype{L}[1]{>{\raggedleft\let\newline\\\arraybackslash\hspace{0pt}}m{#1}}

\begin{frontmatter}
\title{Evacuation trials from a double-deck electric train unit: Experimental data and sensitivity analysis\footnote{This is a preprint version of the article accepted to Safety Science with DOI 10.1016/j.ssci.2021.105523}}

\author[fsv,uceeb]{Hana Najmanov\'a}
\ead{hana.najmanova@cvut.cz}
\author[fsv]{Luk\'a\v s Kukl\'ik}
\ead{}
\author[fsv]{Veronika Pe\v skov\'a}
\ead{}
\author[fjfi]{Marek Buk\'a\v cek}
\ead{marek.bukacek@fjfi.cvut.cz}
\author[fit]{Pavel Hrab\'{a}k\corref{cor1}}
\ead{pavel.hrabak@fit.cvut.cz}
\author[fit]{Daniel Va\v sata}
\ead{daniel.vasata@fit.cvut.cz}
\cortext[cor1]{Corresponding author. Tel. +420224359885 ,email: pavel.hrabak@fit.cvut.cz}
\address[fsv]{Czech Technical University in Prague, Faculty of Civil Engineering, Thakurova 7, 160 00 Prague, Czechia}
\address[uceeb]{Czech Technical University in Prague, University Centre for Energy Efficient Buildings, T\v rineck\'a 1024, 273 43 Bu\v st\v ehrad, Czechia}
\address[fjfi]{Czech Technical University in Prague, Faculty of Nuclear Sciences and Physical Engineering, Trojanova 13, 120 00 Prague, Czechia}
\address[fit]{Czech Technical University in Prague, Faculty of Information Technology, Thakurova 9, 160 00 Prague, Czechia}

\begin{abstract}
 Passenger trains represent a challenging environment in emergencies, with specific evacuation conditions resulting from the typical layout and interior design inherent to public transportation vehicles. This paper describes a dataset obtained in a full-scale controlled experiment emulating the emergency evacuation of a double-deck electric unit railcar carried out in Prague in 2018. 15~evacuation trials involving 91~participants were conducted under various evacuation scenarios considering different compositions of passenger crowd, exit widths, and exit types (e.g. egress to a high platform, to an open rail line using stairs, and a 750~mm jump without any supporting equipment). The study’s main goals were to collect experimental data on the movement conditions in the railcar and to study the impact of various boundary conditions on evacuation process and total evacuation time. Movement characteristics (exit flows, speeds) and human behaviour (pre-movement activities, exiting behaviours) were also analysed.
 The data obtained was used to validate and adjust a Pathfinder model to capture important aspects of evacuation from the railcar. Furthermore, a series of simulations using this model was performed to provide sensitivity analysis of the influence of crowd composition, exit width, and exit type on total evacuation time. As a key finding, we can conclude that for the case of a standard exit path (platform or stairs) the width of the main exit had the greatest impact on total evacuation time, however, crowd composition played the prevailing role in evacuation scenarios involving a jump.
\end{abstract}

\begin{keyword}
Fire Safety \sep
Evacuation \sep
Passenger Train \sep
Movement Characteristics \sep
Evacuation modelling \sep
Sensitivity Analysis
\end{keyword}

\end{frontmatter}

\paragraph{ Highlights }
\begin{itemize}
	\item Controlled evacuation experiment from a double-deck railcar.
    \item Exit width, exit type, and crowd composition varied in 15~trials (30~scenarios).
    \item Total evacuation times, exit flows, travel speeds, 
    jump-induced delays were measured.
    \item Pathfinder simulations were used to validate against the presented experimental dataset.
    \item Sensitivity analysis was used to study the impact of boundary conditions on TET.
\end{itemize}

\newpage


\section{Introduction}
\label{sec:intro}
To achieve expected levels of safety on passenger trains, special attention must be given to fire safety on board. In this context, the successful evacuation of people from a risk area to a place of safety is a crucial issue. The preservation of tenability conditions that allow passengers and crew to exit railcars quickly and easily together with effective evacuation processes are the key instruments for successful evacuations. General regulations related to fire protection should be followed in the whole life cycle of a vehicle, including design, assembly, and operation phases. For example, as noted in various studies \cite{braun_fire_1975,braun_fire_1978,peacock_fire_1984,peacock_fire_1999,peacock_fire_2002,peacock_fire_2004,goransson_fires_1990,briggs_firestarr_2001}, specific requirements for limiting fire and smoke in railcars reduce and control fire hazards. Similarly, national regulations and standards (e.g. TSI LOC\&Pas~1302/2014 \cite{noauthor_commission_2014}, ATOC Standards \cite{noauthor_atoc_2002}) mandate escape routes and exits. 

In any fire safety assessment, adequate knowledge of movement characteristics and aspects of human behaviour are vital for calculating required safe egress times (RSET). The unique environment of passenger train units poses challenging conditions for an egress process. Commonly known and used formulas published in engineering handbooks derived from observations and evacuation research focused on buildings may not be fully appropriate and applicable to railcar settings. First, different estimation methods cause different values for exit flows; second, the widths of aisles and slopes of staircases in a railcar are different than in observations of buildings \cite{markos_passenger_2013}. Therefore, research data describing the specifics of the evacuation process from public transportation vehicles is desirable. Controlled experiments, field experiments, and full-scale evacuation drills may be helpful in extending our understanding of public transportation vehicles, such as railcars, though such studies are expensive and time-consuming. Such methods may also produce datasets which are not ideal; e.g. they may simplify boundary conditions, limit numbers of scenarios, or exhibit an insufficient level of realism. Therefore, such studies should be presented and considered within a full experimental context in order to enable more comprehensive interpretation of results and more accurate analysis of limitations, ideally pointing the way forward for new safety innovations.
Particularly when supported by reliable input data, computer evacuation models are rapidly developing and increasingly useful engineering tools \cite{kuligowski_computer_2016, ronchi_developing_2020}. Moreover, for trained model users, simulated environments may constitute cost-efficient alternatives to standard physical evacuation trials and offer further advantages, such as more diverse options for creating evacuation scenarios without risking the safety of participants \cite{siyam_research_2019}. The key condition for broader adoption of evacuation models remains their ability to realistically simulate the railway operating environment and to validate simulation results with relevant empirical data~\cite{markos_passenger_2013}. The possibility of performing various simulations enables more detailed statistical analysis of the evacuation process investigated. When the simulation model is sufficiently validated against available experimental data (e.g. gained by controlled evacuation trials), the evacuation process can be studied by means of the simulations in more detail by performing simulations for finer sets of boundary conditions then covered by an experiment.
Such extrapolation of experimental trials makes it possible to, in a sufficiently reliable way, study the influence of investigated boundary conditions on total evacuation time by means of sensitivity analysis~\cite{RahDatLov2016FEMTC}. The results of sensitivity analysis may considerably improve the experimental design of subsequent physical trials, thus leading to more detailed and relevant experimental datasets. 

Although being essential for both design and validation processes, comprehensive empirical datasets describing evacuation from trains are currently scarce, when available, and usually focus on certain investigated aspects of train evacuation \cite{huang_experimental_2018,  klingsch_evacuation_2010,kim_experiments_2012,fridolf_fire_2013,fridolf_flow_2014,fridolf_evacuation_2016,cuesta_experimental_2017,noren_modelling_2003, markos_passenger_2015}. This paper aims to add depth to this area of research and describes an original full-scale train evacuation experiment, including a thorough description of measured data and experimental conditions as well as data collection and evaluation methods. The experiment was designed in a controlled environment to study movement conditions in a railcar during evacuation situations, particularly to explore the impact of different crowd compositions, exit widths, and exit types on the evacuation processes inside the railcar represented by the total evacuation time (TET). While most controlled experiments are typically performed with a rather homogeneous group of people without any limitations to their movements (e.g. university students, soldiers, and so on), the experiment described here was primarily focused on the impact of heterogeneity in its group of participants (including children, seniors, or other people with certain movement disabilities) in order to better understand the crowd dynamics of non-standard exit paths typical for emergency train evacuations. Quantitative sensitivity analysis of the influence of different boundary conditions on evacuation pathways deemed to enable deep insight into the train evacuation process. In addition, sensitivity analysis conducted using the experimental datasets was complemented by more refined statistical analysis of simulations performed using the Pathfinder evacuation model~\cite{Pathfinder}. Pathfinder was selected since it is available to the public and because its features and capabilities sufficiently reflected the specifics of simple evacuation scenarios for a unique train environment, including individual perspectives, continuous representation of space and time, and decision-making behaviours. The goals of this paper are to:

\begin{enumerate}[label=\alph*)]
    \item Present a new dataset describing movement conditions in a double-deck rail unit during a full-scale controlled experiment. The experiment provided the parameters needed as inputs for evacuation modelling purposes. 
    \item Evaluate the impact of different boundary conditions on total evacuation time (TET), such as passenger heterogeneity, exit width, and type of exit path.
    \item Validate the Pathfinder model for train evacuation against the experimental dataset presented here. 
    \item Use sensitivity analysis of experimental and simulation data to quantify and compare the influence of variable factors which had the most impact on total evacuation time. 
\end{enumerate}
The remainder of this paper is organised in the following way: Section~2 provides a summary of previous research studies  dealing with evacuation in public transport vehicles. Section~3 presents, in detail, the experimental study of evacuation from a double-deck railcar unit, including setup of the experiment, methods used, results, and discussion. Section~4 shows how the experimental data was used in the Pathfinder evacuation model and for its validation, and describes the sensitivity analysis of the influence of boundary conditions on total evacuation time with support from Pathfinder simulations. Finally, conclusions  are presented in Section~5.
Selected experimental data and simulation results in CSV format are available in~\cite{data}.
%
%
\section{Literature review}
\label{sec:literature}
Railcar environments differ from buildings by having simple but cramped layouts, implying a high concentration of people because of their interior geometries, composed mostly of narrow aisles and a limited number of tight exits. In such vehicles, emergency situations typically occur when vehicles are  underway and when standard egress to platforms is not available. When this happens, passengers have to move to a temporary place of relative safety (e.g. to an adjacent railcar) before a vehicle can be stopped. An emergency stop outside of station areas provided for this purpose may lead to non-standard and increasingly difficult exit paths. Complications may be caused due to a height difference between the track level and the train floor, surface terrain, and extreme stopping point locations such as on a bridge or in a tunnel. Recent research focused on evacuation pathways from public transportation vehicles includes observations, laboratory, and field experiments. 

In order to investigate evacuation times, congestion conditions, evacuation orders, and aisle conflicts typical for train environments, Huang et al. \cite{huang_experimental_2018} carried  out laboratory research focused on evacuation behaviors for railcars having narrow seat aisles (0.4–-0.6~m wide). The movement characteristics of people egressing a railcar (especially during the exit phase) has been observed as a part of projects dealing with evacuation from railway tunnels and subways \cite{klingsch_evacuation_2010,kim_experiments_2012,fridolf_fire_2013,fridolf_flow_2014,fridolf_evacuation_2016,cuesta_experimental_2017}. In these studies, the flow rate of people through exits and its relation to the different kinds of space available at exit points (e.g. limited side or open spaces) were found as some of the most important variables describing the evacuation process. Exit flow rates of passengers getting out of railway vehicles have also been investigated during normal operations and include observation and recording in railway and subway stations in several European metropolises \cite{noren_modelling_2003}. Marcos and Pollard \cite{markos_passenger_2015} provided  detailed results describing movement characteristics related to required egress times from a railcar drawn from different experimental egress trials. Flow rates, egress times, and walking speeds were measured under both normal and emergency lighting conditions and with various exit paths (egress using side door[s] to a high platform/low platform, into an adjacent railcar, or onto an open line, i.e. to the right-of-way, ROW). The exit behaviours of persons with artificially limited movement capabilities were also observed in selected trials. Egress challenges caused by a difference in the level between the exit floor and the surrounding terrain were also discussed by Oswald et al.  \cite{klingsch_evacuation_2010}, who categorised participants into three types (\quotes{Jumper}, \quotes{Sider}, \quotes{Sitter}) according to their manner of overcoming the height of train exits. Furthermore, studies dealing with specific hazardous conditions, such as an overturned railcar or a smoke-filled interior, were performed  \cite{galea_estimating_2000,waldau_full-scale_2007, markos_passenger_2015, fridolf_evacuation_2016}. Table~\ref{tab:rew_flow} presents an overview of results describing exit flow rates available in the literature. 
. \par 

%
{\small
\begin{longtable}{R{1.8cm}|R{2.8cm}|R{10.2cm}}
    \caption{Literature review of exit flow rates observed in trains.} 
    \label{tab:rew_flow} \\

       \hline
       \rowcolor{Gray}
        Reference & Exit flow rate (mean/min--max) [pers/s]  & Background information \\
        \hline
               \multicolumn{3}{l}{\textbf{Egress to the high platform}} \\
        \hline
         \endfirsthead
        \multicolumn{3}{l}{\textit{Continuation of Table \ref{tab:rew_flow}}}\\
        \hline
               \rowcolor{Gray}
        Reference & Exit flow rate (mean/min--max) [pers/s]  & Background information \\
       \hline \endhead
       \hline \endfoot
       \hline \endlastfoot
       
        \multirow{2}{=}{Noren and Winer \cite{noren_modelling_2003}} & 1.588/- & Video recordings at a metro station during normal operation; clear exit width \numprint{1200}~mm; 77~persons observed\\
        \cline{2-3}
        & 0.717/- & Video recordings at a train station during normal operation; clear exit width \numprint{1270}~mm; 119~persons observed \\
        \hline
        Markos and Pollard \cite{markos_passenger_2015} & -/0.85--0.91 & Experimental trials; clear exit width 990~mm; experimental trials under normal lighting conditions; 84~participants without mobility impairments (volunteers from public), exact age range not specified
        (groups \quotes{under~30}, \quotes{30--50}, \quotes{50+})\\ 
        \hline
        \multicolumn{3}{l}{\textbf{Egress using stairs}}\\
        \hline
        Capote et al. \cite{capote_analysis_2012} & 0.57/- & Evacuation trial; rise height of 250~mm (detailed stair geometry not available); exit width 810~mm (effective width 510~mm); 218~participants (further information not available) \\
        \hline
        \multirow{2}{=}{Markos and Pollard \cite{markos_passenger_2015}} & 0.69/0.65--0.73  & Experimental trials; using side-door stairway to a low platform; 5~steps (rise height of 229~mm, the height between the last step and the low platform 380~mm; exit width not available; 17~participants without mobility impairments (staff members), exact age range not specified (groups \quotes{under~30}, \quotes{30--50}, \quotes{50+}) \\
        \cline{2-3}
        & 0.34/0.32--0.35  & Experimental trials using side-door stairway to right-of-way; 5~steps (rise height of 229~mm, the height between the last step and the terrain (640~mm) decreased by using a step box to 410~mm; exit width not available; 15~participants (staff members, one mobility-impaired individual), exact age range not specified (groups \quotes{under~30}, \quotes{30--50}, \quotes{50+})\\
        \hline
        \multicolumn{3}{l}{\textbf{Egress to the open line}}\\ 
        \hline
        Oswald et al. \cite{klingsch_evacuation_2010} & 0.325/- & Experimental trial; distance from the exit to the obstacle 750~mm (clear distance 650~mm), vertical distance from the exit to the terrain \numprint{1150}~mm, exit width \numprint{1300}~mm; 439~participants (staff members ranging 11--60~years, mean age of 40~years) \\
        \hline
        Fridolf et al. \cite{fridolf_flow_2014} & -/0.491--0.536 \newline (VD 700~mm) \newline -/0.228--0.301 \newline (VD 1400~mm)  & Laboratory experiment (small-scale); distance from the exit to the obstacle 850~mm, exit width \numprint{1700}~mm, vertical distance from the exit to the floor (VD) 700~mm or \numprint{1400}~mm; different floor material, lightening, and handles conditions, ladders not used; in total 84~participants (students ranging 18--40~years, mean age of 22.9~years) \\
        \hline
        Markos and Pollard \cite{markos_passenger_2015} & 0.729/- & Evacuation exercise in a smoke-filled tunnel; exit width \numprint{1270}~mm, vertical distance to the lower platform 700~mm; 86 participants without mobility impairments (further information not available) \\
        \hline
        Fridolf et al. \cite{fridolf_evacuation_2016} & -/0.19--0.22  & Evacuation experiment in a smoke-filled metro tunnel; distance from an exit to the tunnel wall \numprint{1000}~mm, exit width \numprint{1400}~mm, vertical distance to the tunnel floor \numprint{1400}~mm, emergency ladders partially used;  135~participants (volunteers from public ranging 19--76~years, mean age of 38~years) \\  
\end{longtable}
}

Concerning passenger train environments, a special challenge is posed by fire safety evacuation modelling. Evacuation computer models originally developed for predicting evacuation processes in buildings and other public spaces (e.g. transit stations) may not be fully capable of simulating the unique conditions of a railcar egress situation. Markos and Pollard \cite{markos_passenger_2013} studied the potential application of evacuation models to anticipate passenger egress times under various emergency conditions. They concluded that the ability of evacuation models to simulate passenger railcar egress situations may be limited due to the absence of the factors representative of passenger trains that have not been experimentally quantified.

Hence, the evacuation models reviewed by Markos and Pollard provide suitable features for designing simple evacuation simulation scenarios (e.g. egress to a high platform or from a railcar end door into an adjacent car) rather than informing non-standard exit paths such as direct egress to the track level or egress from an overturned railcar.

Their study also indicated that building evacuation models mostly lack appropriate means of conducting validation against actual passenger railcar occupant data. Some attempts to use building evacuation models to estimate egress time predictions from passenger trains (Simulex \cite{kangedal_fire_2002}, STEPS \cite{kindler_evacuation_2012}, MassMotion \cite{anvari_toward_2017}) and from rail tunnels (NOMAD \cite{daamen_passengers_2007}, Pahfinder \cite{jevtic_safety_2017}) can be found in the literature. In these applications, the nature of evacuation models (including the lack of features for simulation of the exit to the track level), upper limits for density, and discrete space representation of some models were indicated as the main drawbacks for using such models for modelling train evacuation~\cite{kangedal_fire_2002,kindler_evacuation_2012, anvari_toward_2017}. \par

Separate research has focused on the development of evacuation models designed specifically for railway environments \cite{galea_passenger_2014,alonso_new_2014} and on simulating rail tunnel evacuation \cite{hutchison_agent-based_2009,wang_simulation_2014}. The railEXODUS multi-agent evacuation model has been being developed by the Fire Safety Engineering Group (FSEG) at the University of Greenwich as the rail module for the EXODUS suite of software \cite{galea_passenger_2014,galea_development_2013} to simulate passenger rail car evacuation for various emergency scenarios (including inter-car egress; egress from single and multi-level railcars to high platforms, to low platforms, and to open lines; and egress from a partially overturned railcar). The EvacTrain model is a stochastic, cellular automata model (cell size $0.5\times0.5$~m) presented by the GIDAI Group at the University of Cantabria \cite{alonso_new_2014, peacock_evacuation_2011,capote_stochastic_2012}. The model uses a stochastic approach for representing human behaviour using Monte Carlo methods to simulate the random characteristics of passengers and their delay actions.

When using computer software to simulate evacuation from a specific environment, any model must be provided with key input data in order to sufficiently mimic the real-life environment. It is beneficial to know what parameters and boundary conditions have a prevailing impact on observed quantities. Thus, the influence of geometry on total evacuation time has been investigated in \cite{GreGre2014RDO} using Pathfinder \cite{Pathfinder} simulations and optiSLang \cite{optiSLang} software for sensitivity analysis. The sensitivity of the total evacuation time of a hospital to the ratio of ambulant and non-ambulant patients, for example, was investigated qualitatively in \cite{RahDatLov2016FEMTC}.

Based on the literature review of passenger train evacuation, the following findings can be summarised:
\begin{enumerate}[label=\alph*)]
    \item The environment of railway transport vehicles is unique, and the evacuation process from them depends on many factors including passenger characteristics, railcar geometry, exit configurations, and operating/exiting conditions. 
    \item In most cases, available empirical datasets provide results from full-scale evacuation trials.
    \item Building evacuation models have the ability to simulate simple evacuation scenarios related to passenger train environments if supported and validated using reliable empirical data.
    \item A very limited number of special evacuation models have been (or are being) developed to simulate passenger transport vehicle evacuation conditions and strategies. The availability of these models to the public is currently limited and more data is needed in order to conduct further validation studies. 
    \item To the best of the authors' knowledge, detailed sensitivity analysis studies quantitatively comparing the influence of boundary conditions on evacuation time from a railcar have not yet been presented.
\end{enumerate}
Three research gaps were also identified: first, relevant empirical datasets suitable as inputs for evacuation modelling purposes are scarce and usually refer to exit flow rates. Second, even for simple evacuation scenarios (e.g. when a train is located at a high-platform station), the ability of building evacuation models to reflect the reality of the railcar egress environment must be validated. To facilitate simulation of non-standard exit paths using building evacuation models, experimental data such as transit delays in exits caused by climbing down to the track level are needed.
Third, a quantitative comparison of the influence of exit type and crowd composition on evacuation time with the influence of exit width is needed, since it may reveal problematic configurations of boundary conditions.


\section{Experimental study of evacuation from a double-deck passenger railcar} 

This section presents a dataset collected during a controlled full-scale evacuation experiment using a double-deck electric unit class 471~(CityElefant) trailer car used for passenger service in suburban rail networks. The one-day experiment was the result of a cooperation between Czech Technical University in Prague and the Research Institute of Railway Rolling Stock in Prague in June~2018 and consisted of a series of 15~evacuation trials.
Each evacuation trial was performed under various boundary conditions for a total of 30~evacuation scenarios involving 91~participants (68\% of the seating capacity of the railcar). The experiment aimed to evaluate the influence of the composition of the passenger crowd (heterogeneity), exit width, and exit type on total evacuation time. The study also included an analysis of exit flows, speed of movement of passengers inside the railcar, pre-movement activities, and exiting behaviours in order to understand how passengers exited in relation to the height of a train exit.

\subsection{Experimental setup}
This section describes the experimental layout, participants involved, evacuation scenarios, and experimental procedures. 
\subsubsection{Geometry and layout of the railcar}
\begin{figure}[!htb]
    \centering
    \includegraphics[width = 1.0\linewidth]{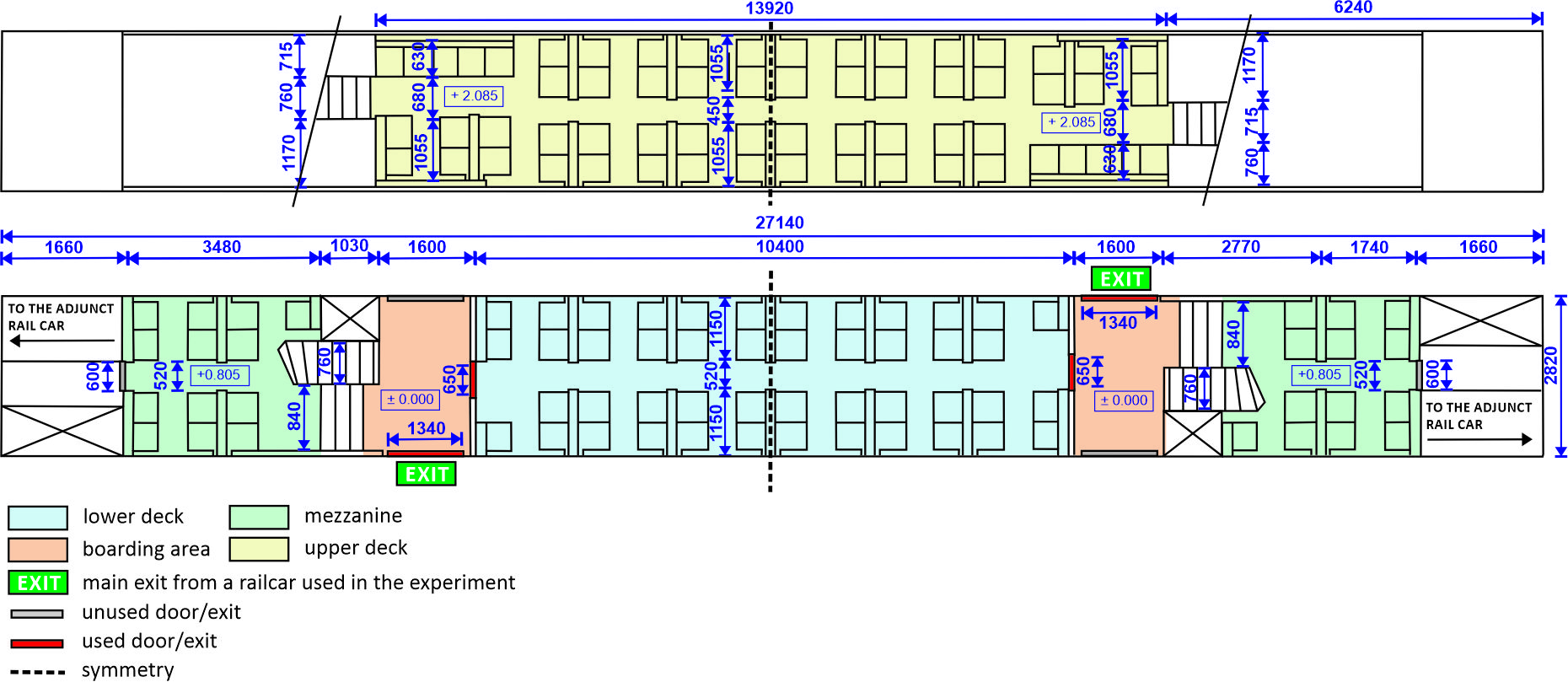}
    \caption{Geometry and layout of the railcar.}
   \label{fig:layout}
\end{figure}

A single unit class 471~(CityElefant) train consists of a power car with first class seating (Class~471), a centre trailer passenger car (Class~071), and a driving trailer (Class~971). In view of its large transport capacity, a Class~071 double-deck centre trailer car was selected for use in this experiment. This railcar has 134~seats (total capacity of 268~passengers) with the following compartments: a lower deck (62~seats), an upper deck (46~seats), and 2~mezzanines (2$\times$13~seats). On the lower deck, 2~boarding areas (the main exits) are located across from one another on each side of the railcar and have double sliding doors (clear width \numprint{1300}~mm). Figure~\ref{fig:layout} shows the layout of the railcar. 

The lower deck is connected to the boarding area by a single sliding door (650~mm clear width). The upper deck leads via a straight staircase (7~steps, 760~mm wide, level difference \numprint{1280}~mm, 36$^\mathrm{\circ}$ slope) to the mezzanine. The mezzanine is connected to the boarding area with a straight flight (840~mm clear width, level difference 805~mm, 40$^\mathrm{\circ}$ slope). From the mezzanines, a single sliding door (600~mm in clear width) enables passage to adjunct railcars. All doors in the railcar open automatically by pressing a button. The main exits were permanently open during this experiment. The vertical distance between the track level and the train floor was 750~mm. 
\subsubsection{Evacuation scenarios and population}
In order to evaluate the impact of various boundary conditions on the evacuation process and on total evacuation time, evacuation scenarios were designed in relation to three main boundary conditions: 
\begin{enumerate}[label=\alph*)]
    \item Crowd composition (heterogeneity),
    \item Width of the main exit,
    \item Type of exit from the railcar.
\end{enumerate}
The influence of boundary conditions was quantitatively studied by means of the variance-based sensitivity analysis (see Section~\ref{sec:sensitivity}). When using this method, it is recommended to cover most of the parametric space to obtain reliable results~\cite{NPLsensitivity, saltelli:2008}. Accordingly, the evacuation trials were conducted for all possible combinations of investigated boundary conditions (5~different widths of main exit and 3 different types of the exit for two different passenger groups), altogether 30~different~evacuation scenarios. Consequently, each trial was conducted only once because of time limitations. Therefore, any conclusions based on the flow or total evacuation time measurements should not be drawn for each scenario separately, but rather from evaluating scenarios with similar conditions as a whole, e.g. from the overall trend captured by the dependence of the observed quantities on exit width.\par

\textbf{Crowd composition (heterogeneity)}\par
Based on our review of prior research, the majority of the experimental evacuation trials from passenger trains to date have been carried out with homogeneous groups such as university students or depot employees who did not have motion impairments or other movement limitations. Passengers with such limitations may considerably impact the evacuation procedures and resulting evacuation times. In order to observe the influence of a diverse population on evacuation, of the 91~participants in the study, 88~participants were able to walk and 3~participants had to be carried. Of the 91~participants, 46~were female and 45~were male, ranging in age from 0.5~to 80~years. Participants were divided into two groups, homogeneous (HOM) and a heterogeneous (HET). 
\begin{itemize}
    \item The homogeneous group (HOM) included participants 18-–38~years of age (mostly students, all in a good physical condition). They represented the most commonly-employed populations for evacuation experiments, according to the literature. 
    \item The composition of the heterogeneous group (HET) was designed to imitate the ordinary suburban railway travelling population. Hence, participants of all age groups were involved, including babies, children, students, seniors, and people with simulated movement disabilities. The composition of this group was based on a combination of statistical data provided by the operator of the train and our own survey of age groups who normally use this type of transportation (details provided in Table~\ref{tab:popu}). 
\end{itemize}
\begin{table}[!htb]
\centering
\small
    \begin{tabular}{R{4cm} R{3,5cm} C{1,5cm} C{2.8cm}} 
        \hline
        \rowcolor{Gray}
         Age group & Movement \newline assumptions & Own \newline survey (\%) & Heterogeneous group (\%)\\
         \hline
         Toddlers (under~2 years) & Carried by others & 2 & 4\\
         \hline
         Children (~3--10 years) & Individual movement under supervision \newline of parents & 8 & 10\\
         \hline
         Adolescents (~11--17 years) & Without limitations, less experienced & 9 & 10\\
         \hline
         Adults (~18--60 years) & Without limitations & 71 & 66\\
         \hline
         Seniors (~60~plus) & Potentially limited movement ability, especially when exiting & 10 & 8\\
         \hline
         Passengers with disabilities & Assistance dependent & 0 & 2$^\mathrm{1)}$\\
         \hline
         \multicolumn{4}{R{11cm}}{\footnotesize{$^\mathrm{1)}$~The passenger with simulated disabilities (1 participant) was equipped with crutches and asked to move without active use of one leg.}}\\
         \hline
    \end{tabular}
    \caption{The composition of the heterogeneous (HET) group.}
    \label{tab:popu}
\end{table}
\noindent\textbf{Width of the main exit}\par
\begin{figure} [!htb]
    \centering
    \includegraphics[width = 1.0\linewidth]{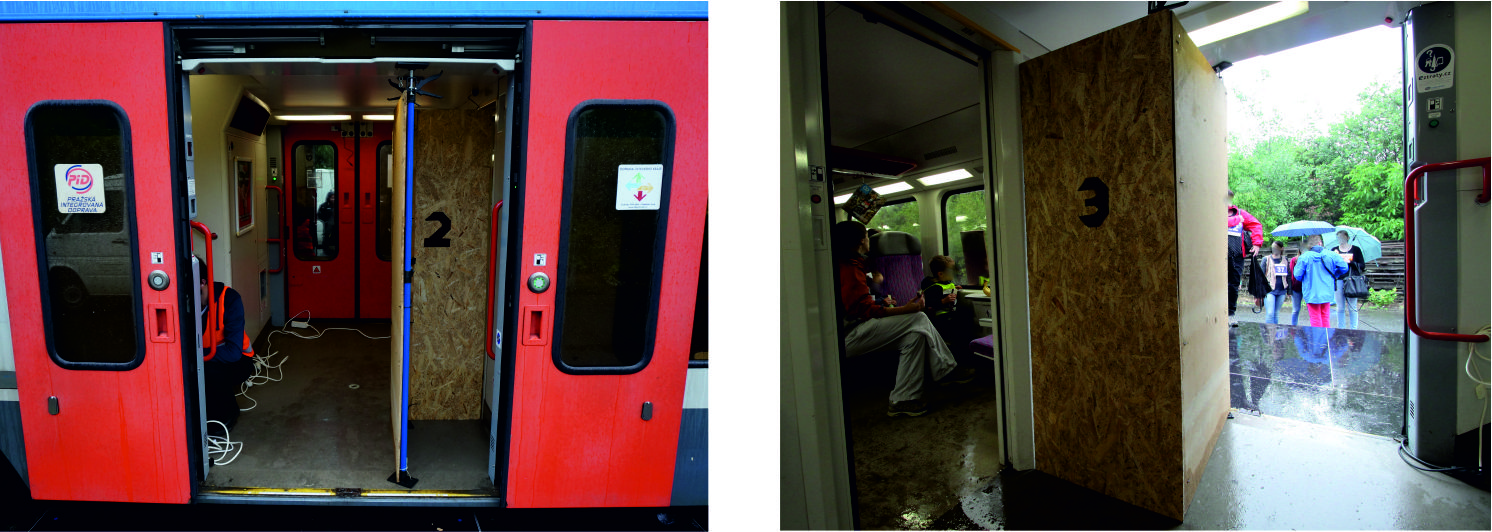}
    \caption{Illustration of different exit/boarding area widths.}
   \label{fig:width}
\end{figure}

Since interior and exit geometries may significantly vary for different types of railcars, 5~values of exit/boarding area width were defined in this experiment: 650~mm, 750~mm, 900~mm, \numprint{1100}~mm and \numprint{1340}~mm. The width of exits in trials was changed randomly in order to limit any possible training effects. The different exit space widths were achieved using barriers in the boarding areas made of rods and boards (OSB), see Figure~\ref{fig:width}. 

\noindent\textbf{Type of exit from the railcar}\par
\begin{figure} [!htb]
    \centering
    \includegraphics[width = 1.0\linewidth]{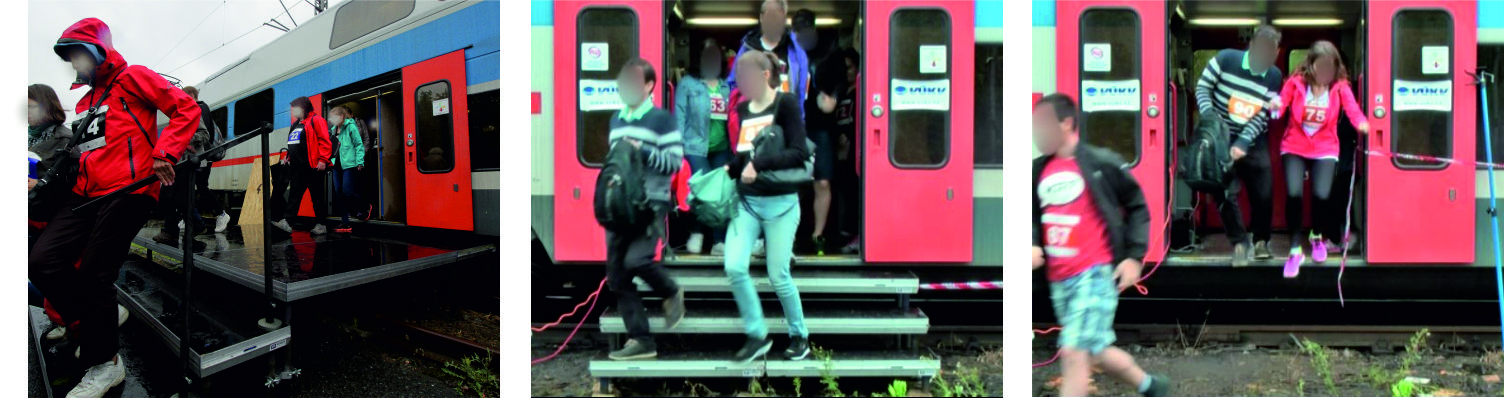}
    \caption{Different exit types considered in the experiment; Left: Egress to a high platform; Middle: Egress to an open line using stairs; Right: Egress to an open line with a drop of 750~mm.}
   \label{fig:exit}
\end{figure}

In order to consider various types of environments where evacuation from a passenger train might be performed, three types of exit from the railcar were emulated in order to consider different emergency scenarios: egress to a high platform, egress to an open line using stairs, and egress to an open line using any devices (details in Table~\ref{tab:exiting}). Figure~\ref{fig:exit} shows the different evacuation paths. %
\begin{table}[!htb]
\small
    \centering
    \begin{tabular}{ C{0.6cm} R{3cm} R{4cm} C{1cm} R{2.4cm}}
        \hline
        \rowcolor{Gray}
        No. & Type of exit & Exit geometry & Drop height & Emulated situation \\
        \hline
        1 & Egress to a high platform & A stage with dimensions of $3\times3$~m ended with 3~stairs equipped with handrails on sides & - & Stopping point in a station\\
        \hline
        2 & Egress to an open line using stairs & 3~stage stairs (rise height of 200~mm, tread depth of 500~mm, slope 22$^\mathrm{\circ}$) & - & Stopping point outside of a station\\
        \hline
        3 & Egress to an open line without any devices & - & 750~mm & Stopping point outside of a station\\
    \end{tabular}
    \caption{Description of experimental exit paths.}
    \label{tab:exiting}
\end{table}
Since the experiment did not take place in a standard train station, a high station platform was 215~replaced with a stage (dimensions $3\times3$~m) located at the level of the railcar's floor (Figure~\ref{fig:exit}, left). To ensure conditions similar to a high platform and to enable continual movement of passengers while they were exiting, the minimum dimensions of the stage were estimated from prior analysis. The auxiliary stairs of the stage were equipped with a handrail to speed up the stage exit process and to avoid the stage having an influence on exit flows. 

\begin{figure} [!htb]
    \centering
    \includegraphics[width = 1.0\linewidth]{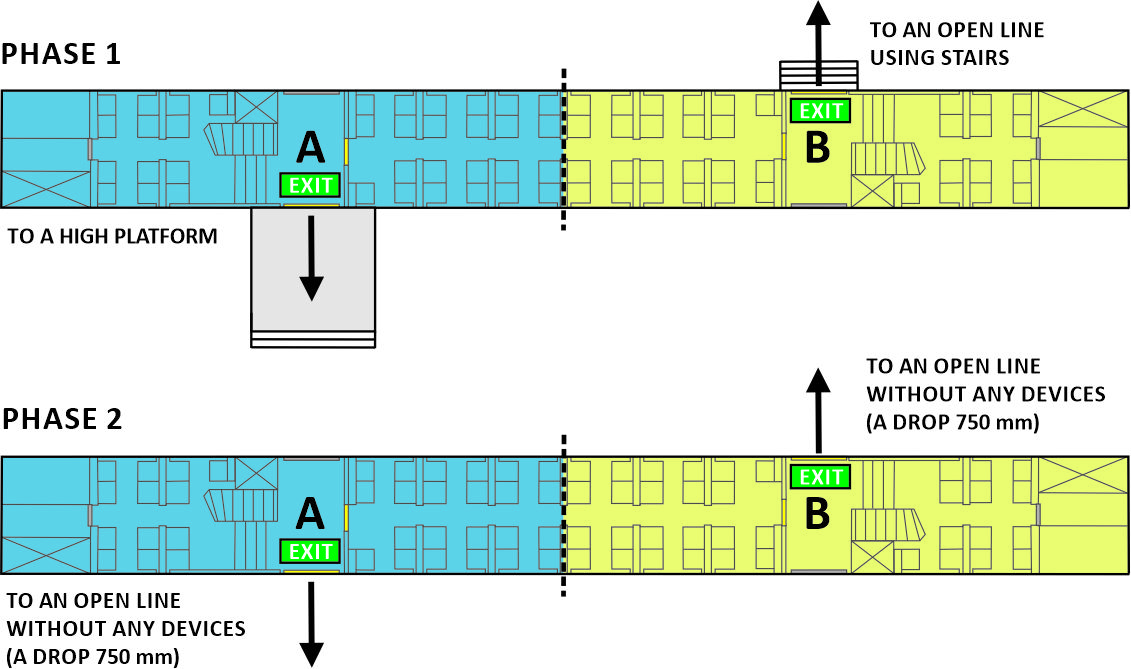}
    \caption{Phases of the experiment considering different types of the exit locations.}
   \label{fig:phases}
\end{figure}

Based on the combination of the characteristics described above, a total of 30~evacuation scenarios were designed. Thanks to the diagonal symmetry of the railcar used in the experiment, it was possible to divide the interior space into identical halves, which allowed two evacuation scenarios to be performed in one trial (see parts A and B of the railcar in Figure~\ref{fig:phases}). All evacuation scenarios were performed in 15~evacuation trials, with the homogeneous (HOM) group located in one half of the train and the heterogeneous (HET) group in the other.
The experiment was divided into two phases (Figure~\ref{fig:phases}): in Phase 1, passengers in part~A of the railcar exited to a high platform while passengers in part~B of the railcar exited via stairs; in Phase~2, all passengers had to navigate a 750~mm drop. In order to minimize the training effect, HET and HOM group locations changed at the beginning of every third trial, and the order of different exit widths was randomly changed (see Table~\ref{tab:scen}). In all trials, the experimental train unit was stopped and the main exits were opened before the evacuation process started. 
\begin{table}[!htb]
    \centering \small
    \begin{tabular}{C{1cm}|c|c|c|c|c|c|c|c|c|c|c|c|c|c|c}
    \hline
    \rowcolor{Gray}
    \multicolumn{16}{c}{\textbf{The homogeneous (HOM) group}}\\
    \hline
    \multicolumn{1}{C{1cm}|}{Exit width} & \multicolumn{5}{C{4cm}|}{High platform} & \multicolumn{5}{C{4cm}|}{Open line using stairs} & \multicolumn{5}{C{4cm}}{Open line without any devices}\\
    \hline
    650 & & 4A & & & & & & & 7B & & & & & 14B & \\
    \hline
    750 & & & & 8A & & & 3B & & & & & & 13A & & \\
    \hline
    900 & 1A & & & & & & & & & 10B & 11B & & & & \\
    \hline
    1100 & & & & & 9A & 2B & & & & & & 12A & & & \\
    \hline
    1340 & & & 5A & & & & & 6B & & & & & & & 15B \\
    \hline
    \multicolumn{16}{c}{}\\
    \hline
    \rowcolor{Gray}
    \multicolumn{16}{c}{\textbf{The hetrogenous (HET) group}}\\
    \hline
    \multicolumn{1}{C{1cm}|}{Exit width} & \multicolumn{5}{C{4cm}|}{High platform} & \multicolumn{5}{C{4cm}|}{Open line using stairs} & \multicolumn{5}{C{4cm}}{Open line without any devices}\\
    \hline
    650 & & 3A & & & & & & & 8B & & & & 13B & & \\
    \hline
    750 & & & & 7A & & & 4B & & & & & & & 14A & \\
    \hline
    900 & 2A & & & & & & & & & 9B & & 12B & & & \\
    \hline
    1100 & & & & & 10A & 1B & & & & & 11A & & & & \\
    \hline
    1340 & & & 6A & & & & & 5B & & & & & & & 15A \\
    \hline
    \end{tabular}
    \caption{Schedule of evacuation scenarios and experimental trials (the number indicates the number of the experimental trial; the letter denotes the part of the railcar where the group was located).}
    \label{tab:scen}
\end{table}
\subsubsection{Instructions given to participants}\label{subsec:instru}

All participants were provided with detailed information about the experiment, including objectives, schedule, safety instructions, and risks, and they signed informed consent forms at an informational meeting organised at CTU in Prague well before the actual experiment. Children under the age of 15~years had to be accompanied by a parent during the experiment. Children under 6~years old were equipped with yellow reflective vests. Each participant received an identification number (pinned visibly on each participant's chest) and a map of the railcar showing its seating plan. Seating locations were not the same for both groups to more adequately reflect real-world conditions and the assumed differences in passenger strategies when selecting seating (e.g. seniors might prefer seating on the lower deck to avoid stairs).
However, the seating position of each participant was recorded and kept unchanged for all evacuation scenarios, i.e. in all evacuation trials, the participants' seating locations were identical. This approach minimized possible variance caused by variations in seating positions.

Before the beginning of the experiment, participants received the following instructions: 
\begin{itemize}
    \item Take your seat in the railcar, store all luggage, belongings, and outer layers of clothing, and wait until the evacuation signal (whistle) is given. 
    \item Leave the vehicle to the open line according to your own possibilities for movement (no competition conditions were evoked).
    \item Make your own decision about what actions to take when leaving your seat. Imagine you are in a real situation in a passenger train (e.g. you can decide whether to assemble your belongings or not). 
    \item If needed, put on extra layers of clothing inside the railcar before the evacuation trial begins. 
    \item Take utmost care for your own safety and health; you can choose to assist others or not. 
    \item After leaving the railcar, do not block the movement of other participants. 
    \item After the trial is finished, pay attention to further instructions and return to your seat in the railcar. 
\end{itemize}
\subsection{Data collection methods}
\begin{figure} [!htb]
    \centering
    \includegraphics[width = 1.0\linewidth]{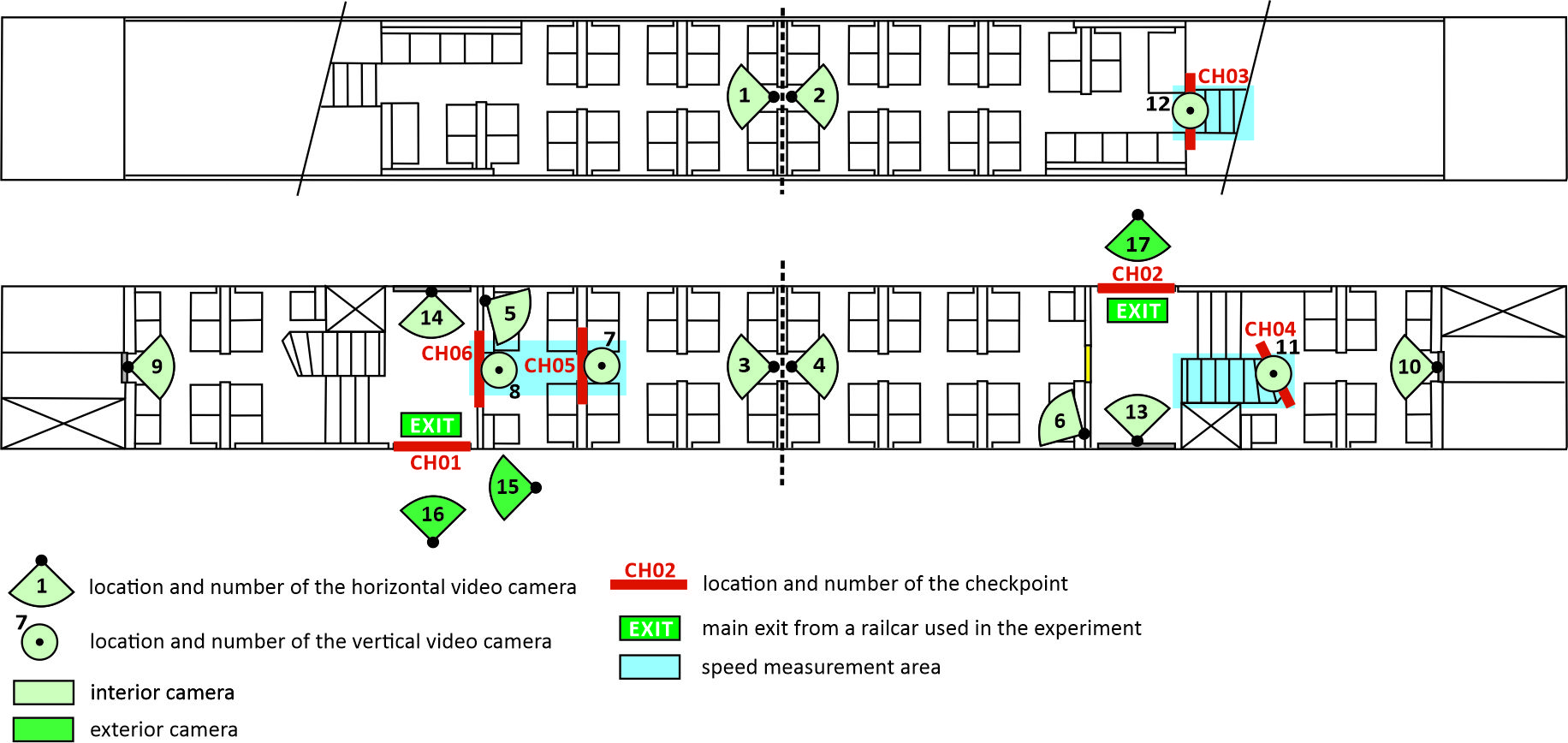}
    \caption{Location of video cameras and predetermined checkpoints in the railcar.}
   \label{fig:cam}
\end{figure}

During the experiment, data was collected using 17~outdoor digital video cameras (resolution 720$\times$480, frame rate 30~fps): 14~video cameras were placed inside the railcar in order to analyse passenger behaviour, flow, and speed; 3~video cameras were placed on the exterior of the railcar to further observe participants’ exiting behaviours and to measure the total evacuation time for each trial. Video cameras were placed so that they captured predetermined checkpoints used for further evaluation of travel speeds and flows. Video recordings were analysed manually (frame by frame) in a video editor; extracted data was transcribed by hand and organised in spreadsheets.
Locations of the video cameras and predetermined checkpoints are provided in Figure~\ref{fig:cam}. 

\subsection{Results and discussion}
The following variables were collected during the experiment and are presented in this section: 
\begin{itemize}
    \item Total evacuation times,
    \item Exit flow rates,
    \item Travel speeds in the aisle and on internal stairs,
    \item Pre-movement activities,
    \item Exiting behaviour (how the height difference at the train exit was overcome and what delay resulted from the associated jump).
\end{itemize}
This section also discusses undesirable effects on the evacuation process and the consequent limitations of this study. 
\subsubsection{Undesirable effects of the evacuation process on results}
\label{sec:undesirable}
During the experiment and the process of data analysis, some issues arose which may have a considerable impact on the results obtained in this experiment and these should be mentioned in order to understand the full context of data presented. First, due to incorrect communication during on-site preparation for the experiment, the number of participants in the homogeneous group (HOM) and the heterogeneous group (HET) was not equal. The HOM group hadd 42~participants and the HET group 46~participants. This 4-participant difference cannot be neglected and should be taken into account when comparing total evacuation times for these groups.

Besides these unplanned circumstances, the strongest impact on the observed evacuation process had the limited time schedule for this large-scale experiment, which led to a repetition of a number of trials in a short time period. Consequently, an undesirable training effect appeared due to gradual streamlining of the evacuation procedure. An attempt was made to counteract this effect by changing exit widths in an irregular order, but the types of exit could not be changed in such a flexible manner. Hence, the most realistic trials in this study were the first trials for each kind of exit type (i.e. 1A, 1B, 2A, 2B, 11A, 11B; see Table~\ref{tab:scen}).
The impact of the training effect was apparent for trial 11A -- 11$^\mathrm{th}$ HET group trial, the first trial requiring participants to jump to an open line. At the beginning of this trial, a group of 3 children expecting a platform or stairs ran to the main exit and, surprised by the unexpected drop, blocked the door for approximately 5~s, until an adult passenger squeezed past them and helped the children down. Since in the first trials the children stayed rather close to their parents, this probably would not have happened if there had not been a preliminary evacuation drill.
Another considerable delay was observed in trial 2A. In this trial, the HET group missed the beginning of the evacuation signal and were delayed in starting their evacuation by 5~s.
This kind of delay belongs to the intervals of TET which were not subject to analysis in this study (recognition in the pre-evacuation phase, see Section~\ref{sec:TET}), and thus this delay was considered as an undesirable effect.

Since this paper aims to discuss and analyze the influence of boundary conditions on total evacuation time, the data needed to be cleaned from the aforementioned undesirable effects. This is particularly important because there was only one trial per each scenario.
First, the egress times for 42~participants in the homogeneous group (HOM) were extended by egress times for 4~artificial passengers in order to arrive at total evacuation time estimates for 46~passengers. The additional times were obtained using linear extrapolation of the egress times for the last 7~passengers in the HOM group, following the assumption that 4~additional passengers would have exited in a similar manner as the 7~participants before them. 
This assumption was supported by time-headway analyses~\cite{BukPesNaj_tgf2019}.
This method of adding artificial times to the HOM group was preferred over the option of discarding four observations from the HET group, because in every trial, participants were egressing in different orders. Thus, in our opinion, deleting egress times for different passenger types in each trial would cause more significant bias than adding four passengers to the homogeneous group. In the remainder of this paper, measured evacuation times for passengers are marked in graphs as full circles $\bullet$, the estimated (artificial) evacuation times are marked by empty squares $\square$ (see e.g. Figures~\ref{fig:TET1}-\ref{fig:TET03}). The estimated total evacuation time for 46~passengers is shown as TET~46. A similar method for recalculating total evacuation times was also used for another case, because 4~participants from the heterogeneous group (HET) did not take part in 2~trials, 11A and 12B.
This TET extrapolation was accompanied by the correction of 5~s delays in the two cases described above (trials 11A and 2A). These corrections are substantial when comparing individual trials; when applied, a corrected total evacuation time is referred to as \quotes{“TET corr”}.

\subsubsection{Total evacuation time}
\label{sec:TET}
\begin{figure} [!htb]
    \centering
    \includegraphics[width =0.8 \linewidth]{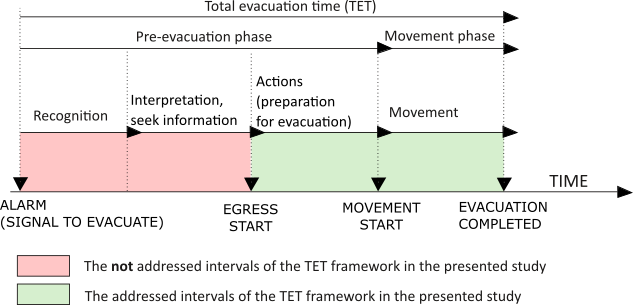}
    \caption{{Intervals of TET considered in the experiment.}}
   \label{fig:rset}
\end{figure}

\quotes{“Total evacuation time”} is defined as the time interval between the signal to evacuate (whistle sound) and the moment when the last person in the group left the railcar through an exit (checkpoints CH01 and CH02 in Figure~\ref{fig:cam}). Since the experiment primarily aimed at investigating movement conditions in the controlled environment and because participants received detailed information about each trial in advance, certain time intervals usually included in the pre-evacuation phase of the TET framework were not addressed in this study (see Figure~\ref{fig:rset}).

\begin{figure} [!htb]
    \centering
    \includegraphics[width = 0.9\linewidth]{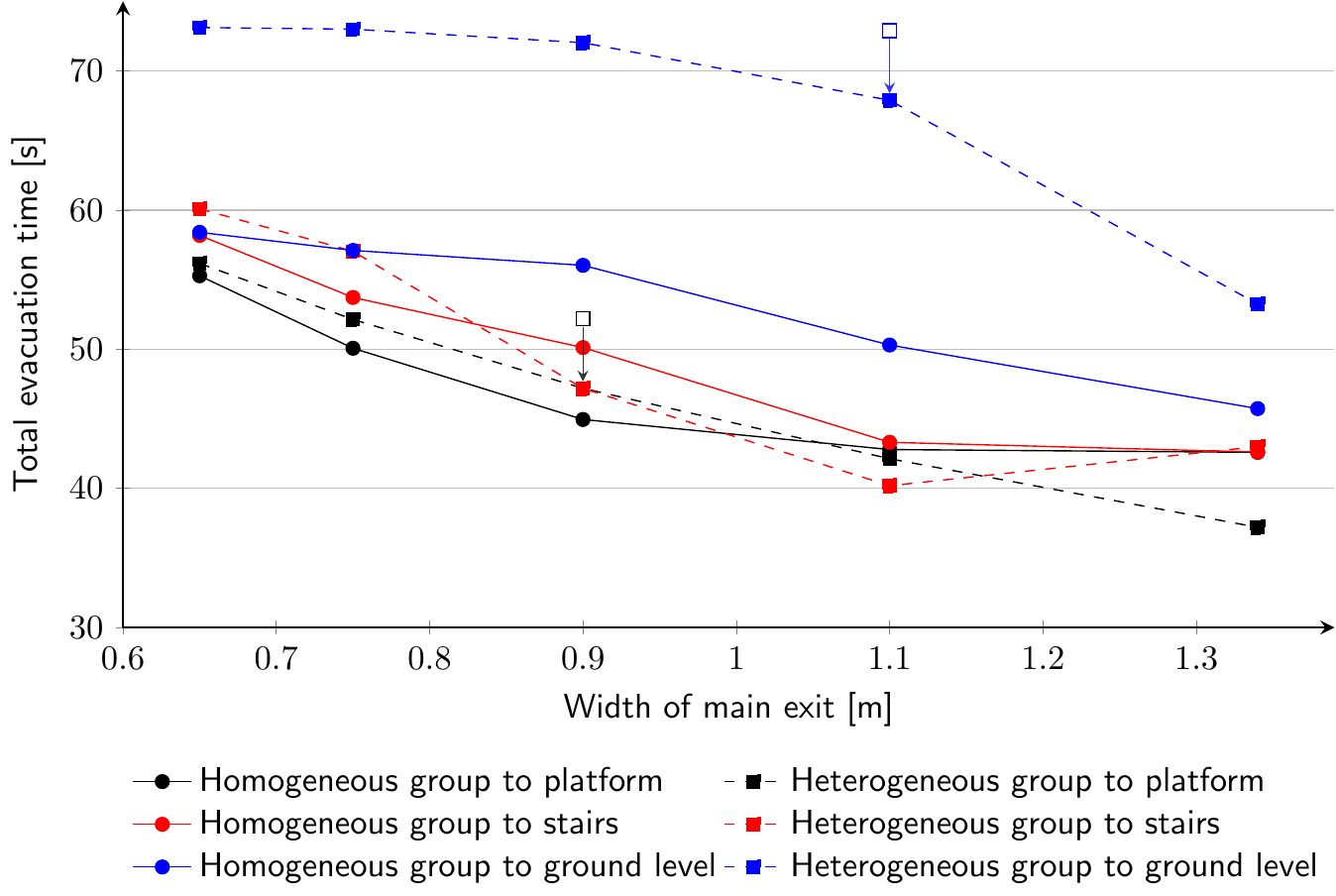}
    \caption{Total evacuation time with respect to main exit width for individual passenger groups and exit types: to a high platform (platform), to an open line using stairs (stairs), and to an open line without any devices (ground level). For comparison reasons, the estimated time for TET~46 is plotted. The delay correction to TET corr in trials 2A and 11A is shown with arrows.}
   \label{fig:TET}
\end{figure}

Total evacuation times for individual trials are given in Table~\ref{tab:TETcorretions} (\ref{app:TET}). Figure~\ref{fig:TET} shows the relation between total evacuation time and the variable conditions. For comparison, the transformed time TET~46 is presented and the corrections to TET~corr are shown with arrows. 

\begin{figure} [!htb]
    \centering
    \includegraphics[width = .48\linewidth]{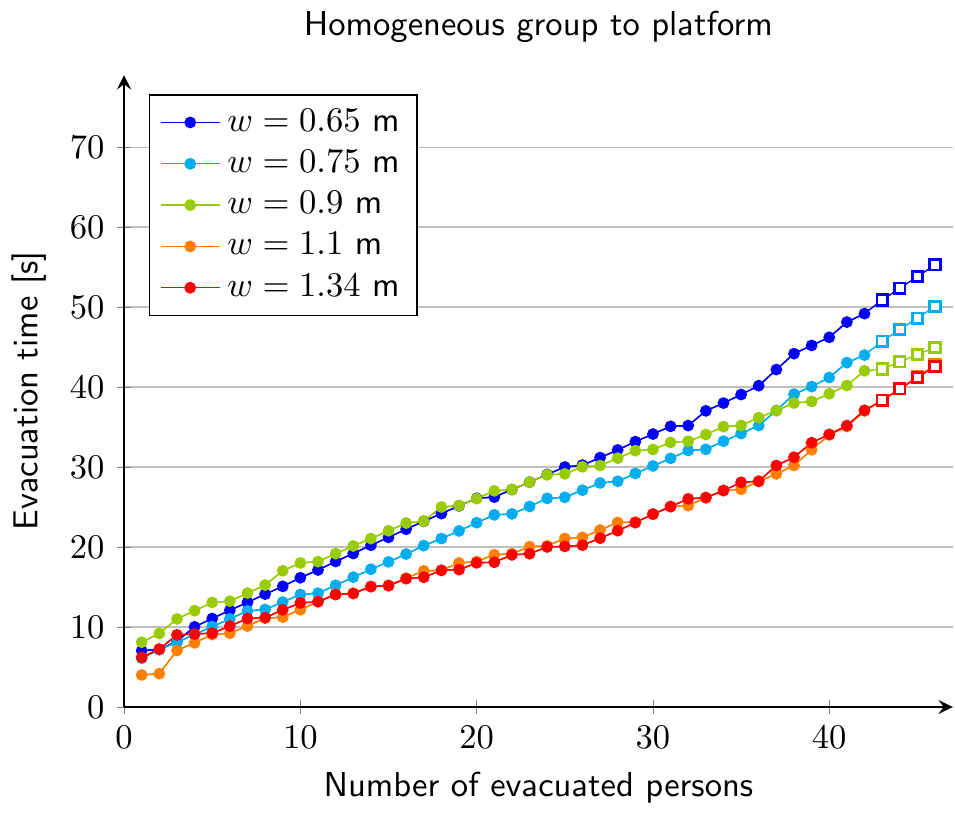}
    \hfill
    \includegraphics[width = .48\linewidth]{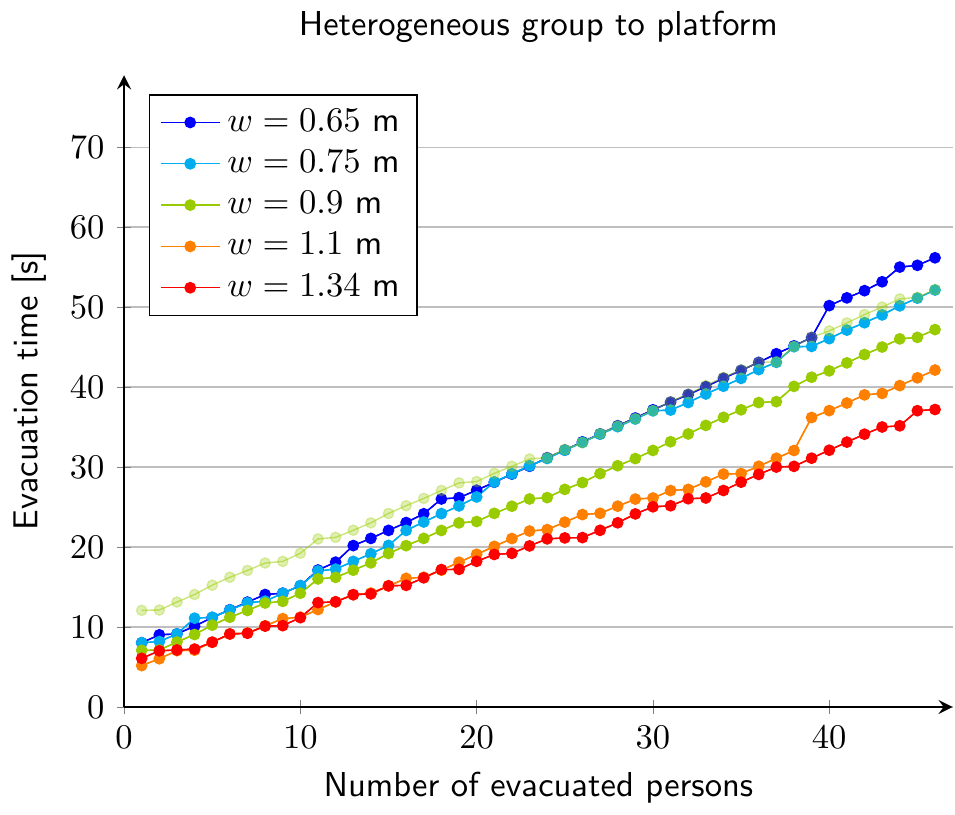}
    \caption{Occupant-evacuation curves for trials with exit to a platform. The original data for the $w=0.9$~m HET group are in pale green; full green shows the curve shifted by 5~s. Empty squares indicate estimated values.}
   \label{fig:TET1}
\end{figure}

\begin{figure} [!htb]
    \centering
    \includegraphics[width = .48\linewidth]{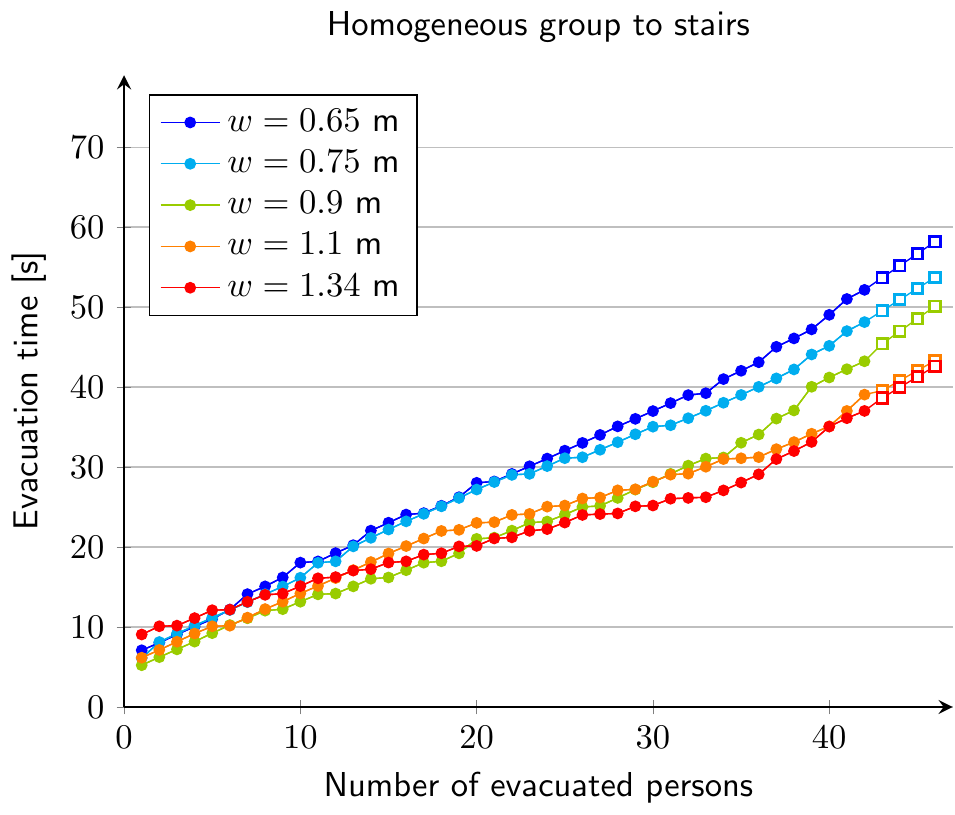}
    \hfill
    \includegraphics[width = .48\linewidth]{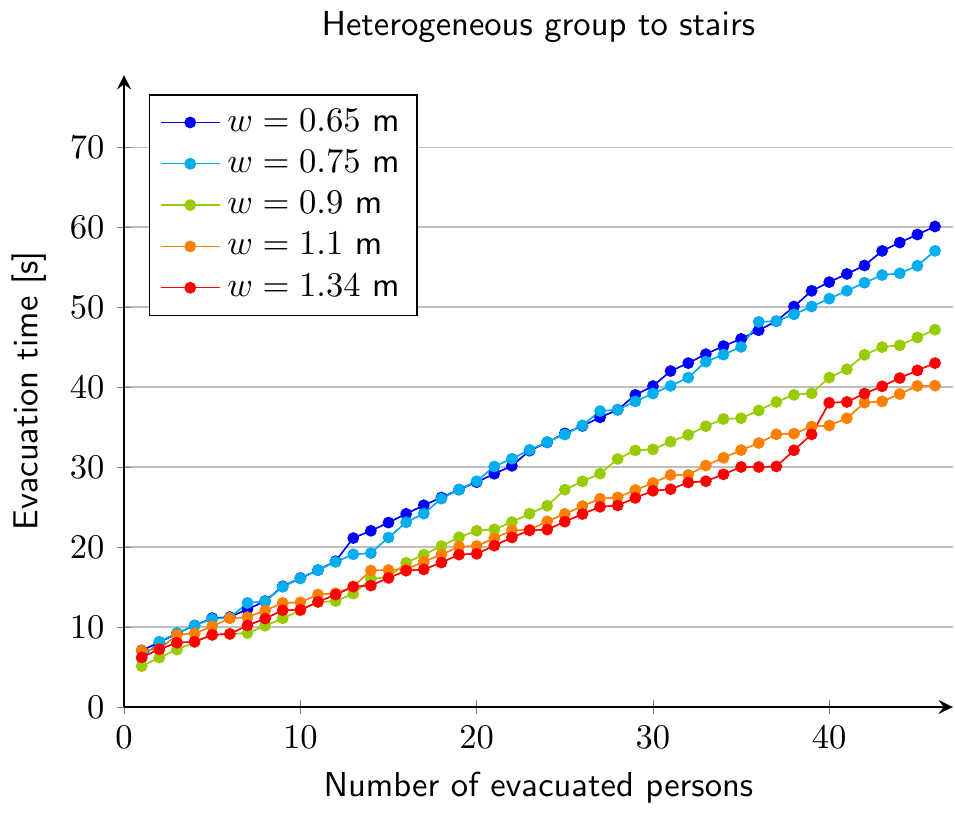}
    \caption{Occupant-evacuation curves for trials with exit using stairs. Empty squares indicate estimated values.}
   \label{fig:TET02}
\end{figure}

\begin{figure} [!htb]
    \centering
    \includegraphics[width = .48\linewidth]{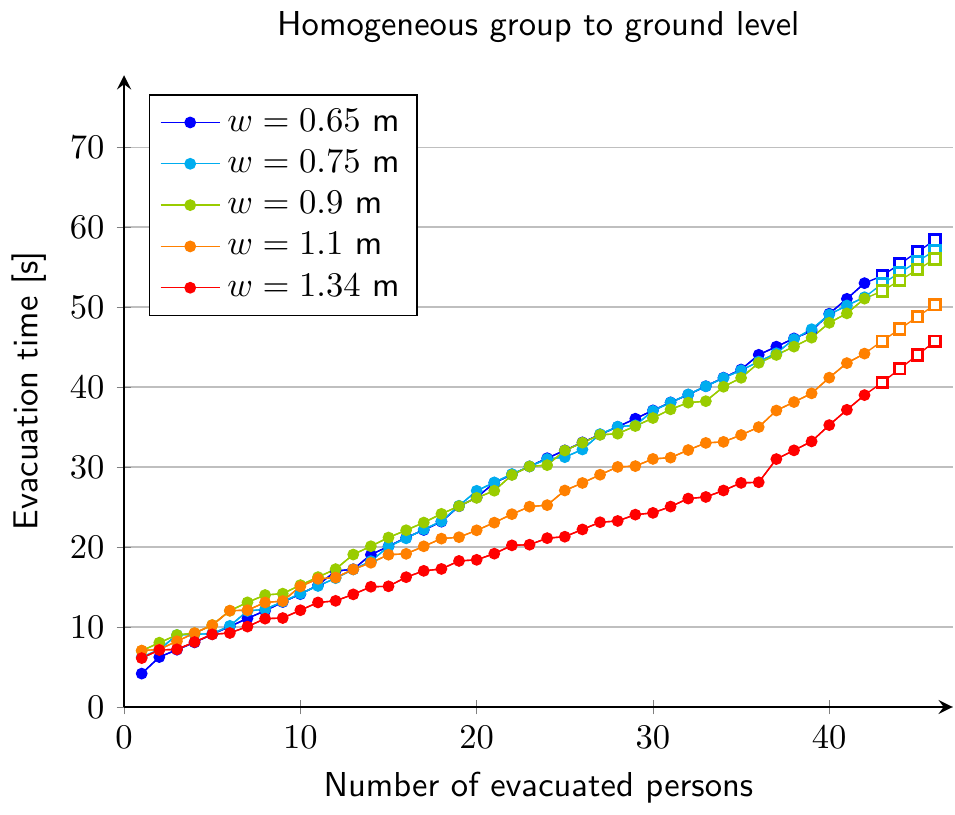}
    \hfill
    \includegraphics[width = .48\linewidth]{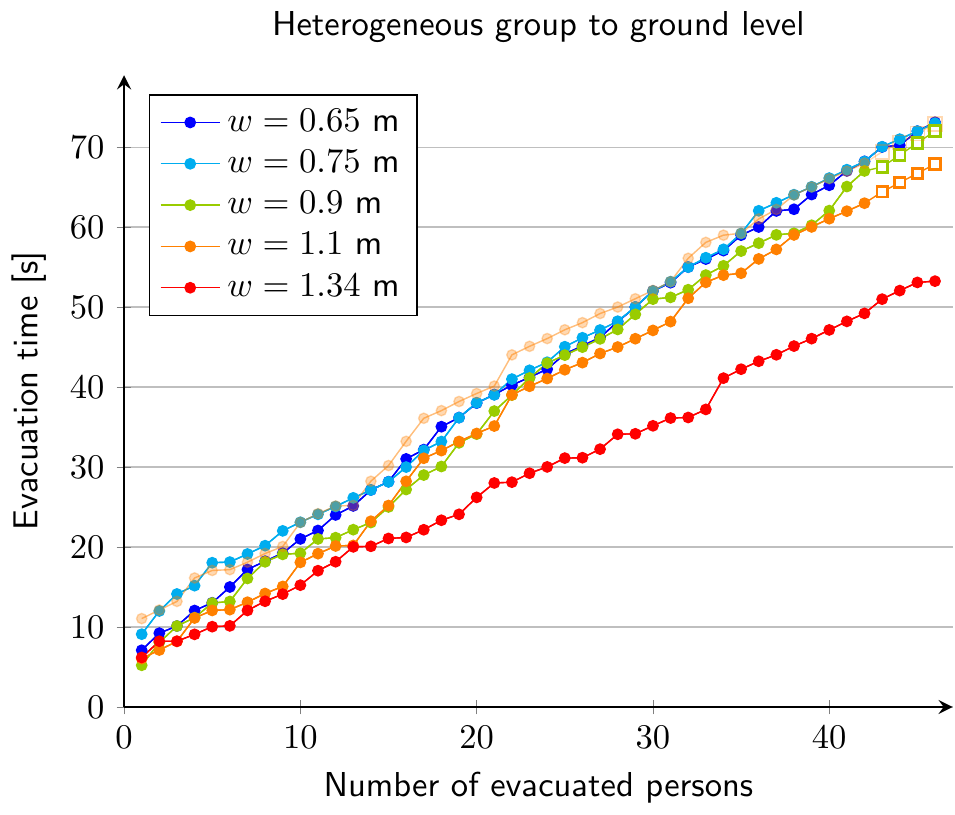}
    \caption{ Occupant-evacuation curves for trials with exit to an open line without any evacuation aid device. The original data for the $w=1.1$~m HET group are in pale orange; full orange shows the curve shifted by 5 s. Empty squares indicate estimated values.}
   \label{fig:TET03}
\end{figure}

Looking at Figure~\ref{fig:TET}, it is apparent that total evacuation time decreased almost linearly with increasing exit width. The greatest difference in total evacuation time for the two groups (HOM and HET) occurred during egress to an open line without any devices, when participants had to navigate a 750~mm drop (the average time difference between the HET and HOM group was 12.8~s).
Considering other exit types, the differences between corresponding total evacuation times for the HOM and HET groups seem to be negligible (average time difference 1.2~s) and no general relation between total evacuation times for these groups can be observed; in some trials with exit to a high platform or using stairs, the HET group evacuated more quickly than the HOM group, and vice versa.
The graphs describing the occupant-evacuation curves for each trial and population group are shown in Figures~\ref{fig:TET1}--\ref{fig:TET03}. 


Based on video analysis of the evacuation process, critical evacuation moments occurred at places where flows from the upper and lower deck merged, namely in the mezzanines and in the boarding/exit areas. Dissimilarities in zipping effects, which directly affected total evacuation time, were evident, particularly when the width of an exit (and simultaneously of the boarding area) changed. 

\subsubsection{Exit flow rates}
\begin{figure} [!htb]
    \centering
    \includegraphics[width = .48\linewidth]{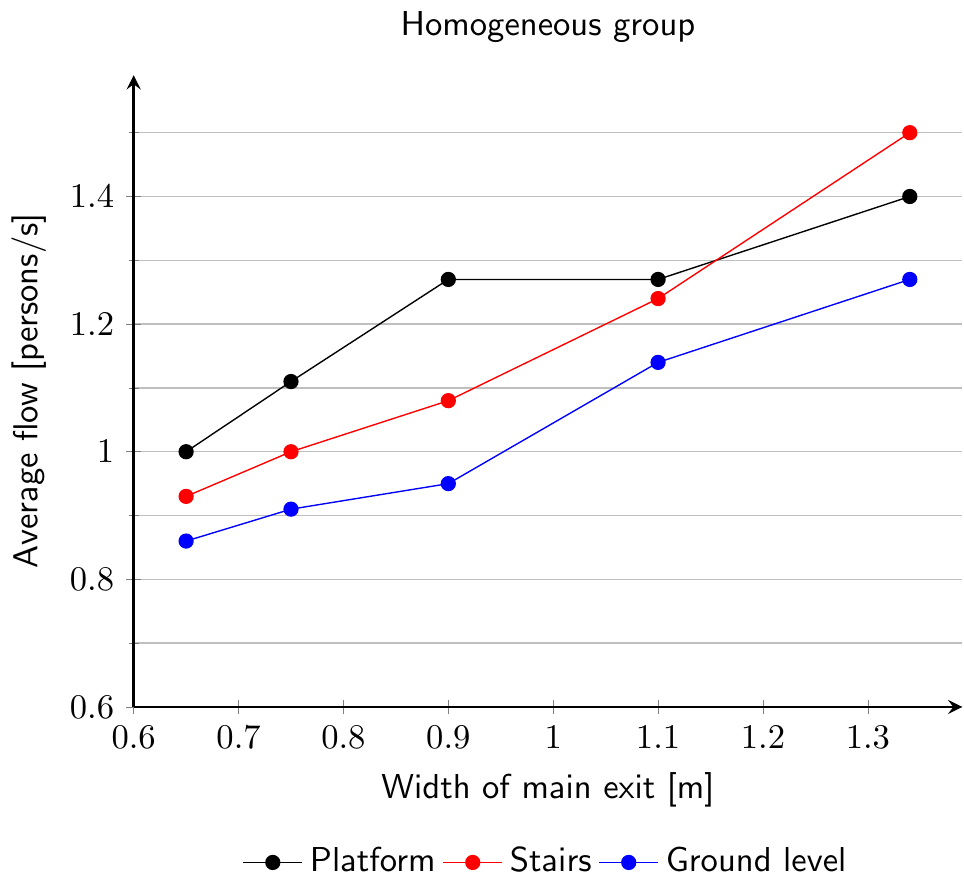}
    \hfill
    \includegraphics[width = .48\linewidth]{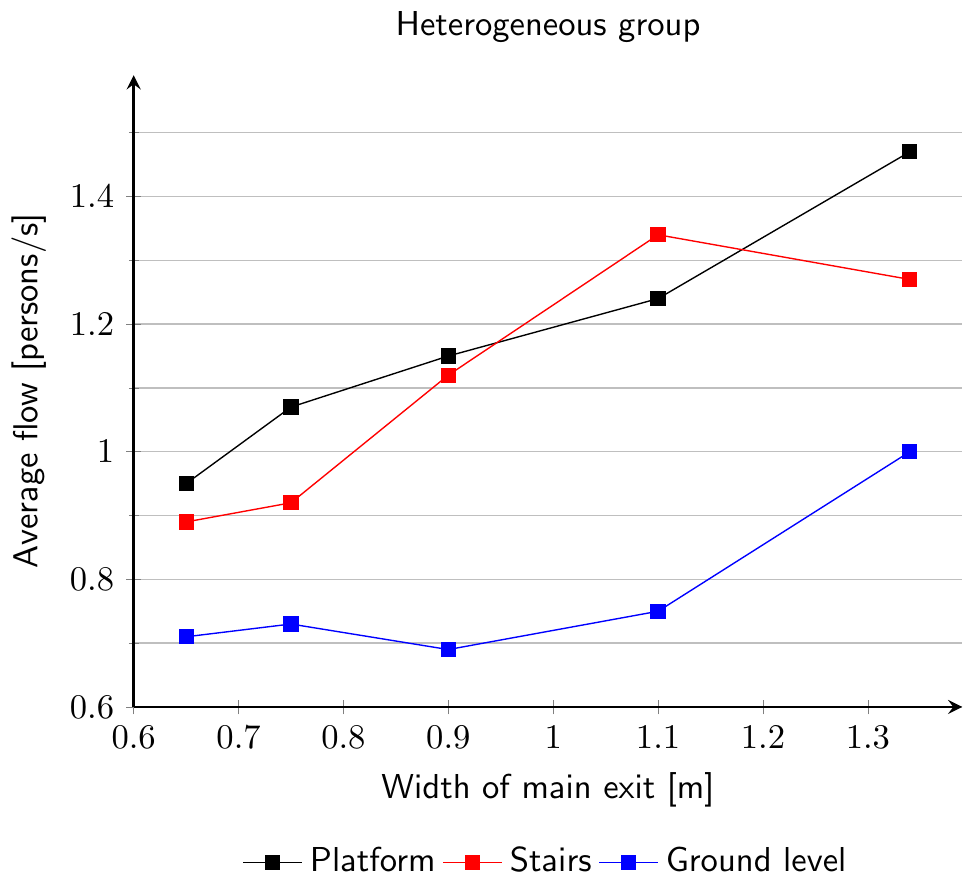}
    \caption{Average flow through the main exit compared to the width of the main exit for individual passenger groups and exit types; to a high platform (platform), to an open line using stairs (stairs), and to an open line without any devices (ground level).}
   \label{fig:flow}
\end{figure}

Flow rates for the main exits were measured during the experiment. Exit flow was calculated as an average value for all participants in a group; the time considered was determined to be the interval between the first and the last participant passing through an exit. Figure~\ref{fig:flow} shows the dependence of exit flow on the variable conditions for both population groups. \par
The linear relationship between exit flow and exit width can be identified for all existing paths. The exit width higher than \numprint{1100}~mm enabled simultaneous exit by two individuals side-by-side. If the exit width was less than 900~mm, participants exited in staggered manner (shoulder-to-shoulder) or in a single line, which explains the linear increase in exit flow with increasing exit width, which is in agreement with~\cite{SeyPasSteBolRupKli2009TS}.
The measurements follow the general assumption that the average flow of individuals should increase with wider exits and with more accessible exit locations.
However, in two cases here, the flow at a certain width was higher for egress to the open line using stairs than exit to a high platform. because this happened for different exit widths for the HOM and HET groups and given the fact that in both cases the situation was related to the deviation of the measurement from the linear trend, this may be considered as a result of the variation in people's behaviour.

Considering the curves associated with the HOM group (Figure~\ref{fig:flow}, left), the curve describing egress to an open line with a 750~mm drop was similar to the other two exit option graphs. A more salient decrease in flow rates can be observed for the HET group and the 750~mm drop (Figure~\ref{fig:flow}, right). This supports the conclusion that navigating a drop is considerably more challenging for a mixed population. For the HET group exiting to an open line, the flow rates were quite similar (around 0.71-–0.75~pers/s) up until an exit width of \numprint{1100}~mm. A sharp increase in flow rate occurred in the trial where a regular exit width (\numprint{1340}~mm) was used (1.00~pers/s). 

Overall, minimum values of flow rates were observed for both population groups at an exit width of 650~mm (roughly the same as a single leaf door), ranging from 0.89-–1.00~pers/s when participants eggressed to the high platform and via stairs, and ranging from 0.86–-0.71~pers/s when a drop had to be navigated.  Maximum flow rates were measured for exits through the largest exit (\numprint{1340}~mm wide, like a double leaf door). For the largest exit, values were similar for both population groups only when the high platform or stairs exit option were available, with values ranging from 1.27-–1.5~pers/s. The flow rate for the HET group when jumping 750~mm through the full width exit was 1.00~pers/s, equal to the flow of the HET group using the 650~mm exit to the platform. Table~\ref{tab:flow} displays results for each trial and population group.

\begin{table}[!htb]
    \centering \small
    \begin{tabular}{C{2cm}|c|c|c|c|c|c}
    \hline
     & \multicolumn{2}{C{3,2cm}|}{High platform} & \multicolumn{2}{C{3,2cm}|}{Open line using stairs} & \multicolumn{2}{C{3,2cm}}{Open line (a drop of 750~mm)}\\
    \cline {2-7}
    \multirow{-2}{\linewidth}{Exit width [mm]} & HOM & HET & HOM & HET & HOM & HET\\
    \hline
    650 & 1.00 (60) & 0.98 (59) & 0.93 (56) & 0.89 (53) & 0.86 (51) & 0.71 (43)\\
    \hline
    750 & 1.11 (66) & 1.07 (64) & 1.00 (60) & 0.92 (55) & 0.91 (55) & 0.73 (44)\\
    \hline
    900 & \textbf{1.27 (76)} & \textbf{1.15 (69)} & 1.08 (65) & 1.12 (67) & \textbf{0.95 (57)} & 0.69 (42)\\
    \hline
    1100 & 1.27 (76) & 1.24 (74) & \textbf{1.24 (74)} & \textbf{1.34 (81)} & 1.14 (68) & \textbf{0.75 (45)}\\
    \hline
    1340 & 1.40 (84) & 1.47 (88) & 1.50 (90) & 1.27 (76) & 1.27 (76) & 1.00 (60)\\
    \hline
    \multicolumn{7}{l}{\footnotesize{Bold numbers represent the first trial for each population group considering a particular exit type}}\\
    \hline
    \end{tabular}
    \caption{Average exit flows [pers/s (pers/min)].}
    \label{tab:flow}
\end{table}
Even though the number of research studies focused on exit flow rates in train environments is limited \cite{ klingsch_evacuation_2010,fridolf_flow_2014, fridolf_evacuation_2016, noren_modelling_2003,markos_passenger_2015, capote_analysis_2012} and the datasets provided vary considerably in terms of experimental methods and setups, the results presented here can still be assessed against the values available in literature (see Table~\ref{tab:rew_flow}).
Considering a high platform exit path, Markos and Pollard \cite{markos_passenger_2015} provided experimental results of flow rates through side doors (990~mm clear width) ranging from 0.85–-0.91~pers/s. Flow rates for a metro station (1.588~pers/m through an exit \numprint{1200}~mm wide) and a train station (0.717~pers/m through an exit \numprint{1270}~mm wide) during normal operation were presented in Noren and Winer \cite{noren_modelling_2003}. Looking at the values measured in this study, the flow rates through exits with similar widths (900~mm and \numprint{1100}~mm) were higher than the results presented in \cite{markos_passenger_2015}. The difference might be attributed to different motivations for participants in the different studies. Comparing the obtained experimental flow rates here to the observations made during normal operations in prior research, our results do not match the observational measurements from prior studies. However, the impact of different environments, data collection methods, motivation of participants, and the small number of datasets compared here limit comparison. The same holds true when comparing flow rates from a railcar using side stairs. The quite consistent values reported by Markos and Pollard  \cite{markos_passenger_2015} and Capote et al.  \cite{capote_analysis_2012} (ranging 0.57-–0.73~pers/s) correspond to a vehicle and staircase geometry different from the railcar used in our study. In both cases, the effective width of the exit was lower (510~mm in \cite{capote_analysis_2012}), the rise height was higher (250~mm \cite{capote_analysis_2012} and 229~mm \cite{markos_passenger_2015}), and the staircase was steeper (i.e. stairs were built into the railcar)  than in this experiment.. In both of those studies, a specific gap between the last step and the terrain (380~mm \cite{markos_passenger_2015} and not determined \cite{capote_analysis_2012}) had to be navigated to reach ground level.

The staircase used in our experiment was not as challenging for participants and the flow rates are similar to the flow rates observed for exiting to a high platform. For the drop of 750~mm in our study, the measured flow rates through the maximum width (\numprint{1340}~mm) exit were roughly twice the valuess in studies performed in simulated tunnel conditions \cite{klingsch_evacuation_2010, fridolf_flow_2014,markos_passenger_2015,fridolf_evacuation_2016}. This is perhaps attributable to the different spaces available at the point of exit in the different studies (e.g. limited side or open space exits) and appears to match the conclusion in \cite{klingsch_evacuation_2010}, that limited space along a pathway in a simulated tunnel layout caused a 50\% decrease in exit flow rate, i.e. the flow rates were half the values when escaping to open/free spaces (unfortunately, the values of flow rates to open spaces were not specified in that study \cite{klingsch_evacuation_2010}). 

\noindent\subsubsection {Travel speeds in the aisle and on the internal staircase}
\label{sec:aislespeed}
In our study, the participants' travel speeds were measured for two segments of space in the railcar (see Figure~\ref{fig:cam} for graphical representations): in the aisle on the lower deck (CH05-CH06, distance \numprint{1830}~mm, width 520~mm), and on the straight staircase connecting the upper deck and the mezzanine (CH03-CH04, distance in slope \numprint{1880}~mm, width 760~mm). Travel speeds were calculated individually for each participant as the ratio of distance travelled to the corresponding time of travel. The distance on the staircase was measured as the inclined length of the flight. \par
\noindent\textbf{Travel speeds in the aisle}\par
Due to the limited space layout in the railcar, the movement of participants was considerably influenced by the high density of people in the railcar and the formation of queues in the aisle. Hence, an evaluation of continuous movement and an estimation of unimpeded (free) travel speeds were not possible. The main disturbing factors were a time delay when opening the sliding door leading to the boarding area (4~s), limited space available for accessing/merging in the aisle, and merging flows coming from the upper and lower decks to the boarding area. Due to technical problems with video recording during the experiment, only a limited amount of relevant data was collected in this segment of the trials. Table~\ref{tab:speed} shows travel speeds, and the average travel speed based on all observations was 0.35~m/s. \par
\begin{table}[!htb]
    \centering 
    \small
    \begin{tabular}{C{2.5cm}|C{4cm}|C{4cm}}
    \hline
    \rowcolor{Gray}
    \multirow{2}{*}{Exit width [mm]} & \multicolumn{2}{c}{Speed (mean/min/max/SD) [m/s] (data-points)} \\
    \cline{2-3}
    & HOM & HET \\
    \hline
    750 & 0.54/0.30/0.92/0.19 (14)  & 0.18/0.12/0.23/0.03 (13) \\
    \hline
    1100 & 0.43/0.21/0.83/0.19 (14) &  0.23/0.09/0.37/0.08 (13) \\
    \hline
    1340 & - &  0.34/0.26/0.47/0.05 (13)\\ 
    \end{tabular}
    \caption{Travel speed measured in the aisle.}
    \label{tab:speed}
\end{table}
\noindent\textbf{Travel speeds on the internal staircase}\par
The cramped railcar layout with a high concentration of participants in the interior of the railcar did not allow unimpeded (free) travel speeds to be measured during all trials. The unrestricted travel speeds for the internal staircase were measured only in a few evacuation trials, when flow from the upper deck was not limited by participants gathering in the mezzanine. For measurable cases, the average unimpeded speed of adults was 0.96~m/s. As a result of flows merging  in the boarding area, a queue was formed on the observed staircase, and the resulting average travel speed based on all trials was 0.41~m/s. Table~\ref{tab:speedst} shows the travel speeds for the staircase during egress to an open line, and Table~\ref{tab:speedstjump}, travel speeds for the staircase during egress to an open line without any equipment. \par 
\begin{table}[!htb]
    \centering 
    \small
    \begin{tabular}{C{2.5cm}|C{4cm}|C{4cm}}
    \hline
    \rowcolor{Gray}
    \multirow{2}{*}{Exit width [mm]} & \multicolumn{2}{c}{Speed (mean/min/max/SD) [m/s] (data-points)} \\
    \cline{2-3}
    & HOM & HET \\
    \hline
    650 & 0.40/0.21/0.65/0.38 (13) & 0.38/0.18/0.89/0.22 (22)  \\
    \hline
    750 & 0.38/0.15/0.91/0.20 (13) & 0.36/0.16/0.88/0.19 (22)  \\
    \hline
    900 & 0.35/0.17/0.66/0.18 (13) & 0.46/0.33/0.98/0.13 (22)  \\
    \hline
    1100 & 0.44/0.31/0.63/0.10 (13) & 0.59/0.38/0.90/0.18 (22) \\
    \hline
    1340 & 0.44/0.27/0.87/0.15 (13) & 0.50/0.31/0.95/0.14 (22)  \\
    \end{tabular}
    \caption{Travel speed for the internal staircase when egressing to an open line using a staircase.}
    \label{tab:speedst}
\end{table}
\begin{table}[!htb]
    \centering 
    \small
    \begin{tabular}{C{2.5cm}|C{4cm}|C{4cm}}
    \hline
    \rowcolor{Gray}
    \multirow{2}{*}{Exit width [mm]} & \multicolumn{2}{c}{Speed (mean/min/max/SD) [m/s] (data-points)} \\
    \cline{2-3}
    & HOM & HET \\
    \hline
    650 & 0.36/0.13/0.94/0.23 (13) & 0.26/0.15/0.38/0.07 (22) \\
    \hline
    750 & - & - \\
    \hline
    900 & 0.30/0.17/0.64/0.12 (13) & 0.46/0.08/1.84/0.42 (18) \\
    \hline
     1100 & - & - \\
    \hline
    1340 & 0.48/0.27/1.02/0.21 (13) & - \\
    \end{tabular}
    \caption{Travel speed for the internal staircase when egressing to an open line without any equipment (drop of 750~mm).}
    \label{tab:speedstjump}
\end{table}
Though there appears to be a trend of increasing travel speed on the staircase in trials with wider exits (boarding areas), the results do not prove any clear interdependence. The main impact on travel speeds for participants appears to be the limited space for movement and the presence of zipping behaviour in the mezzanines and in the boarding areas. \par

\noindent\textbf{\quotes{Non-emergency} movement}\par
Due to the continuous video recording during the entire one-day experiment, it was possible to measure travel speeds at times when participants were spontaneously moving inside the railcar, also in the aisle and on the internal staircase. These values can be considered to be characteristic values for unrestricted movement when passing through an aisle for this railcar model and they are shown in Table~\ref{tab:speednon}.\par
\begin{table}[!htb]
    \centering \small
    \begin{tabular}{c|c|c}
    \hline
    \rowcolor{Gray}
    \multirow{2}{*}{Measured segment} & \multicolumn{2}{c}{Speed (mean/min/max/SD) [m/s] (data-points)} \\
    \cline{2-3}
    & Adults (18$\mathrm{+}$~years) & Children (5--8~years) \\
    \hline    
    In the aisle & 0.94/0.59/1.56/0.24 (28) & 0.59/-/-/- (1) \\ 
    \hline
    Downstairs & 0.75/0.47/0.98/0.23 (5) & - \\
    \hline
    Upstairs & 0.58/0.39/0.93/0.13 (40) & 0.41/0.32/0.46/0.06 (10) \\
    \hline
    \end{tabular}
    \caption{Travel speed measured when experimental trials were not in progress.}
    \label{tab:speednon}
\end{table}
In the available literature, only two studies discuss the travel speeds of passengers inside a railcar, both in relation to horizontal movement (i.e. when people move to an adjacent car). Markos and Pollard \cite{markos_passenger_2015} presented walking speeds for people passing through the vestibule of a railcar (travel distance 1.1 m, width of vestibule 1.2~m), with these ranging from 1.0-–1.5~m/s. Capote et al. \cite{capote_analysis_2012}  mentioned walking speeds for passengers moving in a single queue and not constrained by others within an aisle in a moving train (travel distance and aisle width not available). Walking activities showed a normal distribution with a mean value of 0.99~m/s and a standard deviation of 0.20~m/s. These values are similar to what we observed (0.94~m/s). The travel speed in the aisle during the experimental trials (average value 0.35~m/s; details in Table~\ref{tab:speed}) correspond better to the travel speed of a flow (i.e. group of people) slowed by a cramped layout and high density of others in the railcar space. \par
\subsubsection {Pre-movement activities}
As a part of the evaluation of the experiment, the pre-movement phase (before the movement from the railcar was started) was observed. Although participants were familiar with the schedule and boundary conditions of the experiment and while the exact signal to begin each trial (a whistle blow) was clearly presented, participants had leeway in deciding what actions to perform when leaving their seats (see Section~\ref{subsec:instru}). Some participants spent time when putting on their outer layers of clothing or taking their belongings from the luggage compartments located above the seats. Others skipped such activities and immediately started moving to a safe place. The duration of such activities, which were not influenced by the cramped environment (e.g. limited space for putting on clothing, limited access to storage), were measured and are presented in Table~\ref{tab:pre}. \par 
\begin{table}[!htb]
    \centering \small
    \begin{tabular}{R{8cm}|C{7cm}}
    \hline
    \rowcolor{Gray}
    Activity & Time of duration (mean/min/max/SD) [s] (data-points) \\
    \hline    
    Gripping a backpack from the luggage compartment & 6/2 /10/2 (18) \\
    \hline
    Gripping a toddler in arms & 4/3/5/1 (2) \\
    \hline
    Putting on an outer layer of clothing & 10/9/12/1 (6) \\
    \hline
    Turning off and storing a laptop in a bag & 11/10/11/1 (2) \\
    \end{tabular}
    \caption{Participants' pre-movement activities.}
    \label{tab:pre}
\end{table}
Further video analysis confirmed that activities performed in the pre-movement phase impacted individual evacuation times. The participants who started to evacuate without any other actions reached the railcar exit first. In contrast, participants who were unable to move due to queues in the aisle generally used this waiting time for additional actions such as gathering their belongings or for putting on outdoor clothing. The time intervals observed for pre-movement activities are in good agreement with results from Capote et al. \cite{capote_analysis_2012}, in which times measured for preparing to evacuate from a railcar ranged from 1.5–-26.0~s with a mean value of 12~s and a standard deviation of 8.0~s. \par
\subsubsection {Exiting behaviour}
\begin{figure} [!htb]
    \centering
    \includegraphics[width = 1.0\linewidth]{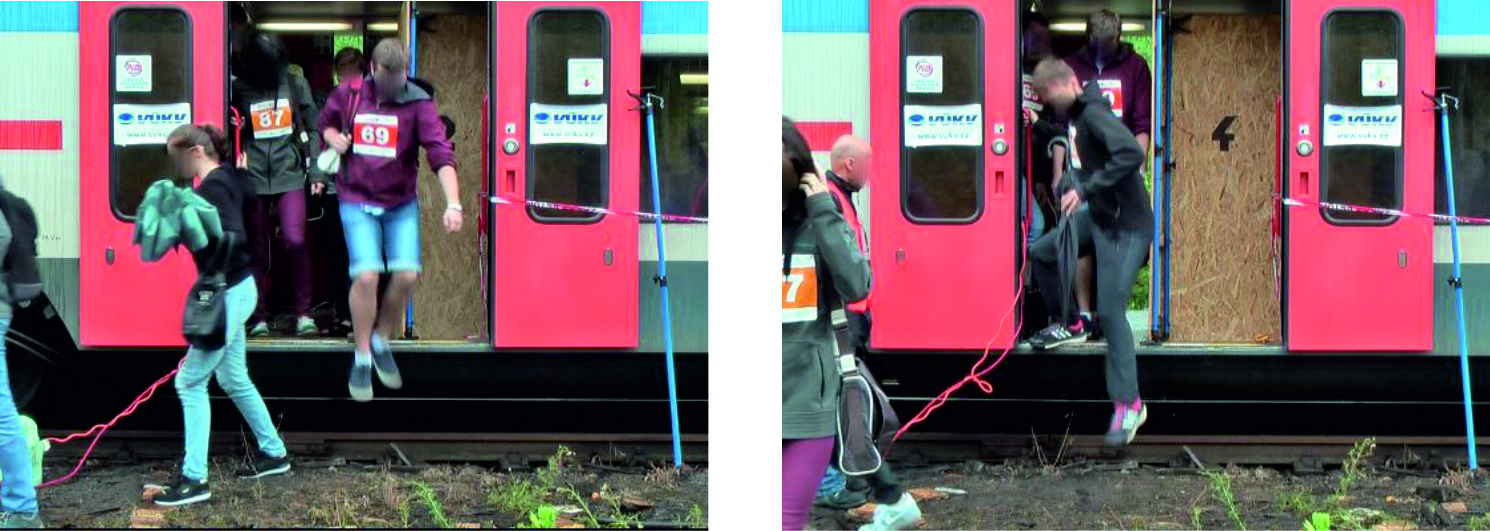}
    \caption{Most HOM group participants exited the railcar using one large step followed by a jump.}
   \label{fig:jump}
\end{figure}

The data collected was also analysed in relation to strategies for navigating the 750~mm jump between the railcar and the surrounding terrain. No complications or delays were observed for the homogeneous (HOM) group. Most participants stepped off the railcar simply using one large step followed by a jump to the surrounding terrain (this strategy is referred to as \quotes{Jumper} in the literature \cite{klingsch_evacuation_2010}; Figure~\ref{fig:jump}). Side handrails were used in approximately 20\% of cases. \par
\begin{figure} [!htb]
    \centering
    \includegraphics[width = 1.0\linewidth]{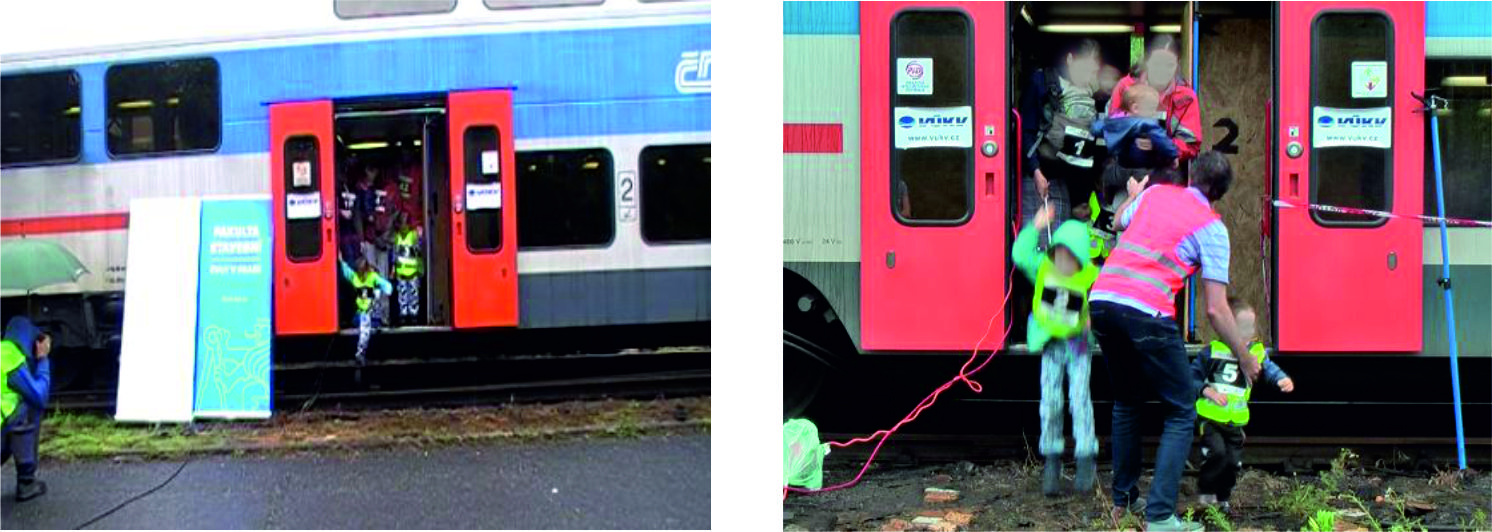}
    \caption{Left: Children exiting without any assistance (\quotes{Sitter} strategy); Right: Exiting children assisted by adults.}
   \label{fig:child}
\end{figure}

\begin{figure} [!htb]
    \centering
    \includegraphics[width = 1.0\linewidth]{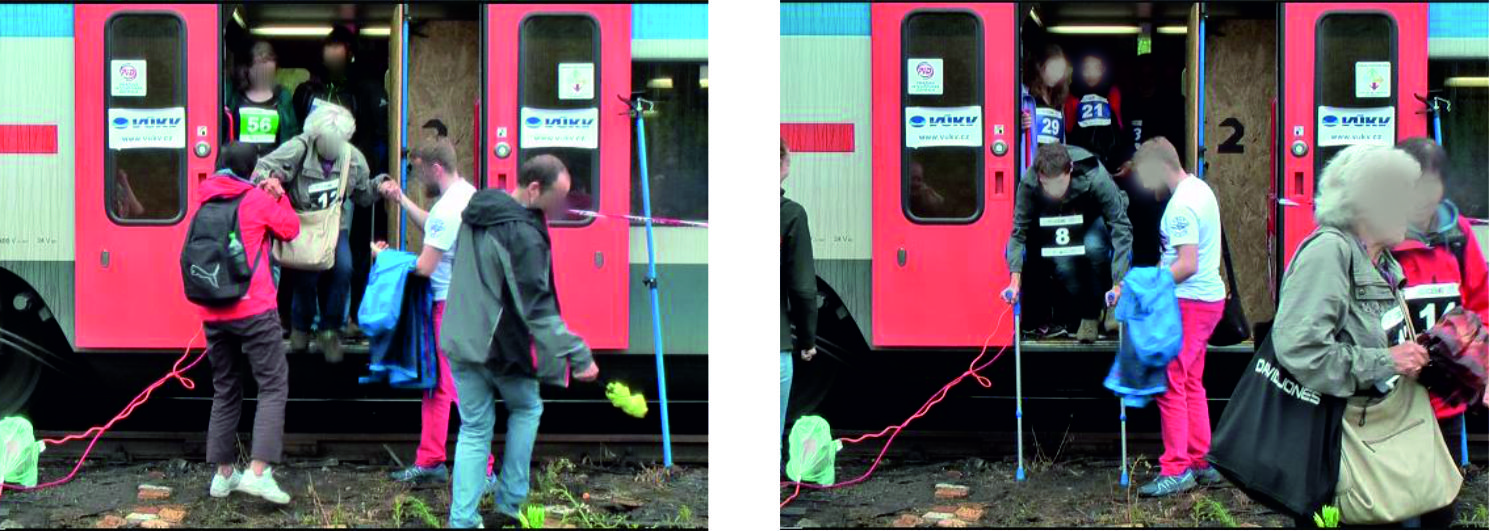}
    \caption{Exiting seniors (left) and the participant with simulated movement disabilities (right) received help from other participants; helping behaviors became more efficient with experience, illustrating the training effect during the experiment.}
   \label{fig:senior}
\end{figure}

Looking at the heterogeneous (HET) group, several critical moments slowed the evacuation process considerably. In the first trial, five-year-old children arrived at the main exit first and blocked the exit as they contemplated what to do about the jump. They ended up selecting a strategy of sitting down on the floor and sliding down onto the terrain (the \quotes{Sitter} strategy \cite{klingsch_evacuation_2010}); Figure~\ref{fig:child}, left). In other trials, the children received assistance from adults (Figure~\ref{fig:child}, right). The 750~mm drop was also challenging for elderly participants, who in the first trial (before receiving assistance from other participants) searched for a device to help them down onto the terrain. As the trials were repeated, participants in the HET group became experienced in helping each other to leave the railcar, with some participants waiting close to the exit in order to provide assistance to others (Figure~\ref{fig:senior}). \par 
The lack of empirical data in the literature for quantitatively describing possible problems presented in navigating evacuation drops from railcars illustrate a possible limitation to future studies. The time delays that occurred when the participants exited the railcar were measured and evaluated here, but because the HOM group navigated the 750~mm drop smoothly, only the behaviour of  the HET group was analysed. Time delays in the boarding area before exiting, in this study, were mostly caused by hesitation and preparation for the jump, and the time delays after exiting due to a balancing process at the terrain level were also observed and are presented in Table~\ref{tab:delay}.\par   
\begin{table}[hbt!]
    \centering \small
    \begin{tabular}{l C{3cm} C{3cm}}
        \hline
        \rowcolor{Gray}
        Age group & Time delay before exiting [s] & Time delay after exiting [s]  \\
        \hline
        Children (3--10~years) & 2.5 & 0.5 \\
        \hline
        Adults carrying a toddler & 1.5 & 0.5 \\
        \hline
        Seniors (60+) & 2.0 & 1.0 \\
        \hline
        Passengers with disabilities$^\mathrm{1)}$ \ & 1.5 & 0.75 \\
        \hline
        \multicolumn{3}{R{11cm}}{$^\mathrm{1)}$\footnotesize{The passenger with simulated disabilities (1~participant) was equipped with crutches and asked to move without active use of one leg.}}\\
         \hline
        \end{tabular}
    \caption{Time delays for participants navigating an exit drop of 750 mm to the terrain level.}
    \label{tab:delay}
\end{table}
\subsection{Summary of the experimental study}
The following conclusions highlight the main findings of the experimental study: 
\begin{itemize}
    \item Total evacuation times measured during the experiment without any correction ranged from 37~s to 73~s, depending on the boundary conditions of each trial, notably the composition of crowd (heterogeneity), exit width (boarding area), and the type of exit path. A salient increase in total evacuation time was observed for the heterogeneous group (HET) navigating an exit jump of 750~mm. This type of exit represented the most challenging condition for this group. Exit widths greater than \numprint{1100}~mm enabled simultaneous exit of two people side-by-side. Conversely, for exit widths less than 900~mm, participants exited in a staggered  manner (shoulder-to-shoulder) or in a single line, which led to a gradual increase in total evacuation times. 

    \item Average exit flow rates ranged from 0.71~pers/s (HET group, minimum exit width of 650~mm, exit path to an open line) to 1.50~pers/s (HOM group, maximum exit width of \numprint{1340}~mm, exit path using stairs). A linear relationship between exit flow rate and exit width can be identified for all types of exit paths. 
  
    \item Travel speeds analysed in the aisle and on the straight internal staircase of the railcar were strongly impacted by the layout of the railcar, which led to cramped conditions in the aisle and on stairs. The average travel speed for flow in the aisle was 0.35~m/s, and the average travel speed on stairs was 0.41~m/s. Unimpeded travel speeds on stairs were observed only in a few of the trials, when flow from the upper deck was not limited by participants gathering in the mezzanine (average speed value: 0.96~m/s) and at times when trials were not being conducted in the aisle, when participants spontaneously moved inside the railcar (average speed value: 0.94~m/s). 

    \item Time intervals describing activities performed by participants in the pre-movement phase were analysed, with the average values of 6~s (getting something from the luggage compartment), 4~s (participants taking  toddlers into their arms), 10~s (putting on clothing for outside conditions), and 11~s (turning off and storing a laptop in a bag). 
  
    \item Considering strategies for navigating the height difference of 750~mm between the railcar floor and the surrounding terrain, participants in the homogeneous group (HOM) mostly decided to step off the railcar simply using one large step followed by a jump to the ground  without any particular delays. Conversely, the 750~mm drop caused some critical moments for the heterogeneous group (HET) since participants with limited movement abilities such as children or seniors required assistance in navigating the height difference. Time delays for the 750~mm drop varied 2–-3~s for different age groups. 
\end{itemize}
For a proper interpretation of the presented outcomes, several limitations of the proposed experimental study should be recalled. The experimental setup was designed in a controlled environment involving informed and instructed participants. Despite its many advantages, this data collection method cannot fully reflect the real-life situations which may occur in emergency conditions. Thus, the limitations associated with the lack of contextual realism (e.g. the absence of fire conditions, emergency lighting, presence of crew staff), limited capacity for simulating more evacuation scenarios, and the presence of the training effect must be considered. Similarly, the representativeness of the presented results must be seen as limited to the particular evacuation scenario including the type and configuration of the railcar, occupancy, exit paths used, and environmental conditions.     

\section{Sensitivity analysis of evacuation from a double-deck passenger railcar}

The main goal of this section is to study the influence of boundary conditions such as main exit width, crowd composition, and exit type on total evacuation time. In order to describe the influence quantitatively, sensitivity analysis was performed using optiSLang software~\cite{optiSLang}.

Since the experimentally measured data is rather sparse and burdened with undesirable effects (see Section~\ref{sec:undesirable}), sensitivity analysis was performed not only for the experimental data, but the conclusions were supported by the sensitivity analysis of results from Pathfinder simulations. In order to consider the simulation results as a valid emulation of the experimental trials, some features of the model were modified according to experimental observations as described in Section~\ref{sec:pathfinder}. Validation of the model against the experiment is described in Section~\ref{sec:validation}. Sensitivity analysis of experimental and simulation data is provided in Section~\ref{sec:sensitivity}.

\subsection{Simulations of the evacuation experiment using Pathfinder}
\label{sec:pathfinder}

Pathfinder, version 2019~\cite{PathfinderMAN, PathfinderTECH}, developed by Thunderhead Engineering Consultants, Inc., was used to perform the simulations. In this section we summarize the most important aspects and parameters of the model that had to be modified in order to obtain sufficient correspondence of simulation results with the experimental data. The Pathfinder simulation results were intended to compensate for the fact that for each experimental setup, there was just one experimental trial. The main goal of the simulations was to support the conclusions drawn from the analysis of the experimental data by the analysis performed for the simulation datasets.

To mimic the experimental conditions as much as possible, one set of agents for the HOM group and one set of agents for the HET group were used for all simulations. This means that all randomly distributed parameters of agents (maximal speed, agent diameter, initial orientation) were generated just once before the simulations and were kept unchanged throughout the simulation process. 
Thus, when changing a geometry-related boundary condition (exit width, exit type), the  simulation was performed with the same set of agents (HOM or HET) having identical properties.
Nevertheless, even when repeating the evacuation with the same initial conditions (same group of participants, same seating position), the course of the evacuation may vary due to variance in human behaviour. In order to achieve such effect in the simulations, the random seeds of all agents (agent property \verb+rseed+) were randomly re-generated at the beginning of every simulation. This property influences the results of randomized processes in the Pathfinder model (e.g. conflict solution), and  the simulations therefore resulted in different trajectories. Following the logic used in the experimental trials, this approach was chosen to assure that the variation of the simulation results was not burdened by the variance possibly caused by randomness in agents parameters, agent types, or their initial positions.

\subsubsection{Geometry and environment}

\begin{figure}[htb!]
\def\h{.32}
	\includegraphics[width = \h\textwidth]{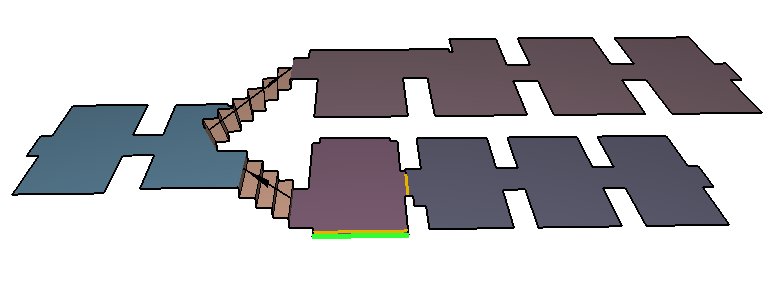}
	\hfill
	\includegraphics[width = \h\textwidth]{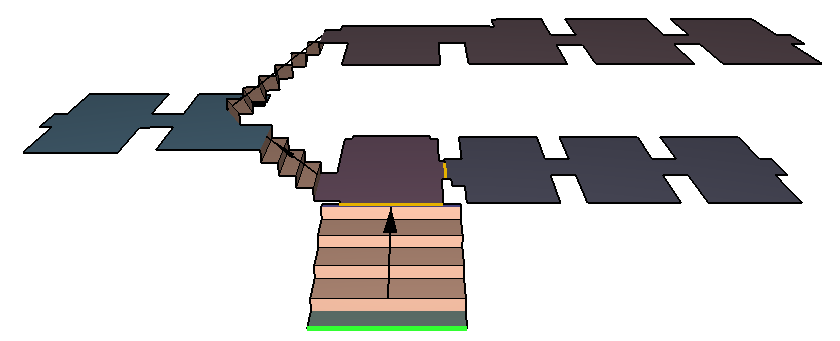}
	\hfill
	\includegraphics[width = \h\textwidth]{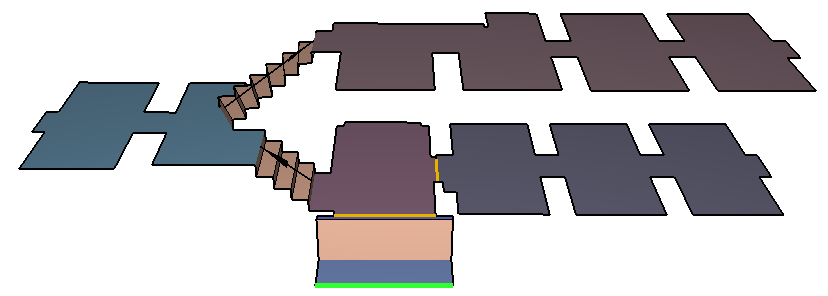}
\caption{Representation of different exit types: to high platform (left), to open line using stairs (middle), to open line without any devices (right).}
\label{fig:exittypes}
\end{figure}

For simulation purposes, a model of one half of a double-deck trailer car  electric unit class 471 (CityElefant) was created in Pathfinder. The geometry of the interior corresponded exactly to the dimensions of the actual railcar. As in the experiment, the measurement checkpoint was located at the main exit door (yellow line in Figure~\ref{fig:exittypes}). Due to the symmetry of the railcar, an identical geometry was used for both groups, HOM and HET. Different exit types were implemented as described below and illustrated in Figure~\ref{fig:exittypes}.

\paragraph{Exit to a high platform} It was observed that after passing the main door to a high platform, pedestrians did not affect the flow through the main exit. Thus there was no additional delay caused by the accumulation of pedestrians behind the exit. For this reason, the platform was simulated by defining the edge of the computational network of the model (green line in Figure~\ref{fig:exittypes}) 0.1~m behind the position of the main exit, i.e an agent who passes the checkpoint is removed from the simulation.

\paragraph{Exit to an open line using stairs} The standard representation of stairs in Pathfinder was used for the simulation of the exit using stairs. The end of the computational network corresponds to the terrain level.

\paragraph{Exit to an open line without any devices} The 750~mm drop was simulated as a step with this height. However, the step itself in Pathfinder does not affect the motion or speed of the agents (such a step serves mainly for illustration purposes). Thus, additional limitations were introduced to the model influencing the motion of agents while passing the step:
\begin{itemize}
	\item The capacity of the step was set to two agents, i.e. at most two agents were allowed to jump down at the same time for both groups, HOM and HET.
	\item Based on its properties, an agent waits a certain time before jumping (for agents with limitations in the HET group).
	\item Based on its properties, an agent waits a certain time after jumping (for agents with limitations in the HET group).
\end{itemize}

These limitations were inspired by the experimental observations. Even for a wide exit, we did not observe more than two participants trying to jump at the same time during the experiment. The delay before and after a jump caused by preparation and balancing was quantitatively implemented according to experiment investigation and summarized in Table~\ref{tab:delay}.

\subsubsection{Composition of the crowd and initial positions}
\label{sec:composition}

Several occupants profiles were defined in the Pathfinder model representing individual agent types introduced in the simulation. The agent types were defined according to the experiment. Agent type ``Without limitations'' representing an adolescent or adult participant without any movement limitations, and four types of agents with limitations were introduced: ``Children'', ``Adults carrying a toddler'', ``Seniors'', and ``With disabilities''.

The simulations were performed in three phases. First, individual experimental trials were reproduced for the purpose of validating the model (see Section~\ref{sec:validation}). The type and initial positions of individual agents corresponded exactly to those in the experiment (participants were seated exactly as they were in all experimental scenarios), i.e. 15~trials for the HOM group with 42~agents; 13~trials for the HET group with 46~agents; and 2~trials for the HET group with 42~agents (for 2~experimental trials, 2 children with their parents did not participate, see the Section~\ref{sec:undesirable}). Here we remind the reader that a person carrying a toddler was considered to be one participant in the analysis and thus was simulated as one agent. 

During the second phase, we investigated the influence of boundary conditions (exit width, crowd composition, exit type) on total evacuation time. It was therefore necessary to have the same number of evacuating agents for each trial. For these purposes, 4~agents of the type ``Without limitations'' were added to the HOM group in order to have 46~evacuees throughout the whole analysis. The positions  of these new agents were chosen to approximate the positions of the 46~passengers in the HET group.

In the third phase, the simulations with a modified ratio of agents with limitations were performed. While in the experiment (and thus, in the second phase of simulations), the HET group contained 28\% of passengers with some limitations, in the third phase, additional simulations for the HET group containing 15\% of passengers with limitations (approximately half of the original sample) and 56\% (approximately twice the original sample) were performed. The agent positions for the HET group stayed the same; only their types were adjusted. The numbers of agents in individual group types were chosen  based on their proportions in the original experimental sample (see Table~\ref{tab:popu}). The parameter related to crowd composition denoted the percentage of passengers with limitations in the crowd, i.e. 0\% for the HOM group and 15\%, 28\%, 56\% for the HET group. Table~\ref{tab:composition} contains numbers of agents for a given agent type involved in the crowd. The positioning of agents in individual trials is presented in Figure~\ref{fig:placement}~(\ref{app:input}).

\begin{table}[htb!]
	\centering
	\begin{tabular}{l*{4}{|C{1.7cm}}}
		\hline
		\multirow{2}{*}{Agent type} & \multicolumn{4}{c}{Number of agents}\\
		\cline{2-5}
		& HOM 0\% & HET 15\% & HET 28\% &  HET 56\%\\
		\hline
		Without limitations & 46 (42) & 39 & 33 (31) & 20\\
		Children & -- & 3 & 5 (3) & 10\\
		Carrying a toddler & -- & 1 & 3 & 6\\
		Seniors & -- & 2 & 4 & 8\\
		With disabilities & -- & 1 & 1 & 2\\
		\hline
		Together & 46 (42)  & 46 & 46 (42) & 46 \\
		\hline
	\end{tabular}
\caption{Composition of agents in the simulation corresponding to the egress of 46 agents. For the case of 42 agents, the numbers are given in parentheses.}
\label{tab:composition}
\end{table}

\subsubsection{Agent parameters}

As explained above, we aimed to imitate the experiment with the simulations as closely as possible. Therefore, key agent parameters were set to correspond to the experimentally measured data whenever possible. In~\cite{GreGre2014RDO}, a sensitivity study of agent parameters on evacuation time from an artificial building was performed using the Pathfinder model. From this study it follows that total evacuation time is primarily sensitive to maximum speed (53\%) and agent radius (36\%), with  acceleration factor (13\%) playing a more minor role. The influence of other parameters is negligible with respect to the variation of evacuation time. Because of this, we focused here on proper adjustment of maximal agent speed in the steering mode~\cite{Rey1999steering} and agent radius corresponding to the shoulder width. For the 750 mm jump case, additional time delay parameters were implemented. Table~\ref{tab:input} summarizes chosen values for the parameters of individual agent types and an explanation of the values presented follows below.

\begin{table}[htb!]
	\begin{tabular}{R{2.7cm}*{5}{|C{2cm}}}
	\hline
	\rowcolor{Gray}
	Agent type	&	Shoulder width [cm]	&	Speed [m/s]	& Speed-Density &	Delay before jump [s]	& Delay after jump [s]\\
	\hline
	Without limitations & $45.7\pm 5.0$ $^\mathrm{1)}$ & $0.94\pm 0.25$ $^\mathrm{2)}$ & constant & -- & --\\
	\hline
	Children & 25 and 32 	& 1.32~\cite{Hamilton2017FireSafety} & SFPE & 2.5 & 0.5\\
	\hline
	Carrying a toddler & 62$^\mathrm{3)}$ 	& 0.94 & constant & 1.5 & 0.5\\
	\hline
	Seniors & 		40	& 0.70~\cite{Kholsevnikov2012}& SFPE & 2.0 & 1.0\\
	\hline
	With disabilities &71$^\mathrm{3)}$ & 0.94~\cite{Boyce1999}& SFPE & 1.5 & 0.75\\
	\hline
	\end{tabular}
	
	\vspace{3mm}
	\footnotesize{$^\mathrm{1)}$~ Truncated normal distribution with mean 45.7, std 5, min 38, max 58.}\\
	\footnotesize{$^\mathrm{2)}$~ Truncated normal distribution with mean 0.95, std 0.25, min 0.64, max 1.56.}\\
	\footnotesize{$^\mathrm{3)}$~ Squeeze factor reduced from default 0.7 to 0.6, minimum diameter increased from default 0.33 to 0.5~m.}
\caption{Parameter values for individual groups of agents/passengers. Most values were directly measured in the experiment. Values taken from the literature referenced by a citation.}
\label{tab:input}
\end{table}

\paragraph{Agent speed} As described in Section~\ref{sec:aislespeed}, recording the entire one-day experiment enabled travel speed in the aisle to be measured. This speed measured for passengers without limitations in the HOM group can be considered to be the speed of unrestricted movement in the aisle for these passengers walking in a line. Thus, the measured values were used as the speed parameter for the agent type ``Without limitations'', i.e. every agent from this group was assigned a random speed from the truncated normal distribution with a mean value of $0.94$~m/s and a standard deviation $0.25$~m/s. Speed was assigned to an agent once and kept constant for all simulations and scenarios. Here we note that this speed already included the influence of surrounding obstacles, queues, and local density, so the SFPE curve~\cite{sfpe2002} (relation between speed and local density) for level terrain movement was turned off for these agents.

Due to lack of data for other passenger types this procedure could not be used to determine the speed for several agent types: ``Children'', ``Adults carrying a toddler'', ``Seniors'', and  ``With disabilities''. Thus, for these agent types, the values for unrestricted free movement speed from the literature were used. For this purpose, the studies and measurements used were chosen based on their similarity to the situation in the experiment studied. Contrary to agent type ``Without limitations'', the maximum speed was not generated randomly but was set to the mean value for all agents of a given type. There were only a few agents of each type, and random speed generation may therefore have caused an undesirable deviation of random velocities from the desired distribution (here we remind the reader that speed was generated only once for all simulations).
\begin{itemize}
	\item ``Children'': The speed of this type of agent was adopted from ~\cite{Hamilton2017FireSafety}, where the mean free walking speed for the group of children aged~4--12 was estimated to be~$1.32$~m/s with a standard deviation $0.40$~m/s. The mean value $1.32$ was chosen as the maximum speed for this agent type. Since this speed is not influenced by the external environment, the SFPE relation of speed and local density was turned ON for these agents.
	\item ``Adults carrying a toddler'': The speed of an adult is not significantly influenced by the carried toddler, thus the agents of this type were assigned a speed of $0.94$~m/s with the SFPE curve turned OFF. The idea was that speed does not decrease because of the dense environment, but rather because of the greater width of an adult-toddler pair, which results in an increase in the number of conflicts.
	\item ``Seniors'': Maximum speed of the agent type ``Seniors'' has been set as $0.7$~m/s following the ~\cite{Kholsevnikov2012}, where the mean free walking speed for the group of seniors without movement aids was estimated to be~$0.7$~m/s. As with ``Children'', the SFPE remained ON.
	\item ``With disabilities'': The free speed of disabled passengers with crutches was chosen according to study~\cite{Boyce1999}, where the  mean speed was estimated as $0.94$~m/s with a standard deviation $0.30$~m/s. As with ``Children'' and ``Seniors'', the SFPE remained ON.
\end{itemize}

\paragraph{Agent diameter} Shoulder width was estimated according to questionnaire data provided by participants involved in the experiment. The width of an adult carrying a toddler and a person with crutches was measured on site. As with speed, to each agent of the type ``Without limitations'' shoulder width was randomly generated according to the estimated distribution. The diameter for other agent types was set as the mean value of the given agent group in order to avoid undesirable randomization effects. 

Due to the narrow and confined environment of the experimental railcar, further spatial parameters related to diameter were modified for the agent types ``With disabilities`` and ``Adults carrying a toddler''. Namely, compressibility due to conflicts between agents (squeeze factor) was changed from a default value of 0.7~to 0.6, since pedestrian width is greater only in the lateral direction for people with crutches, and because body mass is flexible due to the possible change in a child's position for adults carrying a toddler. The minimum diameter to which such agents could reduce their widths due to narrow geometries was set to a minimal value of 0.5~m  instead of the default value of 0.35~m in order to support the fact that  it is complicated for such passengers to move in narrow environments.

\paragraph{Delay before and after a jump} Another agent property introduced to the model was the delay before and after a jump from the 750~mm high step (exit to the open line without any devices), as discussed in the geometry section.
\subsection{Validation of the model against experimental data}
\label{sec:validation}

In order to validate the model against the experiment, simulations were performed exactly according to the real situation for each experiment trial, i.e. the number of passengers, their types, positions and parameters were set to match the experiment. As mentioned above, agent properties were set once for all trials. For every trial, 30~different simulations were performed by running the simulations repetitively via the command line running \verb+testsim.bat+ with the only argument being the input file \verb+sim.txt+, which contains all information necessary to run the simulation (see ``Section Simulating via command-line'' in the User manual~\cite{PathfinderMAN}). To achieve different trajectories,  unique random seeds (\verb+rseed+) for individual agents (Occupants in the input file) were changed directly in the input file before every run of the simulation. Simple convergence study (as in~\cite{RonNil2014BuildSim}) showed that 30 runs of a simulation provide sufficiently precise average total evacuation time.

\begin{figure}[htb!]
\def\w{.4\textwidth}
\begin{tabular}{rcc}
	& HOM  & HET\\
	\rotatebox{90}{\phantom{xxxxxxxxxx}platform} &
	\includegraphics[width = \w]{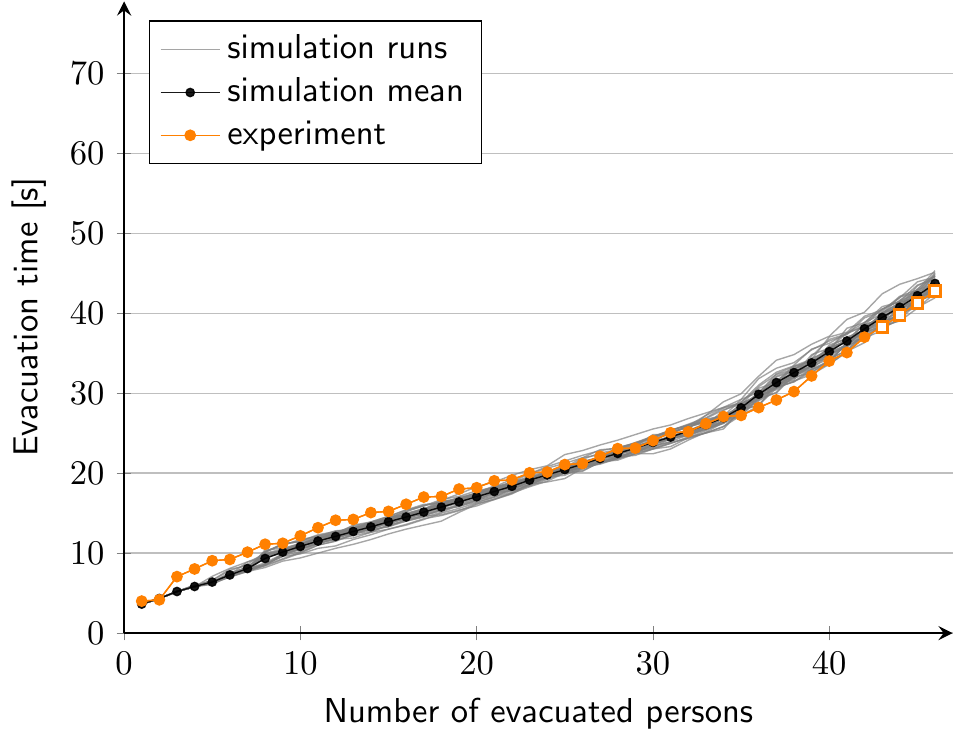} &
	\includegraphics[width = \w]{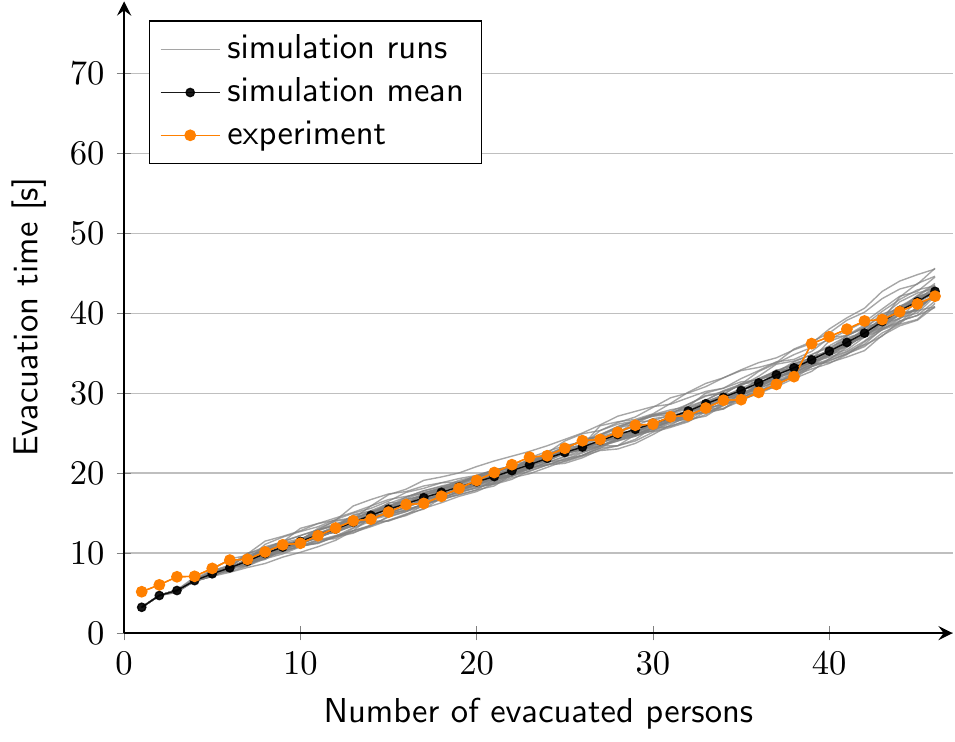}\\

	\rotatebox{90}{\phantom{xxxxxxxxxxx}stairs} &
	\includegraphics[width = \w]{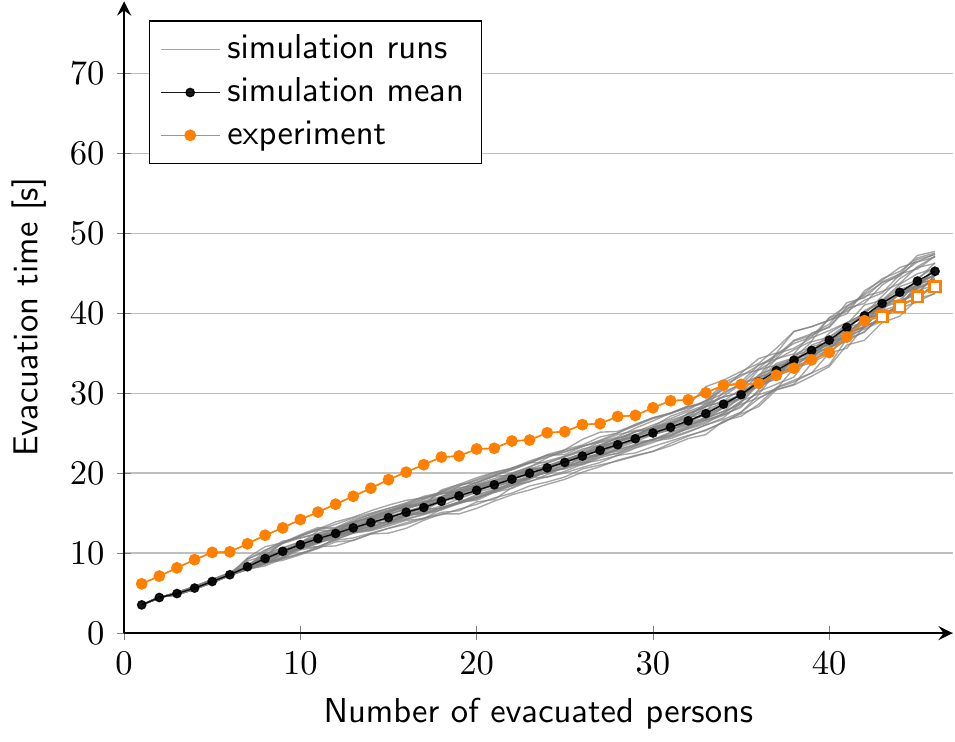} & \includegraphics[width = \w]{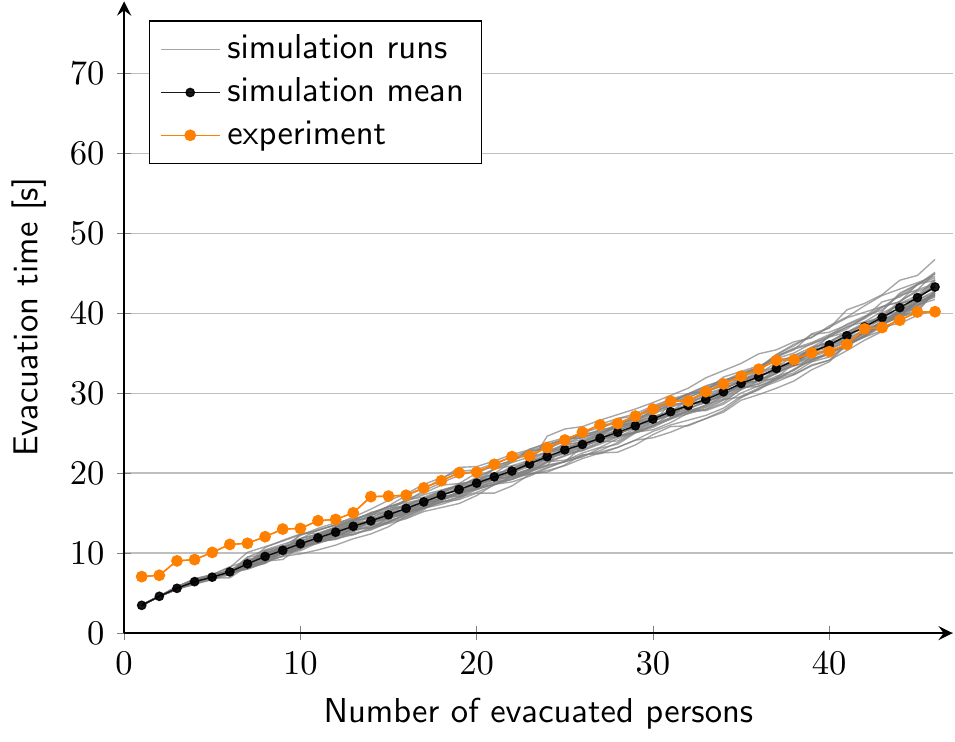}\\

	\rotatebox{90}{\phantom{xxxxxxxxxxx}ground level} &
	\includegraphics[width = \w]{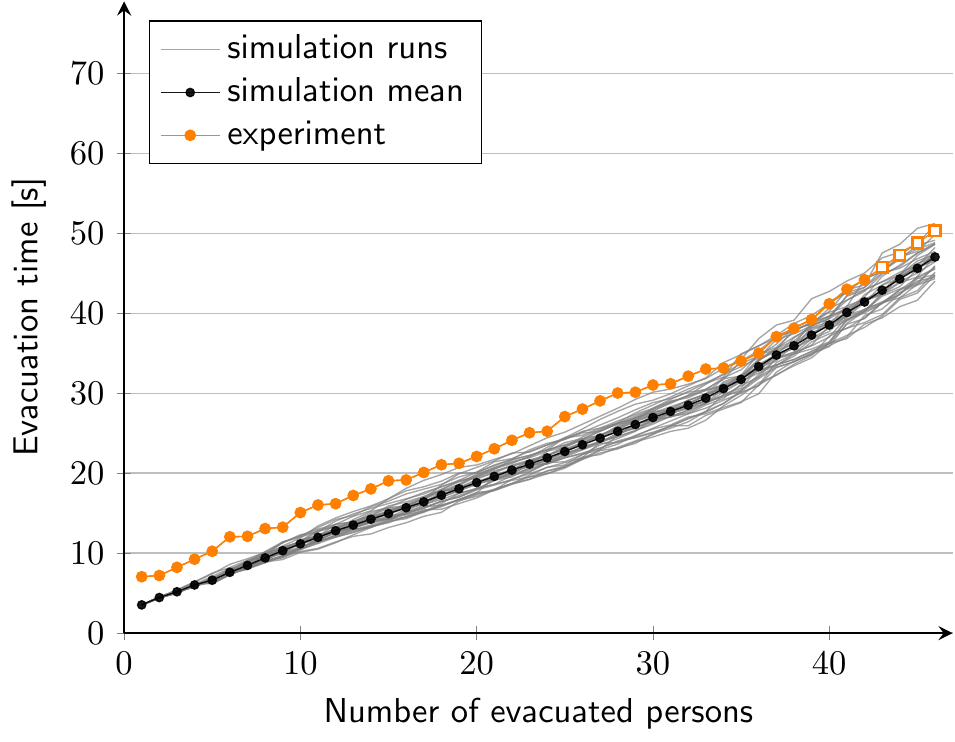} & \includegraphics[width = \w]{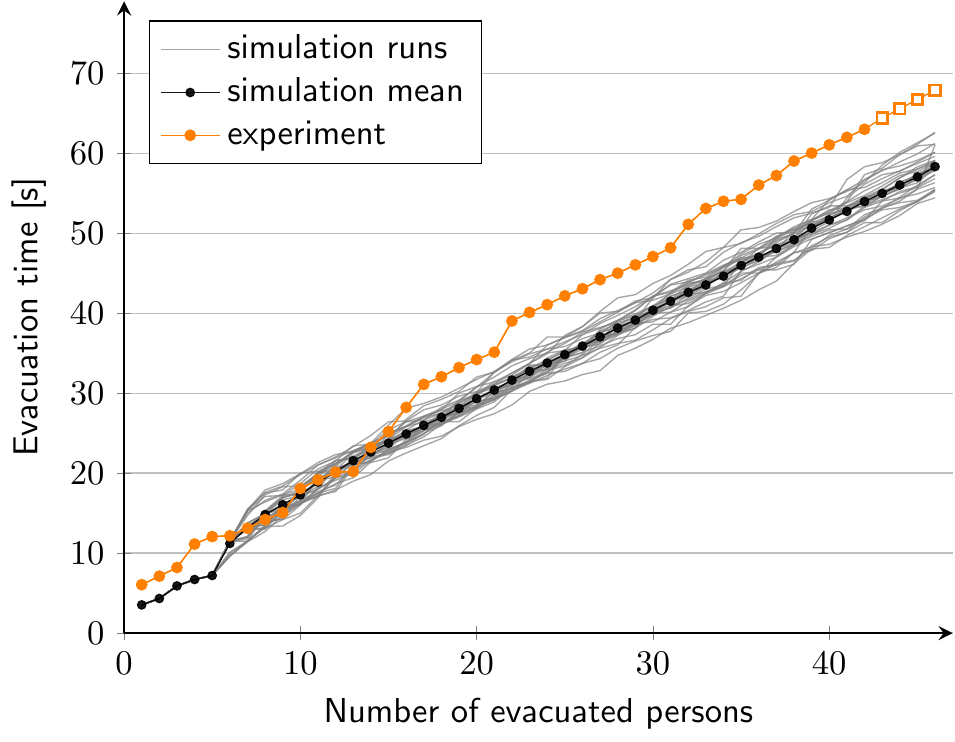}\\
\end{tabular}
\caption{Comparison of the evacuation times for $n$-th participants for a width $1.1$~m for individual passenger groups (HOM, HET), and exit types: to a high platform (platform), to an open line using stairs (stairs), and to an open line without any devices (ground level). Simulation performed for 46 passengers.}
\label{fig:validation}
\end{figure}

For the validation of the model, the sequences of egress times $(t_1,\dots,t_N)$ of the first, second, \dots, $N$-th passenger were chosen. The experimental observations depicted in Figures~\ref{fig:TET1}-\ref{fig:TET03} were compared to individual simulation runs and to the average trajectories. Figure~\ref{fig:validation} contains selected graphs for the exit width 1.1~m illustrating a comparison of simulations and the experiment. An extensive comparison of all setups is provided in~\ref{app:valid}. The experiment curve color corresponds to the colors in Figures~\ref{fig:TET1}--\ref{fig:TET03}.

\subsubsection{Egress to a high platform}

Figure~\ref{fig:valid_platform}~(\ref{app:valid}) shows graphs depicting the egress times for exiting passengers. In the graph for a width of 0.9~m and the HET group, the experimental data was shifted by 5~s because the HET group did not hear the whistle starting this experimental trial, and were delayed. From the graphs we can conclude that for larger widths (0.9~m -- 1.34~m), the simulations correspond very well with the experiment. For smaller widths (0.65~m and 0.75~m), the simulations slightly overestimate the total evacuation time when comparing the simulation average to the experiment. However, the experimental trajectory does not escape the simulation band.
Comparing the graphs for HOM and HET groups, we can see that there is no significant difference in the simulation results, which corresponds to the experimental observations.

\subsubsection{Egress to an open line using stairs}

Figure~\ref{fig:valid_stairs}~(\ref{app:valid}) shows graphs depicting the egress times of exiting passengers are provided in~. The graphs indicate that the simulation results correspond very well to experimental results. The simulations slightly overestimate total evacuation time for narrow exits, but the  difference is even less significant than for the high platform scenario. 

\subsubsection{Egress to an open line without any devices}

Figure~\ref{fig:valid_jump}~(\ref{app:valid}) provides graphs depicting the egress times of exiting passengers. In the graph for an exit width of 1.1~m with the HET group, the experimental data was shifted by 5~s because of delay caused by a temporary dead-lock caused by small children blocking the exit, see Section~\ref{sec:undesirable}. The graphs again show satisfactory correspondence between the simulations and the experiment. For the HOM group, the simulations illustrate the same aspects as for the high platform and stairs scenarios. For the HET group, a high variance in simulation results can be observed even for larger exit widths. Except for exit widths of 0.9~m and 1.1m, where the simulations underestimate the total evacuation time, the experimental egress times are very close to the simulation averages. The slope of the egress time is significantly higher for the HET group for the 750 mm simulated jump.

\subsubsection{Summary of the validation}

From a comparison of the simulations and the experiment, we can conclude that it is possible to simulate railcar evacuation using Pathfinder, keeping in mind that several aspects of the model must be modified. While exiting to a platform or using stairs can be simulated using standard Pathfinder tools, the delay before and after a jump to an open line must be set because it plays a significant role when passengers with movement limitations are present. Based on this validation, we may regard the model as being appropriately set up for the evacuation scenarios investigated by the experimental trials. Therefore, the simulations of those scenarios can be considered to present the likely performance of a real evacuation.

\subsection{Sensitivity Analysis}
\label{sec:sensitivity}

The main goal of sensitivity analysis is to determine the impact of a variable on the key outcomes of a model being investigated~\cite{Saltelli2018}. For evacuation dynamics, this kind of analysis can be used as a tool for investigating the sensitivity of observed variables to variance in internal parameters~\cite{DuiDaaHoo2016PhysA} or to investigate the impact of external parameters (boundary conditions) on key outcomes of an evacuation model (usually total evacuation time)~\cite{LinWu2018AME}. A useful tool for quantification of the impact of parameters is variance-based sensitivity analysis, which consists in probabilistic description of the dependence of output on input variables, usually using regression analysis or other statistical models~\cite{BodRon2019CD}. The idea is then to explain the variance in the output by means of the given variable in the statistical model. This concept is implemented in the optiSLang software~\cite{optiSLang}, which was used in this study.

\subsubsection{Method}

Sensitivity analysis in optiSLang uses the Meta-model of optimal prognosis to choose the optimal input variable set and the most appropriate approximation statistical model~\cite{Most2008Dynardo} for the investigated problem. The final model is chosen using the Coefficient of Prognosis ($CoP$), which is a measure of model quality based on cross-validation. Its value expresses ``what percentage of variation in the output data can be explained by the model.''  The variance contribution of a single input is quantified by the product of the $CoP$ and the sensitivity index of variable $X_i$ \cite{Most2011Dynardo}
\begin{equation}
	CoP(X_i) = CoP\cdot S_T^\mathrm{MOP}(X_i)\,.
\end{equation}

In our study, there was only one dependent variable, total evacuation time $TET$ (in seconds). The explanatory input variables to the statistical model were: main exit width $W$ (in meters); heterogeneity $H$ represented by the percentage of people with limited movement abilities (in percent); exit type $E$ given as $0$ -- exit to a high platform, $1$ -- exit to an open line using stairs, $2$ -- exit to an open line without any devices. The values were sorted according to their expected influence on delay times; no linear dependence was assumed.

Sensitivity analysis was performed using the complete battery of models in the Meta-model of optimal prognosis provided by optisLang. Presented $CoP$s come from this full model. The parameters of polynomial regression with a degree of 2 (part of the MOP) are presented here as well. Despite the $CoP$ of the polynomial regression being lower than the $CoP$ of MOP including more complex models, it provides more illustrative insight into the dependence of total evacuation time on the variables investigated.

The polynomial model can be written in the form 
\begin{equation}
\label{eq:polyfit}
	\begin{array}{cccclclclc}
		TET &=& \alpha_0 &+& \alpha_1\cdot W &+& \alpha_2\cdot H &+& \alpha_3 \cdot E &+\\
		&&&+& \alpha_4\cdot W^2 &+& \alpha_5\cdot H^2 &+& \alpha_6\cdot E^2&+\\
		&&&+& \alpha_7\cdot W\cdot H& +& \alpha_8\cdot H \cdot E& + &\alpha_9\cdot W \cdot E&,
	\end{array}
\end{equation}
where $\alpha_0,\dots,\alpha_9$ are model parameters estimated using polynomial regression.

\subsubsection{Sensitivity analysis of experimental data}

The experiment data include 30~measurements of total evacuation time (Table~\ref{tab:TETcorretions}). It is worth mentioning that, for reasons of comparison, the measured total evacuation times were transformed to TET 46~(and further, to TET corr in two trials) in order to obtain the estimate of TET for a comparable group of 46~passengers (see Section~\ref{sec:undesirable}).

Sensitivity analysis for this data using optiSLang results are: the overall coefficient of prognosis $CoP = 84.4\%$, the contributions of individual variables are $CoP(W) = 43.4\%$, $CoP(H) = 13.9\%$, $CoP(E) = 34.7\%$. Thus, all variables have significant contribution to the TET variance. The influence of exit type $E$ and exit width $W$ is comparable ($43.4\%$ of the TET variance can be explained by the variation of exit width, $34.7\%$ of TET variance can be explained by the variation in exit type). Lower, but still significant influence is observed for the ratio of people with movement disabilities $H$ ($13.9\%$ of TET variance can be explained by the variation of crowd heterogeneity). 

In order to gain deeper insight into the influence of crowd composition, sensitivity analysis was performed separately for data corresponding to each exit type. The values of estimated $CoP$ are provided in Table~\ref{tab:copexp}. This analysis corresponds to the experimental observations summarized in Section~\ref{sec:TET}: For exits in the high platform or stairs scenarios, the influence of crowd composition is negligible (for the stairs scenario, the variable $H$ was even excluded from the model as absolutely insignificant). Contrarily, for the scenario with exit to open line without any devices, the crowd composition parameter $H$ plays a comparable, even more significant, role than the exit width parameter $W$. This corresponds to the observation that the difference between evacuation times for HOM and HET groups was negligible for the high platform and stairs scenarios.

\begin{table}[htb!]
\centering
	\begin{tabular}{l|C{1.3cm}||*{3}{C{1.4cm}|}}
	\hline\rowcolor{Gray}
	$(\%)$ & $CoP$ & $CoP(W)$ & $CoP(H)$ & $CoP(E)$ \\
	\hline
	All variables & 84.4 & 43.4 & 13.9 & 34.7\\
	\hline
	High platform $E = 0$ & 85.7 & 85.7 & 3.4 & --\\
	Stairs $E = 1$ & 92.3 & 92.3 & excluded & --\\
	Open line $E = 2$ & 82.5 & 34.5 & 47.8 & --\\
	\hline
	\end{tabular}
\caption{Coefficients of Prognosis ($CoP$) for MoP in optiSLang for corrected experimental data (TET corr), see Table~\ref{tab:TETcorretions}.}
\label{tab:copexp}
\end{table}

Here it is necessary to remark that the amount of data generated by the experiment was relatively small. There was always just one observation for a given combination of values $W$, $H$, and $E$. Thus, this dataset does not contain values for trials under the same or similar conditions. For this reason, simulations were used to gain more detailed insight into the influence of individual variables.

\subsubsection{Sensitivity analysis of simulated data}

Using the Pathfinder model, 30~simulation runs for the evacuation of 46~passengers were performed for each combination of parameters $W \in  \{0.65, 0.75, 0.90, 1.10, 1.34\}$, $H \in  \{0, 26\}$, and $E \in  \{0, 1, 2\}$ (i.e. same set as in the experimental trials) in order to obtain an extended dataset for total evacuation time. This is described as the second simulation phase in Section~\ref{sec:composition}, and sensitivity analysis was performed on $30\times30 = 900$ data points.

The results of sensitivity analysis performed using optiSLang are presented in Table~\ref{tab:cop}. The first row of the table contains $CoP$  values for sensitivity analysis of all variables. We can see that, unlike in the experiment, TET is mainly sensitive to exit width $W$: 60.1\% of the TET variance can be explained by the variation of exit width $W$. Exit type contributes to TET variance by 29.1\%. In comparison to exit type, crowd composition $H$ has a rather low significance of influence on TET variance, $10.9\%$.

\begin{table}[htb!]
\centering
	\begin{tabular}{l|C{1.3cm}||*{3}{C{1.4cm}|}}
	\hline\rowcolor{Gray}
	$(\%)$ & $CoP$ & $CoP(W)$ & $CoP(H)$ & $CoP(E)$ \\
	\hline
	All variables & 91.9 & 60.1 & 10.9 & 29.1\\
	\hline
	High platform $E = 0$ & 88.4 & 88.4 & excluded & --\\
	Stairs $E = 1$ & 91.6 & 91.6 & 1.3 & --\\
	Open line $E = 2$ & 94.4 & 62.4 & 31.6 & --\\
	\hline
	\end{tabular}
\caption{Coefficients of Prognosis ($CoP$) for polynomial-fit regression using optiSLang. Simulated data in the second phase (basic set of parameters).}
\label{tab:cop}
\end{table}

Sensitivity analysis was also performed for each exit type separately, i.e. considering $W$ and $H$ as explanatory variables with the value $E$ being fixed. From the results, we can read that for the high platform ($E = 0$) and stairs ($E = 1$) scenarios, the contribution of crowd heterogeneity $H$ is marginal or even excluded from the polynomial model as insignificant. This corresponds to observations from the experiment. However, for the scenario with exit to an open line without any devices ($E = 2$), the contribution of heterogeneity $H$ is very significant: 31.6\% of TET variance can be explained by the variance in heterogeneity. We can notice that the ratio of the CoP for heterogeneity $H$ and exit type $E$ is similar for the experimental and simulated data (about one third). A conclusion can be drawn that crowd composition (with respect to movement abilities) plays an important role in evacuation, particularly for then scenario with an exit to open line without any devices, which quantitatively supports the experimental observations in Section~\ref{sec:TET}.

\subsubsection{Sensitivity analysis of simulation with ``finer'' parameters}

In a third phase, Pathfinder simulations were used to generate additional data points for parameter values not involved in the experiment. The simulations thus served as an extrapolation of the  experimental setup. The exit type scenarios remained unchanged. Exit widths of 0.85, 0.95, 1.00, and 1.20~m were added to gain a finer understanding of the dependence on the exit width. Most importantly, however, different percentages of passengers with disabilities were added. 

As described in Section~\ref{sec:composition}, a crowd with 15\% (about one half the experiment) and 56\% (about twice the experiment) of passengers with movement limitations was introduced in simulations. The numbers of agents of given types and their positions were chosen proportionally to the experimental set-up, i.e. we tried to keep the mutual ratio of agents of a given type the same as in the experiment, taking into account the statistics provided by the operator of the selected train as well.

Results of sensitivity analysis are provided in Table~\ref{tab:copfine}. The main factor influencing TET is again exit width $W$. In comparison to the previous sensitivity analysis phases, there is a significant increase in the significance of heterogeneity (or crowd composition) $H$. This can be observed even for cases of partial sensitivity analysis for the high platform ($E = 0$) and stairs ($E = 1$) scenarios, where $CoP(H)\approx 20\%$. In the scenario with an exit to an open line without any devices, the influence of heterogeneity is quantitatively higher than the influence of exit width (in the sense of explaining the variance in TET). 

\begin{table}[htb!]
\centering
	\begin{tabular}{l|C{1.3cm}||*{3}{C{1.4cm}|}}
	\hline\rowcolor{Gray}
	$(\%)$ & $CoP$ & $CoP(W)$ & $CoP(H)$ & $CoP(E)$ \\
	\hline
	Complete set & 92.9 & 42.1 & 25.9 & 31.5\\
	\hline
	high platform $E = 0$ & 89.5 & 68.8 & 20.1 & --\\
	Stairs $E = 1$ & 93.6 & 75.4 & 20.0 & --\\
	Open line $E = 2$ & 93.8 & 39.7 & 53.8 & --\\
	\hline
	\end{tabular}
\caption{Coefficients of Prediction ($COP$) for polynomial-fit regression using optiSLang. Third phase -- finer set of parameters.}
\label{tab:copfine}
\end{table}

The greater influence of heterogeneity for the fine set of scenarios is to be expected, since the sample space is significantly larger (2 values of heterogeneity were extended to 4). Despite this, we can draw similar conclusions as for the basic simulation phase: crowd composition plays an important role in the scenario with the exit to an open line without any evacuation device. There is no significant difference between the high platform and stairs scenarios in terms of crowd composition.

\subsubsection{Summary}

Sensitivity analysis performed using purely experimental data was supplemented by sensitivity analysis of simulated data using a Pathfinder model adjusted to fit the experimental observations. Several conclusions about the influence of the parameters investigated (exit width $W$, crowd composition $H$, and exit type $E$) on TET can be drawn from this study:
\begin{itemize}
    \item As expected, the width of the main exit $W$ ($CoP$ ranging from 40\% to 60\%) has a major influence on TET. This is not surprising, since exit width ranged from 0.65~m to 1.34~m. Thus, TET is significantly influenced by the maximal flow through the main exit, which increases almost linearly with exit width.
    
    \item In addition to exit width, exit type plays a very important role ($CoP$ ranging from 30\% to 40\%). However, the introduction of the variable $E$ with values $0$ for high platform, $1$ for stairs, and $2$ for exit without devices was rather artificial. Because of this, the sensitivity analysis was evaluated separately for individual exit types (i.e. investigating impact of exit width $W$ and crowd composition $H$ on TET).
    
    \item At first glance, crowd composition $H$ seems to have a rather small impact on TET ($CoP$ ranging from 10\% to 20\%). However, a separate investigation of individual exit types tells a different story. Crowd composition has a marginal or an even insignificant influence on TET in high platform and stairs scenarios (i.e. standard exit paths), but for the scenario with an exit to an open line without devices, the influence of crowd composition is comparable or even stronger than the influence of exit width.
\end{itemize}

Since the coefficients of prognosis for the polynomial models of the simulation data are sufficiently high (around 90\%), we can assume that equation~(\ref{eq:polyfit}) with estimated coefficients from Table~\ref{tab:coeff} describes the simulated data well. This may help compare the experiment with simulations from other perspectives.

\begin{table}[htb!]
\centering
	\begin{tabular}{l|*{11}{>{$}r<{$}|}}
	\hline\rowcolor{Gray}
		coefficient &
		\alpha_0 & 
		\alpha_1 & 
		\alpha_2 & 
		\alpha_3 & 
		\alpha_4 &
		\alpha_5 & 
		\alpha_6 & 
		\alpha_7 & 
		\alpha_8 & 
		\alpha_9 &
		R^2 \\
	\hline
		term &
		- & 
		W & 
		H & 
		E & 
		W^2 &
		H^2 & 
		E^2 & 
		WH & 
		HE & 
		WE &
		\\
	\hline
		experiment 46 &
		 80.82 & 
		-49.76 & 
		  0.17 & 
		 -4.96 & 
		 15.29 & 
		 & 
		  4.32 & 
		 -0.27 & 
		  0.26 & 
		 -0.53 & 
		 0.914 \\ 
	\hline
		simulation 46&
		 106.62 & 
		 -99.02 & 
		   0.02 & 
		   0.66 & 
		  38.90 & 
		 & 
		   2.60 & 
		  -0.11 & 
		   0.21 & 
		  -3.95 & 
		  0.921 \\ 
	\hline
		fine simul. 46&
		 106.36 & 
		 -98.09 & 
		  -0.06 & 
		  -0.23 & 
		  37.97 & 
		  0.004 & 
		   3.52 & 
		   0.07 & 
		   0.14 & 
		  -4.39 & 
		  0.930 \\ 
	\hline
	\end{tabular}
\caption{Values of coefficients of polynomial fit from optiSLang based on experimental data, basic simulation data, and fine simulation data. Last column contains the determination coefficient $R^2$.}
\label{tab:coeff}
\end{table}

\begin{figure}[htb!]
\centering
\includegraphics[width = 0.32\textwidth]{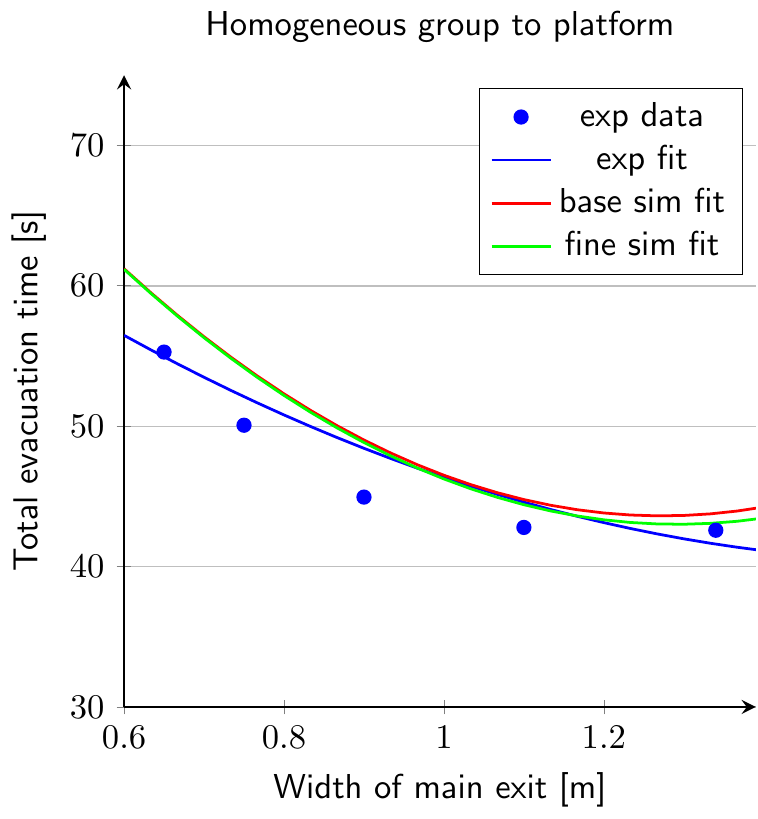}
\includegraphics[width = 0.32\textwidth]{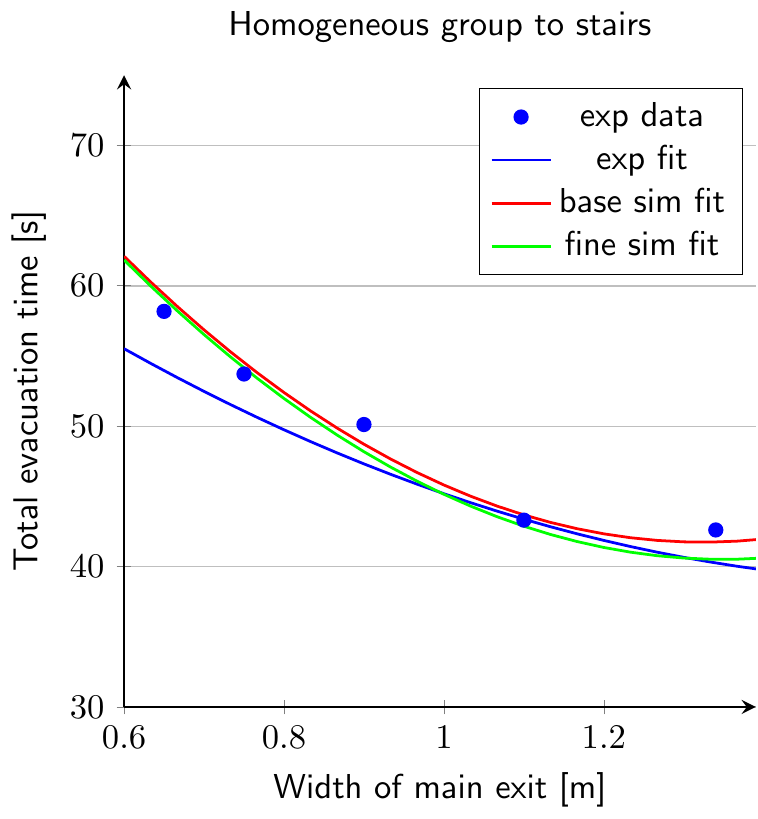}
\includegraphics[width = 0.32\textwidth]{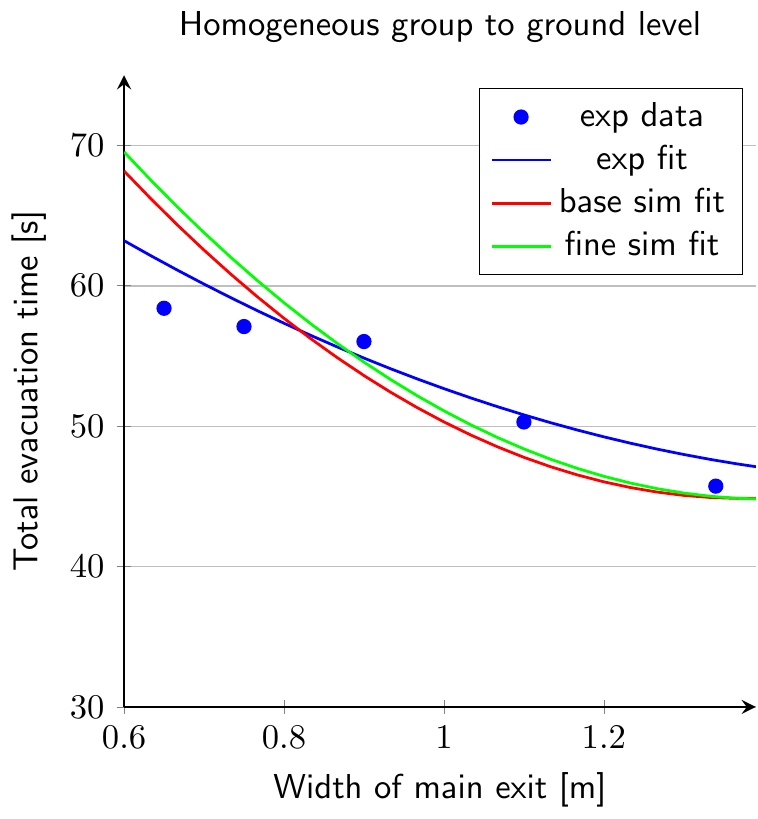}
\\
\includegraphics[width = 0.32\textwidth]{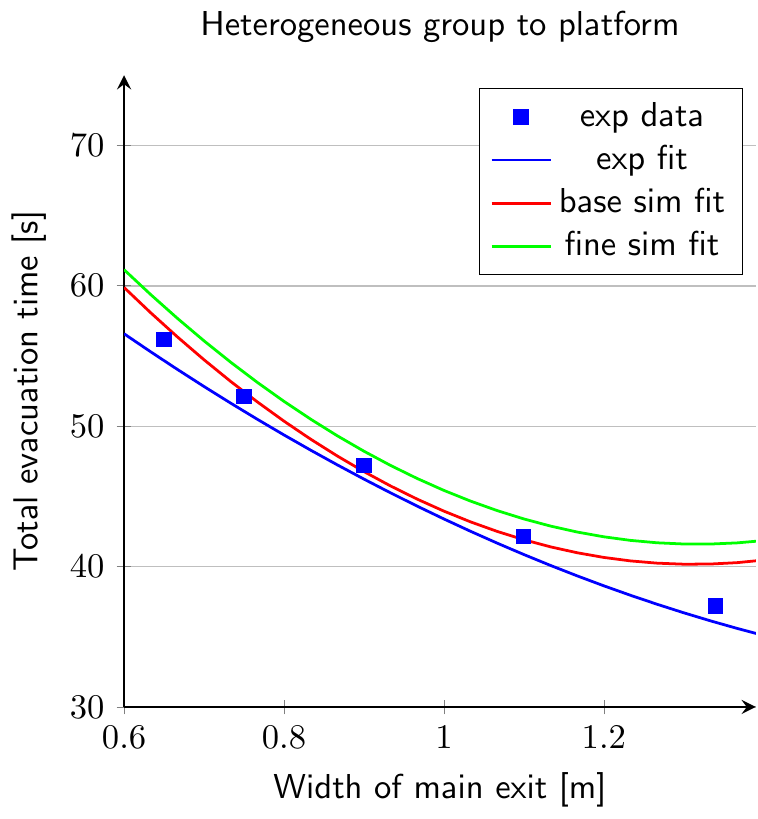}
\includegraphics[width = 0.32\textwidth]{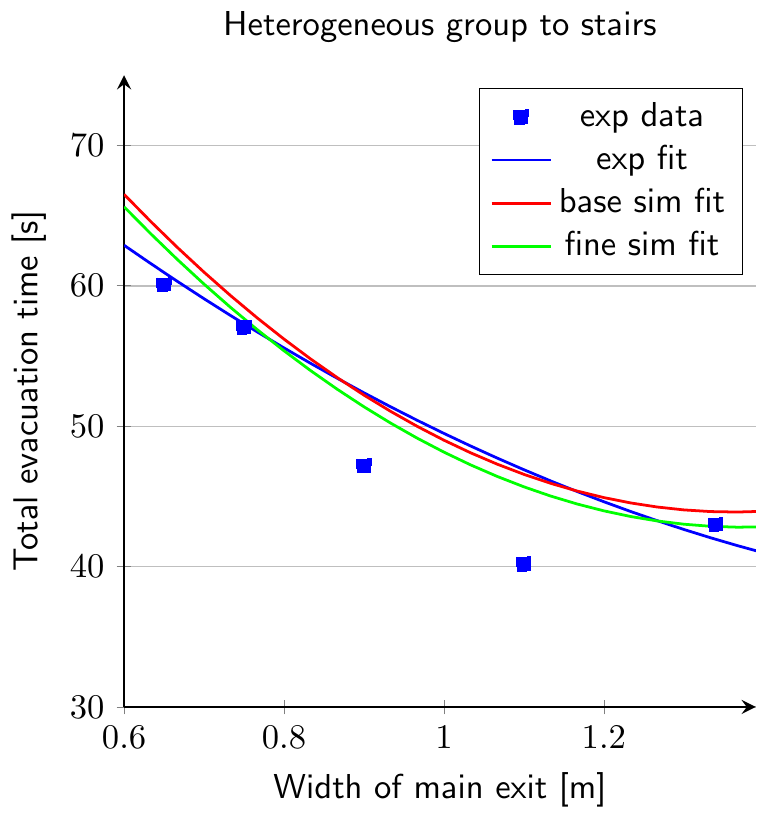}
\includegraphics[width = 0.32\textwidth]{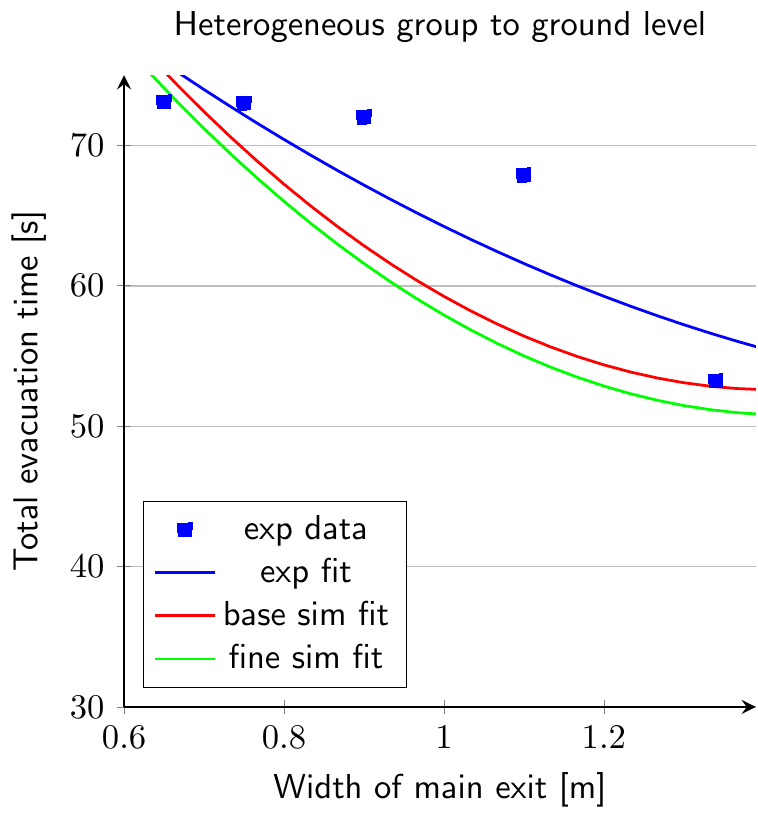}
\caption{Graphical comparison of corrected experimental data (TET corr) with  polynomial regression models~(\ref{eq:polyfit}) with estimated parameters given by Table~\ref{tab:coeff} using experiment (blue), basic simulation (red) and fine simulation (green) data.}
\label{fig:polyfit}
\end{figure}

Visual comparison of experimental measurements with the polynomial models~(\ref{eq:polyfit}) is provided in Figure~\ref{fig:polyfit}. The dependence of TET on exit width $W$ is presented for all combinations of crowd composition $H\in\{0,28\}$ and exit type $E\in\{0,1,2\}$ investigated in the experiment, including experimental data (blue points), the polynomial model derived from experimental data (blue line), the polynomial models derived from simulation data for the basic parameter set (red line), and the fine parameter set (green line).

From this comparison, we can draw the following conclusions:
\begin{itemize}
    \item The polynomial model derived from the simulations using the fine set of parameters is very similar to the basic model, i.e. simulations with more values of heterogeneity (crowd composition) did not reveal new contribution of the parameter $H$ to  TET. Thus, we believe that equation~(\ref{eq:polyfit}) with coefficients from Table~\ref{tab:coeff} describes the dependence of TET using Pathfinder simulations well for an arbitrary ratio of passengers with movement limitations ranging from 0\% to 60\%, while keeping the proportions of agents of given types among the types with limitations (``Children'', ``Adults carrying a toddler'', ``Seniors'', ``With disabilities'') in line with Table~\ref{tab:popu}.
    \item The polynomial model derived from simulations describes the experimental data well. This is another proof for validation of the Pathfinder simulation model, since the polynomial model represents the average TET obtained using simulations.
\end{itemize}

\clearpage
\newpage

\section{Conclusions}

A full-scale controlled evacuation experiment from a railcar (double-deck electric unit class 471, CityElefant) was carried out in Prague in 2018. The experiment consisted of 15~trials with homogeneous (HOM) and heterogeneous (HET) groups conducted in parallel for a total of 30~different evacuation scenarios. Evacuation scenarios included different main exit widths (650~mm, 750~mm, 900~mm, 1100~mm, and 1340~mm), different exit types (to a high platform, to an open line using stairs, to an open line without any device), and different compositions of types of passengers.
A total of 91~participants were divided into two groups, with the HOM group composed of participants without movement limitations (i.e. healthy adolescents or adults aged 11--60), and the HET group containing 26\% of participants with movement limitations (children ages 3-10 and able to move on their own; parents carrying toddlers; seniors over 60 with potential movement limitations; and those with other disabilities, such as an adult using crutches). Each evacuation scenario was conducted only once. In order to minimize possible training effects, scenario settings were changed in random order. 

17~cameras recorded the entire experimental evacuation process. Although total evacuation time (TET) was the main object of focus, the recordings also enabled measurement of other important variables such as exit flow rates, travel speeds, pre-movement activities, and exiting behaviours (notably the delay caused by navigating a 750~mm jump). The data also provided detailed information about the evacuation dynamics inside the narrow environment of a railcar. Quantification of exit behaviours represented by measured delays filled in some gaps in our understanding of how train evacuations unfold. The main and most original contribution of this study lies in quantification of the time delay caused by passengers contemplating how to navigate a 750~mm jump (and how to balance themselves afterwards) for all passenger types studied. Table~\ref{tab:delay} provides this information paired with movement limitations and the qualitative observation that, for main exit widths up to \numprint{1340}~mm, more than two passengers were not seen jumping together. 

The TET (key quantity for analysis) measured covers the time interval between a signal to evacuate (whistle blow) to the egress of the last passenger through the main exit. Since there were unequal numbers of evacuees in the groups considered (46~in the HET group, 42~in the HOM group), corrections for comparative purposes were made. For the case with only 42~passengers, an evacuation time for 46~passengers was estimated using a linear extrapolation of the egress times for the last 7~passengers from the group of 42. This enabled us to compare TET for all evacuation scenarios and to conduct further analysis and to draw conclusions. Our observations allowed us to conclude that: as expected, TET decreased as the width of the main exit increased, regardless of participant group or exit type. For exit to high platforms or using stairs, there did not appear to be a significant difference between the HOM and HET groups or these two exit types. However, in the case of an exit to an open line involving a drop 750~mm, a slight increase for TET was observed for the HOM group and TET increased significantly for the HET group. With the extended and detailed description of the experiment in this paper, the data provided may be useful for validating train evacuation models for possible use or as a case study for projects such the RiMEA Project~\cite{RiMEA} investigating the regulatory environment.
Nevertheless, the presented datasets should be always regarded in the full context of the nature of the experimental methods employed.

Based on the experimental study, a simulation model of the evacuation for one half of the railcar was created using Pathfinder. The agents were divided into groups according to the types of passengers represented. Key parameters for the agents in the model were derived directly from experimental measurements (unrestricted movement speed for the type ``Without limitations'', agent diameter based on experimental shoulder width measurements, and so on). The most important and original experimental data used for customizing  agent parameters in Pathfinder were the values 
925 for the time delays before and after a 750~mm jump. Thanks to this data as well as further modification of default values (the compressibility parameter and width reduction), we were able to capture the specific nature of railcar evacuation through a narrow environment using Pathfinder. The model was validated in the first simulation phase using a comparison of the sequences of egress times for individual agents with the experimental data. We believe that the experience of performing such simulations using experimental data may significantly assist other modellers when simulating emergency railcar evacuation. 

The simulation results, by means of the analysis performed on the simulation dataset, were used as a support for the conclusions drawn from sensitivity analysis of the experimental data. In the second phase of simulation, 30~railcar evacuation simulations with 46~agents were performed for each scenario. In order to mimic the experiment, agent parameters were kept unchanged and only the randomization seeds were modified. Thus, the simulations can truly be considered as being additional realizations of the experiment. In this phase, two ratios of passengers with limitations were considered: 0\% and 28\%.

Using experimental and simulation data, sensitivity analysis for total evacuation time (TET) was performed for main exit width, crowd composition, and exit type. Sensitivity analysis results quantitatively support the observations made during the experiment. As a result of sensitivity analysis, we can conclude that main exit width has a major influence on TET, as expected. More important, however, is the investigation of the influence of crowd composition (percentage of people without limitations) on TET. The sensitivity analysis indicated that for standard exit types (high platform, using wide staircase), crowd composition (heterogeneity) played a marginal role. On the contrary, for egress to an open line without any device including a 750~mm jump, crowd composition played a role comparable to main exit width.

The influence of crowd heterogeneity (presence of passengers with limitations) is even more evident from sensitivity analysis conducted for the third phase of simulations, in which the ratio of agents of the types with limitations (``Children'', ``Carrying a toddler'', ``Seniors'', and ``With disabilities'',  ranged from 0\% to 60\%. The key message we can draw from this analysis is that further research must take the presence of passengers with movement limitations into account, especially in regard to situations involving non-standard exit types. 

The performed analysis indicates that a non-standard exit path during a railcar evacuation is an important factor influencing the evacuation time, particularly when the railcar is occupied by a heterogeneous population of passengers with respect to their movement abilities. As the equipment and shapes of railcar exit paths are very variable including steep stairs, foot boards, different handrails for assisting egress, among other variations, further studies should evaluate the influence of such variable configurations on the evacuation process. The possible correlation with passenger crowd composition on evacuation time should also be the subject of future investigations. 

\section*{Acknowledgments}
The authors acknowledge all participants and team members involved in organizing the experiment for their effort and enthusiasm, which made this research project possible. The authors also thank Dr. Stephanie Krueger for English language review and editing. This project was supported by the Grant Agency of the Czech Technical University in Prague, grant No.~SGS18/107/OHK1/2T/11, and grant No.~SGS18/188/OHK4/3T/14.\\

\section*{References}
\bibliography{references}

\begin{thebibliography}{10}
\expandafter\ifx\csname url\endcsname\relax
  \def\url#1{\texttt{#1}}\fi
\expandafter\ifx\csname urlprefix\endcsname\relax\def\urlprefix{URL }\fi
\expandafter\ifx\csname href\endcsname\relax
  \def\href#1#2{#2} \def\path#1{#1}\fi

\bibitem{braun_fire_1975}
E.~Braun, A {Fire} {Hazard} {Evaluation} of the {Interior} of {WMATA}
  {Metrorail} {Cars}, Tech. Rep. NBSIR 75-971, Center for Fire Research,
  Institute for Applied Technology, National Bureau of Standards, Washington
  (1975).

\bibitem{braun_fire_1978}
E.~Braun, Fire {Hazard} {Evaluation} of {BART} {Vehicles}, Tech. Rep. NBSIR
  78-1421, Center for Fire Research, Institute for Applied Technology, National
  Bureau of Standards, Washington (1978).

\bibitem{peacock_fire_1984}
R.~D. Peacock, E.~Braun, Fire {Tests} of {Amtrak} {Passenger} {Rail} {Vehicle}
  {Interiors}, Tech. Rep. NBS TN 1193, Center for Fire Research, Institute for
  Applied Technology, National Bureau of Standards, Washington (1984).

\bibitem{peacock_fire_1999}
R.~D. Peacock, E.~Braun, Fire {Safety} of {Passenger} {Trains} {Phase} {I}:
  {Material} {Evaluation} ({Cone} {Calorimeter}), Tech. Rep. NISTIR 6132,
  National Institute of Standards and Technology, Gaithersburg, (1999).

\bibitem{peacock_fire_2002}
R.~D. Peacock, P.~A. Reneke, J.~D. Averill, R.~W. Bukowski, J.~H. Klote, Fire
  {Safety} of {Passenger} {Trains} {Phase} {II}: {Application} of {Fire}
  {Hazard} {Analysis} {Techniques}, Tech. Rep. NISTIR 6525, National Institute
  of Standards and Technology, Gaithersburg (2002).

\bibitem{peacock_fire_2004}
R.~D. Peacock, J.~D. Averill, R.~W. Bukowski, D.~W. Stroup, D.~Madrzykowski,
  P.~A. Reneke, Fire {Safety} of {Passenger} {Trains} {Phase} {III}:
  {Evaluation} of {Fire} {Hazard} {Analysis} {Using} {Full}-{Scale} {Passenger}
  {Rail} {Car} {Tests}, Tech. Rep. NISTIR 6563, National Institute of Standards
  and Technology, Gaithersburg (2004).

\bibitem{goransson_fires_1990}
U.~G\"{o}ransson, A.~Lundkvist, {Statens provningsanstalt (Sweden)}, Fires in
  buses and trains, fire test methods, Statens provningsanstalt, Borås, 1990,
  oCLC: 24432835.

\bibitem{briggs_firestarr_2001}
P.~Briggs, S.~Métral, P.~Gil, Y.~Le~Tallec, V.~Le~Sant, D.~Troïano,
  S.~Marrucci, S.~Messa, C.~Baiocchi, H.~Bruelet, {FIRESTARR} {Final} report,
  Tech. Rep. Contract SMT4 - CT 97 - 2164, FIRESTARR Consortium (2001).

\bibitem{noauthor_commission_2014}
{COMMISSION} {REGULATION} ({EU}) {No} 1302/2014 of 18 {November} 2014
  concerning a technical specification for interoperability relating to the
  ‘rolling stock locomotives and passenger rolling stock’ subsystem of the
  rail system in the {European} {Union} (2014).

\bibitem{noauthor_atoc_2002}
{ATOC} {Vehicles} {Standard} {AV}/{ST}9002, {Vehicle} {Interiors} {Design} for
  {Evacuation} and {Fire} {Safety} (2002).

\bibitem{markos_passenger_2013}
S.~Markos, H., J.~K. Pollard, Passenger {Train} {Emergency} {Systems}: {Review}
  of {Egress} {Variables} and {Egress} {Simulation} {Models}, Tech. Rep.
  DOT/FRA/ORD-13/22, Research and Innovative Technology Administration John A.
  Volpe National Transportation Systems Center, Cambridge, MA (2013).

\bibitem{kuligowski_computer_2016}
E.~D. Kuligowski, \href{https://doi.org/10.1007/978-1-4939-2565-0_60}{Computer
  {Evacuation} {Models} for {Buildings}}, in: M.~J. Hurley, D.~Gottuk, J.~R.
  Hall, K.~Harada, E.~Kuligowski, M.~Puchovsky, J.~Torero, J.~M. Watts,
  C.~Wieczorek (Eds.), {SFPE} {Handbook} of {Fire} {Protection} {Engineering},
  Springer, New York, NY, 2016, pp. 2152--2180.
\newblock \href {https://doi.org/10.1007/978-1-4939-2565-0_60}
  {\path{doi:10.1007/978-1-4939-2565-0_60}}.
\newline\urlprefix\url{https://doi.org/10.1007/978-1-4939-2565-0_60}

\bibitem{ronchi_developing_2020}
E.~Ronchi,
  \href{https://www.sciencedirect.com/science/article/pii/S0379711220300679}{Developing
  and validating evacuation models for fire safety engineering}, Fire Safety
  Journal (2020) 103020\href {https://doi.org/10.1016/j.firesaf.2020.103020}
  {\path{doi:10.1016/j.firesaf.2020.103020}}.
\newline\urlprefix\url{https://www.sciencedirect.com/science/article/pii/S0379711220300679}

\bibitem{siyam_research_2019}
N.~Siyam, O.~Alqaryouti, S.~Abdallah, Research {Issues} in {Agent}-{Based}
  {Simulation} for {Pedestrians} {Evacuation}, IEEE Access PP (2019) 1--1.
\newblock \href {https://doi.org/10.1109/ACCESS.2019.2956880}
  {\path{doi:10.1109/ACCESS.2019.2956880}}.

\bibitem{RahDatLov2016FEMTC}
A.~Rahouti, S.~Datoussa\"{i}d, R.~Lovreglio,
  \href{https://www.thunderheadeng.com/2017/03/d1-05-datoussaid/}{A sensitivity
  analysis of a hospital evacuation in case of fire: The impact of the
  percentage of people with reduced mobility and the staff to occupant’s
  ratio}, in: Proceedings of Fire and Evacuation Modelling Conference
  (FEMTC2016), Torremolinos, Spain, 2016, pp. 1--15.
\newline\urlprefix\url{https://www.thunderheadeng.com/2017/03/d1-05-datoussaid/}

\bibitem{huang_experimental_2018}
S.~Huang, S.~Lu, S.~Lo, C.~Li, Y.~Guo, Experimental study on occupant
  evacuation in narrow seat aisle, Physica A: Statistical Mechanics and its
  Applications 502 (2018) 506--517.
\newblock \href {https://doi.org/10.1016/j.physa.2018.02.032}
  {\path{doi:10.1016/j.physa.2018.02.032}}.

\bibitem{klingsch_evacuation_2010}
M.~Oswald, H.~Kirchberger, C.~Lebeda,
  \href{http://link.springer.com/10.1007/978-3-642-04504-2_5}{Evacuation of a
  {High} {Floor} {Metro} {Train} in a {Tunnel} {Situation}: {Experimental}
  {Findings}}, in: W.~W.~F. Klingsch, C.~Rogsch, A.~Schadschneider,
  M.~Schreckenberg (Eds.), Pedestrian and {Evacuation} {Dynamics} 2008,
  Springer Berlin Heidelberg, Berlin, Heidelberg, 2010, pp. 67--81.
\newline\urlprefix\url{http://link.springer.com/10.1007/978-3-642-04504-2_5}

\bibitem{kim_experiments_2012}
J.-H. Kim, W.-H. Kim, D.-H. Lee, W.-S. Jung,
  \href{http://koreascience.or.kr/journal/view.jsp?kj=HJSBCY&py=2012&vnc=v26n3&sp=1}{Experiments
  of {Egress} {Behavior} {When} {Subway} {Car} {Stops} on {Railroad}}, Journal
  of Korean Institute of Fire Science and Engineering 26~(3) (2012) 1--7.
\newblock \href {https://doi.org/10.7731/KIFSE.2012.26.3.001}
  {\path{doi:10.7731/KIFSE.2012.26.3.001}}.
\newline\urlprefix\url{http://koreascience.or.kr/journal/view.jsp?kj=HJSBCY&py=2012&vnc=v26n3&sp=1}

\bibitem{fridolf_fire_2013}
K.~Fridolf, D.~Nilsson, H.~Frantzich,
  \href{http://link.springer.com/10.1007/s10694-011-0217-x}{Fire {Evacuation}
  in {Underground} {Transportation} {Systems}: {A} {Review} of {Accidents} and
  {Empirical} {Research}}, Fire Technology 49~(2) (2013) 451--475.
\newblock \href {https://doi.org/10.1007/s10694-011-0217-x}
  {\path{doi:10.1007/s10694-011-0217-x}}.
\newline\urlprefix\url{http://link.springer.com/10.1007/s10694-011-0217-x}

\bibitem{fridolf_flow_2014}
K.~Fridolf, D.~Nilsson, H.~Frantzich,
  \href{http://linkinghub.elsevier.com/retrieve/pii/S0925753513002336}{The flow
  rate of people during train evacuation in rail tunnels: {Effects} of
  different train exit configurations}, Safety Science 62 (2014) 515--529.
\newblock \href {https://doi.org/10.1016/j.ssci.2013.10.008}
  {\path{doi:10.1016/j.ssci.2013.10.008}}.
\newline\urlprefix\url{http://linkinghub.elsevier.com/retrieve/pii/S0925753513002336}

\bibitem{fridolf_evacuation_2016}
K.~Fridolf, D.~Nilsson, H.~Frantzich,
  \href{http://link.springer.com/10.1007/s10694-015-0471-4}{Evacuation of a
  {Metro} {Train} in an {Underground} {Rail} {Transportation} {System}: {Flow}
  {Rate} {Capacity} of {Train} {Exits}, {Tunnel} {Walking} {Speeds} and {Exit}
  {Choice}}, Fire Technology 52~(5) (2016) 1481--1518.
\newblock \href {https://doi.org/10.1007/s10694-015-0471-4}
  {\path{doi:10.1007/s10694-015-0471-4}}.
\newline\urlprefix\url{http://link.springer.com/10.1007/s10694-015-0471-4}

\bibitem{cuesta_experimental_2017}
A.~Cuesta, O.~Abreu, A.~Balboa, D.~Alvear,
  \href{https://linkinghub.elsevier.com/retrieve/pii/S0886779816308781}{An
  experimental data-set on merging flows in rail tunnel evacuation}, Tunnelling
  and Underground Space Technology 70 (2017) 155--165.
\newblock \href {https://doi.org/10.1016/j.tust.2017.08.001}
  {\path{doi:10.1016/j.tust.2017.08.001}}.
\newline\urlprefix\url{https://linkinghub.elsevier.com/retrieve/pii/S0886779816308781}

\bibitem{noren_modelling_2003}
A.~Norén, J.~Winér, Modelling {Crowd} {Evacuation} from {Road} and {Train}
  {Tunnels} – {Data} and design for faster evacuations, Tech. Rep. Report
  5127, Department of Fire Safety Engineering Lund University, Sweden, Lund
  (2003).

\bibitem{markos_passenger_2015}
S.~Markos, H., J.~K. Pollard, Passenger {Train} {Emergency} {Systems}:
  {Single}-{Level} {Commuter} {Rail} {Car} {Egress} {Experiments}, Tech. Rep.
  DOT/FRA/ORD-15/04, Research and Innovative Technology Administration John A.
  Volpe National Transportation Systems Center, Cambridge, MA (2015).

\bibitem{Pathfinder}
T.~Engineering, \href{https://www.thunderheadeng.com/pathfinder/}{Pathfinder},
  [online] (2019).
\newline\urlprefix\url{https://www.thunderheadeng.com/pathfinder/}

\bibitem{data}
H.~Najmanov\'a, L.~Kukl\'ik, V.~Pe\v{s}kov\'a, M.~Buk\'a\v{c}ek, P.~Hrab\'ak,
  D.~Va\v{s}ata,
  \href{https://data.mendeley.com/datasets/w577r33mxz/draft?a=1fe02b51-e1ee-4372-8342-fe8f865045de}{Data
  for: Evacuation trials from a double-deck electric train unit: Experimental
  data and sensitivity analysis}, Mendeley Data, Draft(Version 1).
\newline\urlprefix\url{https://data.mendeley.com/datasets/w577r33mxz/draft?a=1fe02b51-e1ee-4372-8342-fe8f865045de}

\bibitem{galea_estimating_2000}
E.~R. Galea, S.~Gwynne,
  \href{https://onlinelibrary.wiley.com/doi/abs/10.1002/1099-1018%28200011/12%2924%3A6%3C291%3A%3AAID-FAM750%3E3.0.CO%3B2-6}{Estimating
  the flow rate capacity of an overturned rail carriage end exit in the
  presence of smoke}, Fire and Materials 24~(6) (2000) 291--302.
\newblock \href
  {https://doi.org/10.1002/1099-1018(200011/12)24:6<291::AID-FAM750>3.0.CO;2-6}
  {\path{doi:10.1002/1099-1018(200011/12)24:6<291::AID-FAM750>3.0.CO;2-6}}.
\newline\urlprefix\url{https://onlinelibrary.wiley.com/doi/abs/10.1002/1099-1018%28200011/12%2924%3A6%3C291%3A%3AAID-FAM750%3E3.0.CO%3B2-6}

\bibitem{waldau_full-scale_2007}
M.~Oswald, C.~Lebeda, U.~Schneider, H.~Kirchberger,
  \href{http://link.springer.com/10.1007/978-3-540-47064-9_4}{Full-{Scale}
  {Evacuation} {Experiments} in a smoke filled {Rail} {Carriage} — a detailed
  study of passenger behaviour under reduced visibility}, in: N.~Waldau,
  P.~Gattermann, H.~Knoflacher, M.~Schreckenberg (Eds.), Pedestrian and
  {Evacuation} {Dynamics} 2005, Springer Berlin Heidelberg, Berlin, Heidelberg,
  2007, pp. 41--55.
\newline\urlprefix\url{http://link.springer.com/10.1007/978-3-540-47064-9_4}

\bibitem{capote_analysis_2012}
J.~Capote, D.~Alvear, O.~Abreu, A.~Cuesta,
  \href{http://linkinghub.elsevier.com/retrieve/pii/S0379711211001664}{Analysis
  of evacuation procedures in high speed trains fires}, Fire Safety Journal 49
  (2012) 35--46.
\newblock \href {https://doi.org/10.1016/j.firesaf.2011.12.008}
  {\path{doi:10.1016/j.firesaf.2011.12.008}}.
\newline\urlprefix\url{http://linkinghub.elsevier.com/retrieve/pii/S0379711211001664}

\bibitem{kangedal_fire_2002}
P.~Kangedal, D.~Nilsson, Fire safety on intercity and interregional multiple
  unit trans, Tech. Rep. 5117, Department of Fire Safety Engineering, Lund
  University, Lund (2002).

\bibitem{kindler_evacuation_2012}
C.~Kindler, J.~Sørensen, A.~Dederichs, Evacuation of mixed populations from
  trains on bridges, in: Bridge {Maintenance}, {Safety}, {Management},
  {Resilience} and {Sustainability} : {Proceedings} of the {Sixth}
  {International} {Conference} on {Bridge} {Maintenance}, {Safety} and
  {Management}, C R C Press LLC, Stresa, Lake Maggiore, Italy, 2012, pp.
  1573--1579.

\bibitem{anvari_toward_2017}
B.~Anvari, C.~K. Nip, A.~Majumdar,
  \href{https://trid.trb.org/view/1439293}{Toward an {Accurate} {Microscopic}
  {Passenger} {Train} {Evacuation} {Model} {Using} {MassMotion}}, in: TRB 96th
  Annual Meeting Compendium of Papers, 2017, p. 16p.
\newline\urlprefix\url{https://trid.trb.org/view/1439293}

\bibitem{daamen_passengers_2007}
W.~Daamen, S.~P. Hoogendoorn, N.~Lundgren,
  \href{https://trid.trb.org/view/876648}{Passengers {Evacuating} in {Rail}
  {Tunnels}: {A} {Simulation} {Study} to {Predict} {Evacuation} {Times}}, in:
  11th World Conference on Transport Research, 2007, p. 26p.
\newline\urlprefix\url{https://trid.trb.org/view/876648}

\bibitem{jevtic_safety_2017}
R.~Jevtic,
  \href{http://scindeks.ceon.rs/Article.aspx?artid=0409-29531701098J}{The
  safety in tunnels: {An} example of simulated evacuation from a railway
  tunnel}, Bezbednost, Beograd 59~(1) (2017) 98--114.
\newblock \href {https://doi.org/10.5937/bezbednost1701098J}
  {\path{doi:10.5937/bezbednost1701098J}}.
\newline\urlprefix\url{http://scindeks.ceon.rs/Article.aspx?artid=0409-29531701098J}

\bibitem{galea_passenger_2014}
E.~Galea, D.~Blackshields, K.~Finney, D.~Cooney, Passenger {Train} {Emergency}
  {Systems}: {Development} of {Prototype} {railEXODUS} {Software} for {U}.{S}.
  {Passenger} {Rail} {Car} {Egress}, Tech. Rep. DOT/FRA/ORD-14/35, Fire Safety
  Engineering Group, University of Greenwich, Washington (2014).

\bibitem{alonso_new_2014}
V.~Alonso, O.~V. Abreu, A.~Cuesta, D.~Silió,
  \href{http://linkinghub.elsevier.com/retrieve/pii/S1877042814062417}{A {New}
  {Approach} for {Modelling} {Passenger} {Trains} {Evacuation} {Procedures}},
  Procedia - Social and Behavioral Sciences 160 (2014) 284--293.
\newblock \href {https://doi.org/10.1016/j.sbspro.2014.12.140}
  {\path{doi:10.1016/j.sbspro.2014.12.140}}.
\newline\urlprefix\url{http://linkinghub.elsevier.com/retrieve/pii/S1877042814062417}

\bibitem{hutchison_agent-based_2009}
F.~Kl\"{u}gl, G.~Klubertanz, G.~Rindsf\"{u}ser,
  \href{http://link.springer.com/10.1007/978-3-642-04617-9_79}{Agent-{Based}
  {Pedestrian} {Simulation} of {Train} {Evacuation} {Integrating}
  {Environmental} {Data}}, in: D.~Hutchison, T.~Kanade, J.~Kittler, J.~M.
  Kleinberg, F.~Mattern, J.~C. Mitchell, M.~Naor, O.~Nierstrasz,
  C.~Pandu~Rangan, B.~Steffen, M.~Sudan, D.~Terzopoulos, D.~Tygar, M.~Y. Vardi,
  G.~Weikum, B.~Mertsching, M.~Hund, Z.~Aziz (Eds.), {KI} 2009: {Advances} in
  {Artificial} {Intelligence}, Vol. 5803, Springer Berlin Heidelberg, Berlin,
  Heidelberg, 2009, pp. 631--638.
\newline\urlprefix\url{http://link.springer.com/10.1007/978-3-642-04617-9_79}

\bibitem{wang_simulation_2014}
W.~Wang, T.~J. Lo,
  \href{https://linkinghub.elsevier.com/retrieve/pii/S1877705814004974}{A
  {Simulation} {Study} on {Passenger} {Escape} in {Rail} {Tunnels}}, Procedia
  Engineering 71 (2014) 552--557.
\newblock \href {https://doi.org/10.1016/j.proeng.2014.04.079}
  {\path{doi:10.1016/j.proeng.2014.04.079}}.
\newline\urlprefix\url{https://linkinghub.elsevier.com/retrieve/pii/S1877705814004974}

\bibitem{galea_development_2013}
E.~R. Galea, The development and validation of a rail car evacuation model, in:
  {INTERFLAM} 2013: proceedings of the thirteenth international conference ;
  [13th {International} {Fire} {Science} \& {Engineering} {Conference} ;
  {Royal} {Holloway} {College}, {University} of {London}, {UK}, 24th - 26th
  {June} 2013], Interscience Communications, London, 2013, pp. 1023--1034.

\bibitem{peacock_evacuation_2011}
J.~Capote, D.~Alvear, O.~Abreu, M.~Lázaro, A.~Cuesta,
  \href{http://link.springer.com/10.1007/978-1-4419-9725-8_38}{An {Evacuation}
  {Model} for {High} {Speed} {Trains}}, in: R.~D. Peacock, E.~D. Kuligowski,
  J.~D. Averill (Eds.), Pedestrian and {Evacuation} {Dynamics}, Springer US,
  Boston, MA, 2011, pp. 421--431.
\newline\urlprefix\url{http://link.springer.com/10.1007/978-1-4419-9725-8_38}

\bibitem{capote_stochastic_2012}
J.~A. Capote, D.~Alvear, O.~Abreu, A.~Cuesta, V.~Alonso,
  \href{http://link.springer.com/10.1007/s10694-012-0251-3}{A {Stochastic}
  {Approach} for {Simulating} {Human} {Behaviour} {During} {Evacuation}
  {Process} in {Passenger} {Trains}}, Fire Technology 48~(4) (2012) 911--925.
\newblock \href {https://doi.org/10.1007/s10694-012-0251-3}
  {\path{doi:10.1007/s10694-012-0251-3}}.
\newline\urlprefix\url{http://link.springer.com/10.1007/s10694-012-0251-3}

\bibitem{GreGre2014RDO}
G.~Grewolls, K.~Grewolls,
  \href{https://www.dynardo.de/bibliothek/rdo-journal.html}{Sensitivity
  analysis of evacuation simulations}, RDO-Journal (2014) 22--26An optional
  note.
\newline\urlprefix\url{https://www.dynardo.de/bibliothek/rdo-journal.html}

\bibitem{optiSLang}
{Dynardo - dynamic software \& engineering},
  \href{https://www.dynardo.de/en/software/optislang.html}{optislang}, [online]
  (2019).
\newline\urlprefix\url{https://www.dynardo.de/en/software/optislang.html}

\bibitem{NPLsensitivity}
T.~J. Esward, C.~E. Matthews, L.~Wright, X.-S. Yang,
  \href{http://eprintspublications.npl.co.uk/4783/}{Sensitivity analysis,
  optimisation, and sampling methods applied to continuous models.}, {NPL
  Report}, National Physical Laboratory, Teddington, Middlesex (November 2010).
\newline\urlprefix\url{http://eprintspublications.npl.co.uk/4783/}

\bibitem{saltelli:2008}
A.~Saltelli, Global sensitivity analysis: the primer, John Wiley, 2008.

\bibitem{BukPesNaj_tgf2019}
M.~Buk{\'a}{\v{c}}ek, V.~Pe{\v{s}}kov{\'a}, H.~Najmanov{\'a}, Double-deck rail
  car egress experiment: Microscopic analysis of pedestrian time headways, in:
  I.~Zuriguel, A.~Garcimartin, R.~Cruz (Eds.), Traffic and Granular Flow 2019,
  Springer International Publishing, Cham, 2020, pp. 449--455.

\bibitem{SeyPasSteBolRupKli2009TS}
A.~Seyfried, O.~Passon, B.~Steffen, M.~Boltes, T.~Rupprecht, W.~Klingsch, {New
  Insights into Pedestrian Flow Through Bottlenecks}, Transportation Science
  43~(3) (2009) 395--406.
\newblock \href {https://doi.org/10.1287/trsc.1090.0263}
  {\path{doi:10.1287/trsc.1090.0263}}.

\bibitem{PathfinderMAN}
{Thunderhead Engineering},
  \href{https://www.thunderheadeng.com/wp-content/uploads/dlm_uploads/2011/07/users_guide-7.pdf}{User
  manual pathfinder}, [online] (2019).
\newline\urlprefix\url{https://www.thunderheadeng.com/wp-content/uploads/dlm_uploads/2011/07/users_guide-7.pdf}

\bibitem{PathfinderTECH}
{Thunderhead Engineering},
  \href{https://www.thunderheadeng.com/wp-content/uploads/dlm_uploads/2011/07/tech_ref-5.pdf}{Technical
  reference pathfinder}, [online] (2019).
\newline\urlprefix\url{https://www.thunderheadeng.com/wp-content/uploads/dlm_uploads/2011/07/tech_ref-5.pdf}

\bibitem{Rey1999steering}
C.~Reynolds, Steering behaviors for autonomous characters, in: proceedings of
  Game Developers Conference, 1999, pp. 763--782.

\bibitem{Hamilton2017FireSafety}
G.~N. Hamilton, P.~F. Lennon, J.~O’Raw,
  \href{http://www.sciencedirect.com/science/article/pii/S037971121730111X}{Human
  behaviour during evacuation of primary schools: Investigations on
  pre-evacuation times, movement on stairways and movement on the horizontal
  plane}, Fire Safety Journal 91 (2017) 937 -- 946, fire Safety Science:
  Proceedings of the 12th International Symposium.
\newblock \href {https://doi.org/https://doi.org/10.1016/j.firesaf.2017.04.016}
  {\path{doi:https://doi.org/10.1016/j.firesaf.2017.04.016}}.
\newline\urlprefix\url{http://www.sciencedirect.com/science/article/pii/S037971121730111X}

\bibitem{Kholsevnikov2012}
V.~Kholsevnikov, D.~Samoshin, I.~R., The problems of elderly people safe
  evacuation from senior citizen health care buildings in case of fire, in:
  Human Behaviour In Fire: Proceeding of 5th International Symposium, 2012, p.
  587–593.

\bibitem{Boyce1999}
K.~Boyce, T.~Shields, G.~Silcock, Toward the characterization of building
  occupancies for fire safety engineering: Capabilities of disabled people
  moving horizontally and on an incline, Fire Technology 35~(1) (1999) 51--67.
\newblock \href {https://doi.org/10.1023/A:1015339216366}
  {\path{doi:10.1023/A:1015339216366}}.

\bibitem{sfpe2002}
{National Fire Protection Association. and Society of Fire Protection
  Engineers.}, The title of the work, {Quincy, Mass. : National Fire Protection
  Association ; Bethesda, Md. : Society of Fire Protection Engineers}, The
  address, 2002.

\bibitem{RonNil2014BuildSim}
E.~Ronchi, D.~Nilsson,
  \href{https://doi.org/10.1007/s12273-013-0132-9}{Modelling total evacuation
  strategies for high-rise buildings}, Building Simulation 7~(1) (2014) 73--87.
\newblock \href {https://doi.org/10.1007/s12273-013-0132-9}
  {\path{doi:10.1007/s12273-013-0132-9}}.
\newline\urlprefix\url{https://doi.org/10.1007/s12273-013-0132-9}

\bibitem{Saltelli2018}
A.~Saltelli, M.~Ratto, T.~Andres, F.~Campolongo, J.~Cariboni, D.~Gatelli,
  M.~Saisana, S.~Tarantola, Global Sensitivity Analysis: The Primer, Wiley
  Publishing, 2008.

\bibitem{DuiDaaHoo2016PhysA}
D.~C. Duives, W.~Daamen, S.~P. Hoogendoorn,
  \href{http://www.sciencedirect.com/science/article/pii/S0378437115010134}{Continuum
  modelling of pedestrian flows — part 2: Sensitivity analysis featuring
  crowd movement phenomena}, Physica A: Statistical Mechanics and its
  Applications 447 (2016) 36 -- 48.
\newblock \href {https://doi.org/https://doi.org/10.1016/j.physa.2015.11.025}
  {\path{doi:https://doi.org/10.1016/j.physa.2015.11.025}}.
\newline\urlprefix\url{http://www.sciencedirect.com/science/article/pii/S0378437115010134}

\bibitem{LinWu2018AME}
C.~S. Lin, M.~E. Wu, A study of evaluating an evacuation time, Advances in
  Mechanical Engineering 10~(4) (2018) 1687814018772424.
\newblock \href {https://doi.org/10.1177/1687814018772424}
  {\path{doi:10.1177/1687814018772424}}.

\bibitem{BodRon2019CD}
N.~Bode, E.~Ronchi,
  \href{https://collective-dynamics.eu/index.php/cod/article/view/A20}{Statistical
  model fitting and model selection in pedestrian dynamics research},
  Collective Dynamics 4 (2019) 1--32.
\newblock \href {https://doi.org/10.17815/CD.2019.20}
  {\path{doi:10.17815/CD.2019.20}}.
\newline\urlprefix\url{https://collective-dynamics.eu/index.php/cod/article/view/A20}

\bibitem{Most2008Dynardo}
T.~Most, J.~Will, Meta-model of optimal prognosis -- an automatic approach for
  variable reduction and optimal meta-model selection, in: Proceedings of 5th
  Weimar Optimization and Stochastic Days 2008, 2008, p. 21p.

\bibitem{Most2011Dynardo}
T.~Most, J.~Will, Sensitivity analysis using the metamodel of optimal
  prognosis, in: Proceedings of 8th Weimar Optimization and Stochastic Days
  2011, 2011, p. 17p.

\bibitem{RiMEA}
\href{https://rimea.de/}{{Richtlinie f\"{u}r Mikroskopische
  EntfluchtungsAnalysen}}.
\newline\urlprefix\url{https://rimea.de/}

\end{thebibliography}

\appendix
\clearpage\newpage
\section{Table with measured and corrected TET for individual trials}
\label{app:TET}

\begin{table}[htb!]
\begin{center}
\begin{tabular}{|*{9}{c|}}
\hline
Exit & Comp. & Width [m]   & ID & number & TET [s]  & TET 46 [s] & delay [s] & TET corr \\
\hline\hline
\multirow{10}{*}{\rotatebox{90}{High platform}}
	& \multirow{5}{*}{HOM}
	  & 0.65 &  4A & 42 & 49.19           & 55.28  &       & 55.28    \\
    & & 0.75 &  8A & 42 & 44.00           & 50.07  &       & 50.07    \\
    & & 0.90 &  1A & 42 & \textbf{42.04}  & 44.95  &       & 44.95    \\
    & & 1.10 &  9A & 42 & 37.01           & 42.79  &       & 42.79   \\
    & & 1.34 &  5A & 42 & 37.10           & 42.59  &       & 42.59    \\
    \cline{2-9}
    & \multirow{5}{*}{HET}
      & 0.65 &  3A & 46 & 56.17           & 56.17  &       & 56.17    \\
    & & 0.75 &  7A & 46 & 52.14           & 52.14  &       & 52.14    \\
    & & 0.90 &  2A & 46 & \textbf{52.19}  & 52.19  & 5.00$^{\ast}$ & 47.19    \\
    & & 1.10 & 10A & 46 & 42.13           & 42.13  &       & 42.13    \\
    & & 1.34 &  6A & 46 & 37.20           & 37.20  &       & 37.20    \\
\hline
\multirow{10}{*}{\rotatebox{90}{Open line using stairs}}
    & \multirow{5}{*}{HOM}
      & 0.65 &  7B & 42 & 52.16           & 58.18  &       & 58.18    \\
    & & 0.75 &  3B & 42 & 48.13           & 53.72  &       & 53.72    \\
    & & 0.90 & 10B & 42 & 43.23           & 50.12  &       & 50.12    \\
    & & 1.10 &  2B & 42 & \textbf{39.06}  & 43.31  &       & 43.31    \\
    & & 1.34 &  6B & 42 & 37.02           & 42.61  &       & 42.61    \\
    \cline{2-9}
    & \multirow{5}{*}{HET}
      & 0.65 &  8B & 46 & 60.09           & 60.09  &       & 60.09    \\
    & & 0.75 &  4B & 46 & 57.04           & 57.04  &       & 57.04    \\
    & & 0.90 &  9B & 46 & 47.18           & 47.18  &       & 47.18    \\
    & & 1.10 &  1B & 46 & \textbf{40.19}  & 40.19  &       & 40.19    \\
    & & 1.34 &  5B & 46 & 43.00           & 43.00  &       & 43.00    \\
\hline
\multirow{10}{*}{\rotatebox{90}{Open line without any devices}} 
    & \multirow{5}{*}{HOM} 
      & 0.65 & 14B & 42 & 53.00           & 58.40  &       & 58.40    \\
    & & 0.75 & 13A & 42 & 51.24           & 57.10  &       & 57.10    \\
    & & 0.90 & 11B & 42 & \textbf{51.06}  & 56.03  &       & 56.03    \\
    & & 1.10 & 12A & 42 & 44.19           & 50.30  &       & 50.30    \\
    & & 1.34 & 15B & 42 & 39.00           & 45.73  &       & 45.73    \\
    \cline{2-9}
    & \multirow{5}{*}{HET}
      & 0.65 & 13B & 46 & 73.12           & 73.12  &       & 73.12    \\
    & & 0.75 & 14A & 46 & 73.00           & 73.00  &       & 73.00    \\
    & & 0.90 & 12B & 42 & 67.03           & 72.03  &       & 72.03    \\
    & & 1.10 & 11A & 42 & \textbf{68.00}  & 72.89  & 5.00$^{\dagger}$ & 67.89    \\
    & & 1.34 & 15A & 46 & 53.25           & 53.25  &       & 53.25    \\
\hline
\end{tabular}
\end{center}
$^{\ast}$ Delayed start of evacuation; passengers did not hear the whistle.\\
$^{\dagger}$ Delay caused by small children blocking the main exit due to fearing the jump.
\caption{Overview of measured total evacuation time (TET), extrapolated evacuation time for 46 passengers (TET 46), and corrected TET 46 with observed delay (TET corr). First trials for each given exit type in bold font.}
\label{tab:TETcorretions}
\end{table}

\clearpage\newpage
\section{Agent placement in Pathfinder simulations.}
\label{app:input}


\begin{figure}[htb!]
	\def\h{.45}
	\begin{tabular}{cc}
		HOM 42 & HOM 46\\
		\includegraphics[trim= 5pt 180pt 5pt 175pt, clip,width=\h\textwidth]{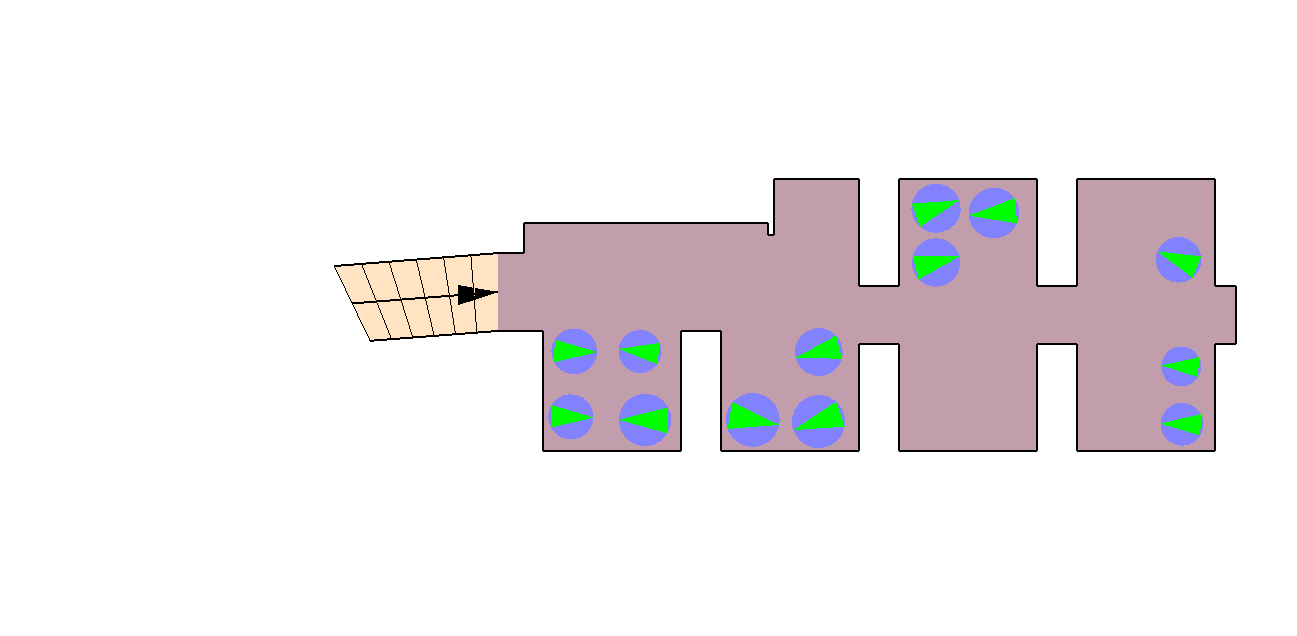} &
		\includegraphics[trim= 5pt 180pt 5pt 175pt, clip,width=\h\textwidth]{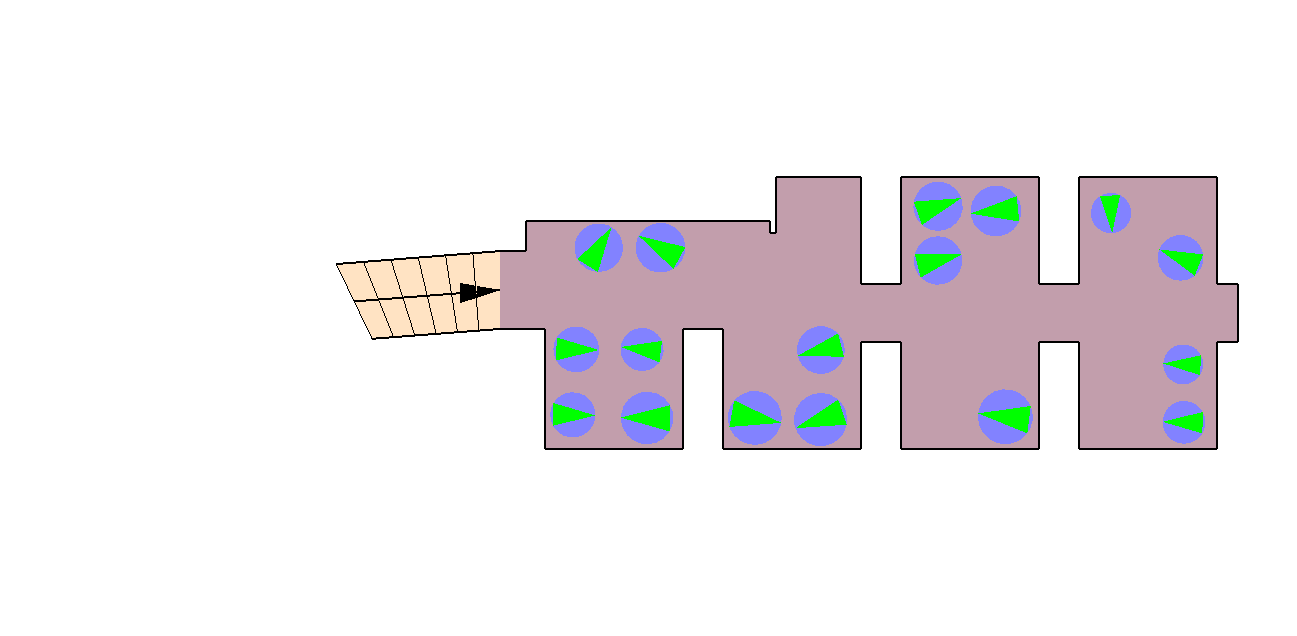} \\
		\includegraphics[trim= 5pt 170pt 5pt 175pt, clip,width=\h\textwidth]{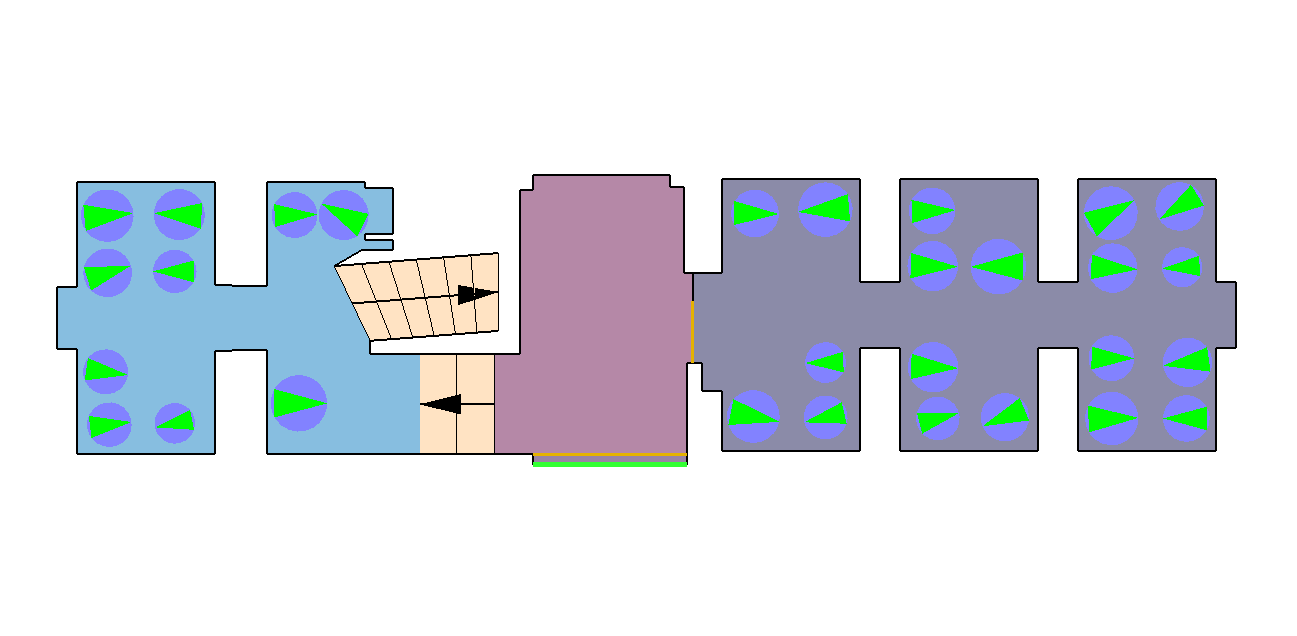} &
		\includegraphics[trim= 5pt 170pt 5pt 175pt, clip,width=\h\textwidth]{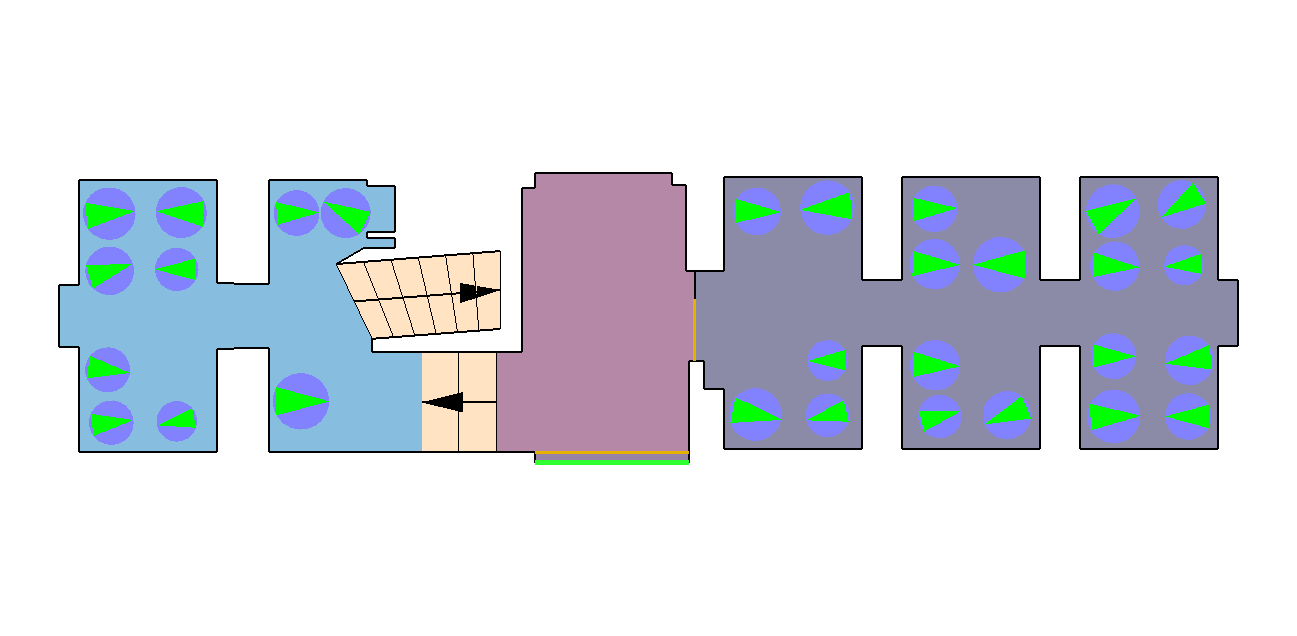} \\
		HET 42 (28\%) & HET 46 (28\%)\\
		\includegraphics[trim= 5pt 180pt 5pt 175pt, clip,width=\h\textwidth]{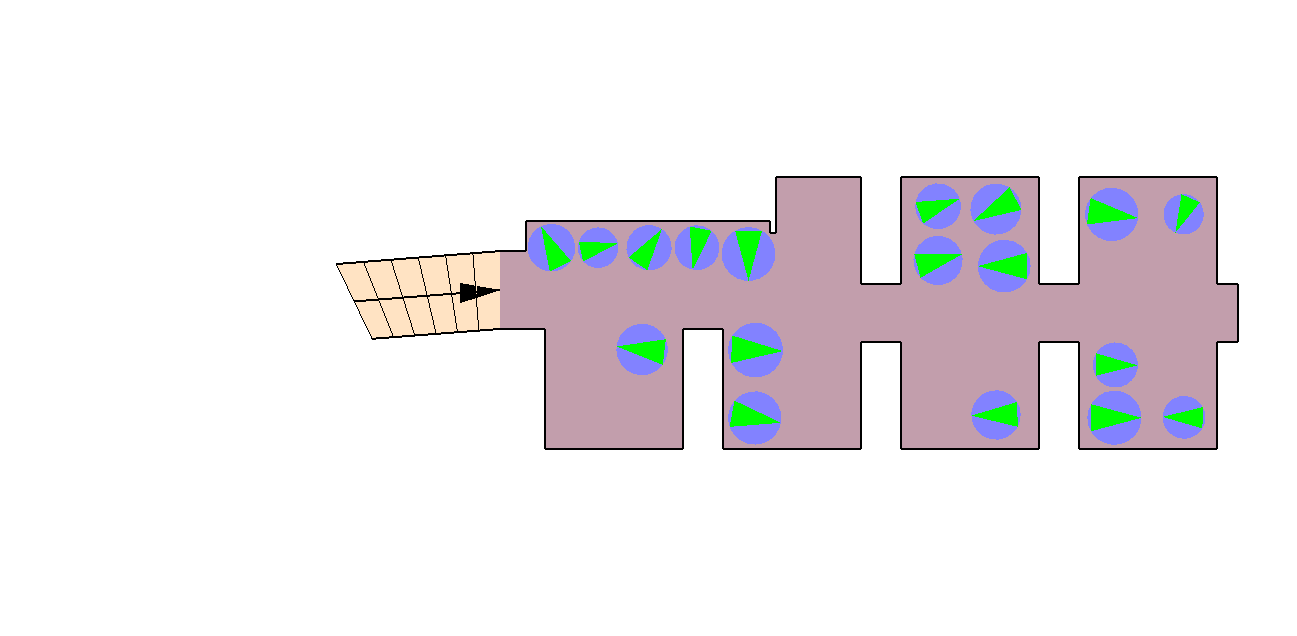} &
		\includegraphics[trim= 5pt 180pt 5pt 175pt, clip,width=\h\textwidth]{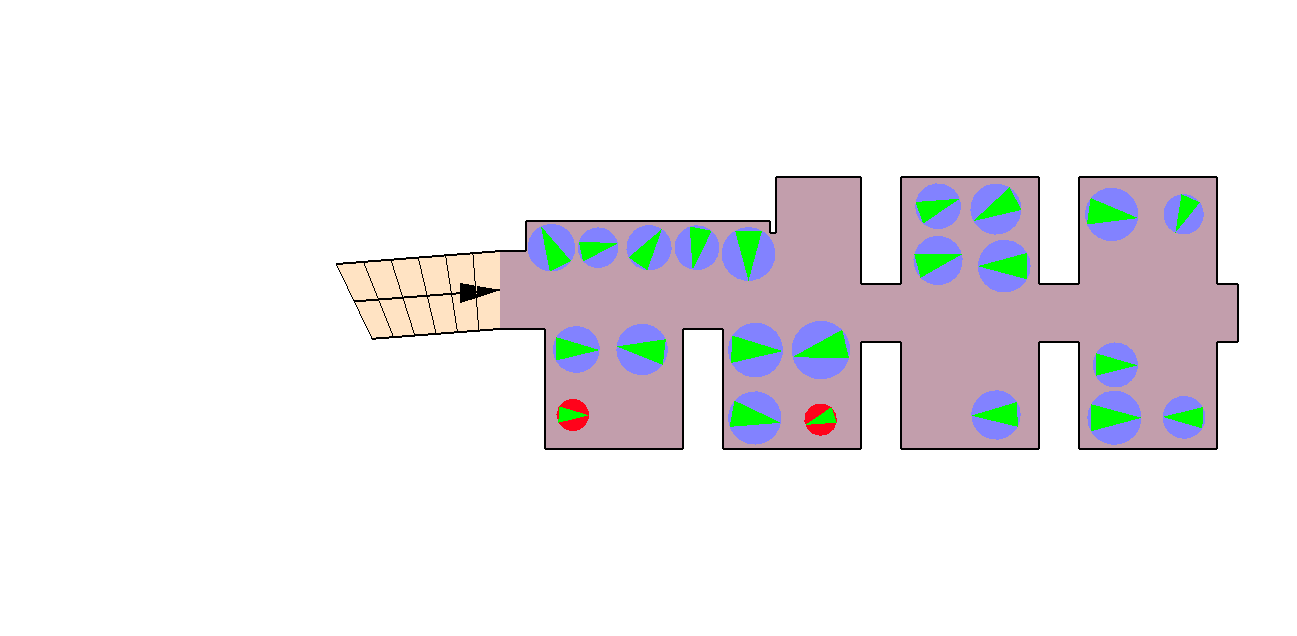} \\
		\includegraphics[trim= 5pt 170pt 5pt 175pt, clip,width=\h\textwidth]{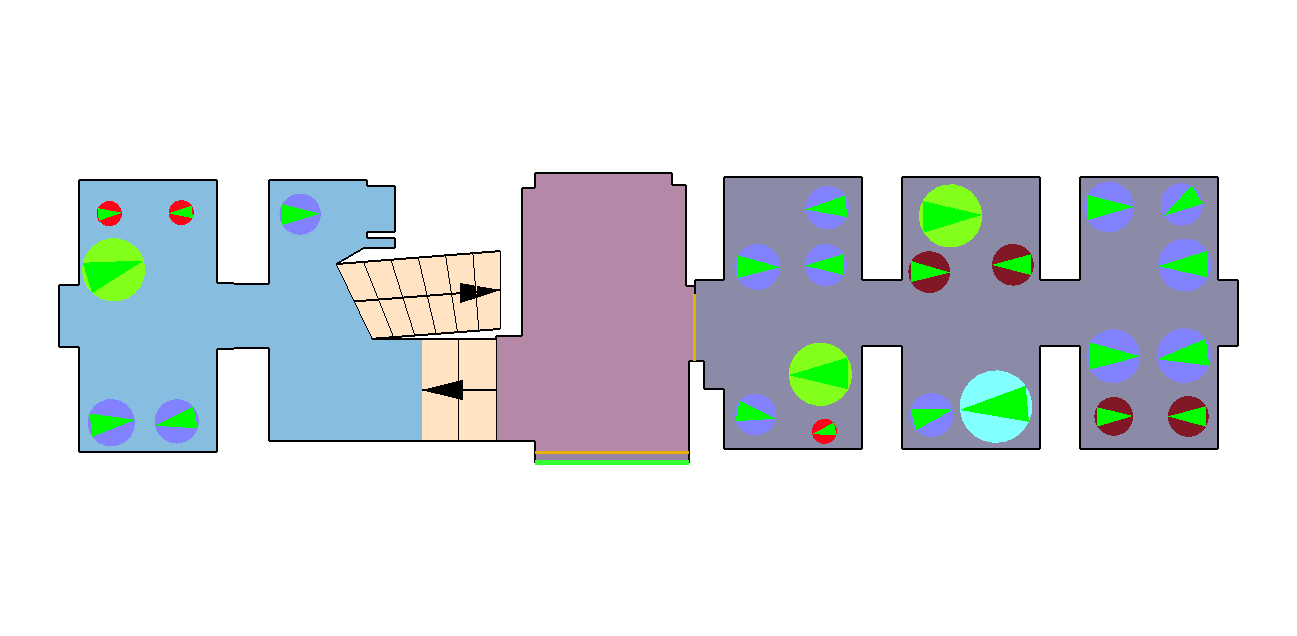} &
		\includegraphics[trim= 5pt 170pt 5pt 175pt, clip,width=\h\textwidth]{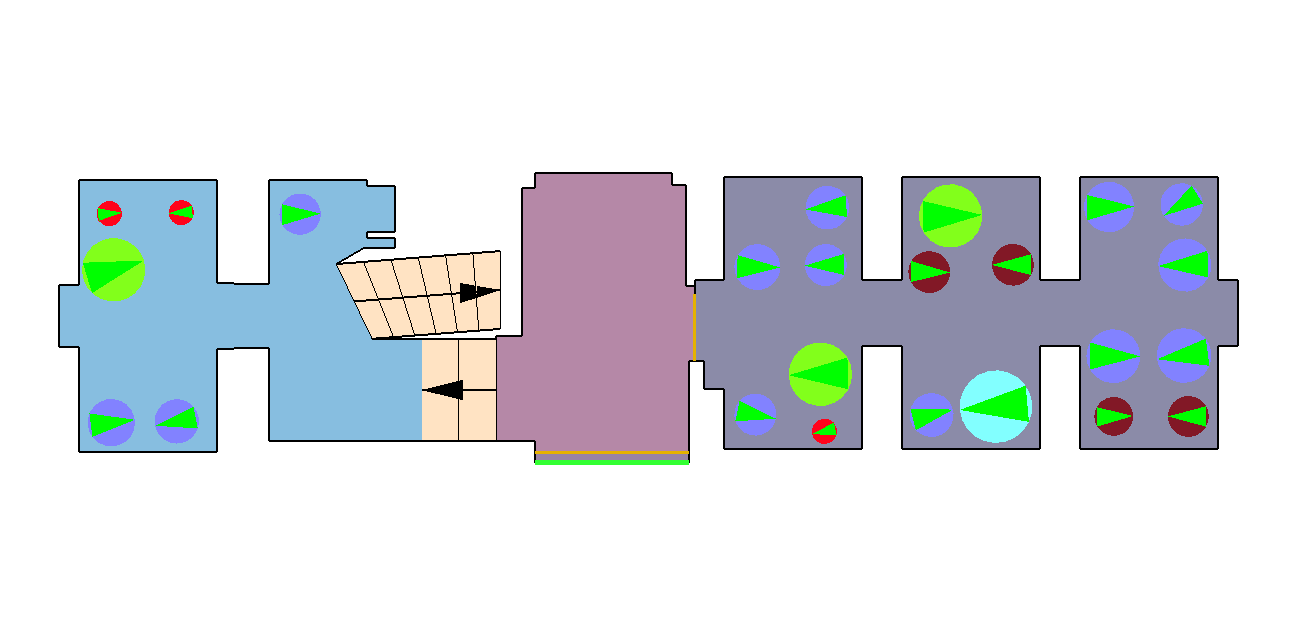} \\
		HET 46 (15\%) & HET 46 (56\%)\\
		\includegraphics[trim= 5pt 180pt 5pt 175pt, clip,width=\h\textwidth]{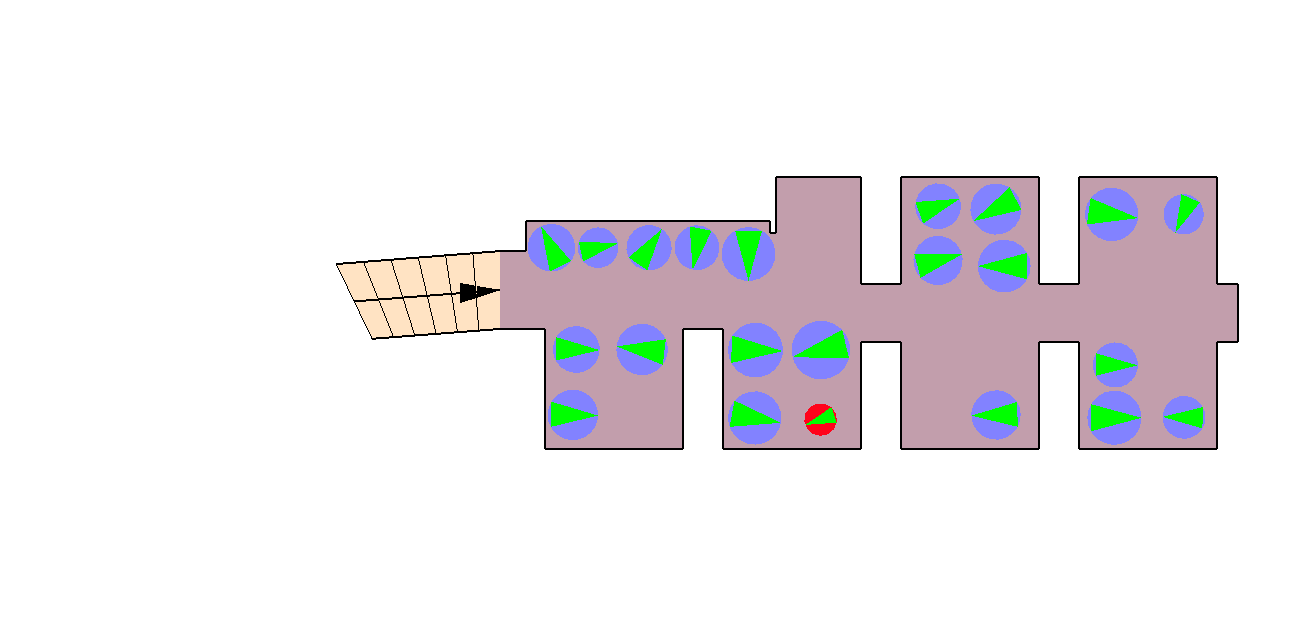} &
		\includegraphics[trim= 5pt 180pt 5pt 175pt, clip,width=\h\textwidth]{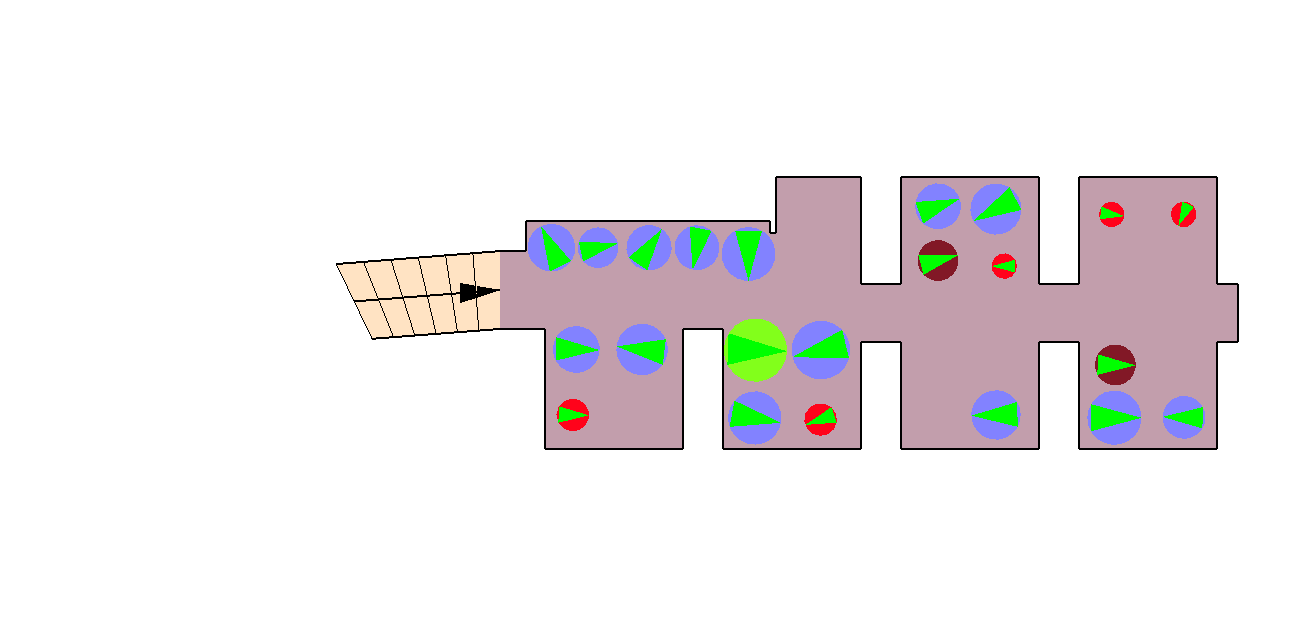} \\
		\includegraphics[trim= 5pt 170pt 5pt 175pt, clip,width=\h\textwidth]{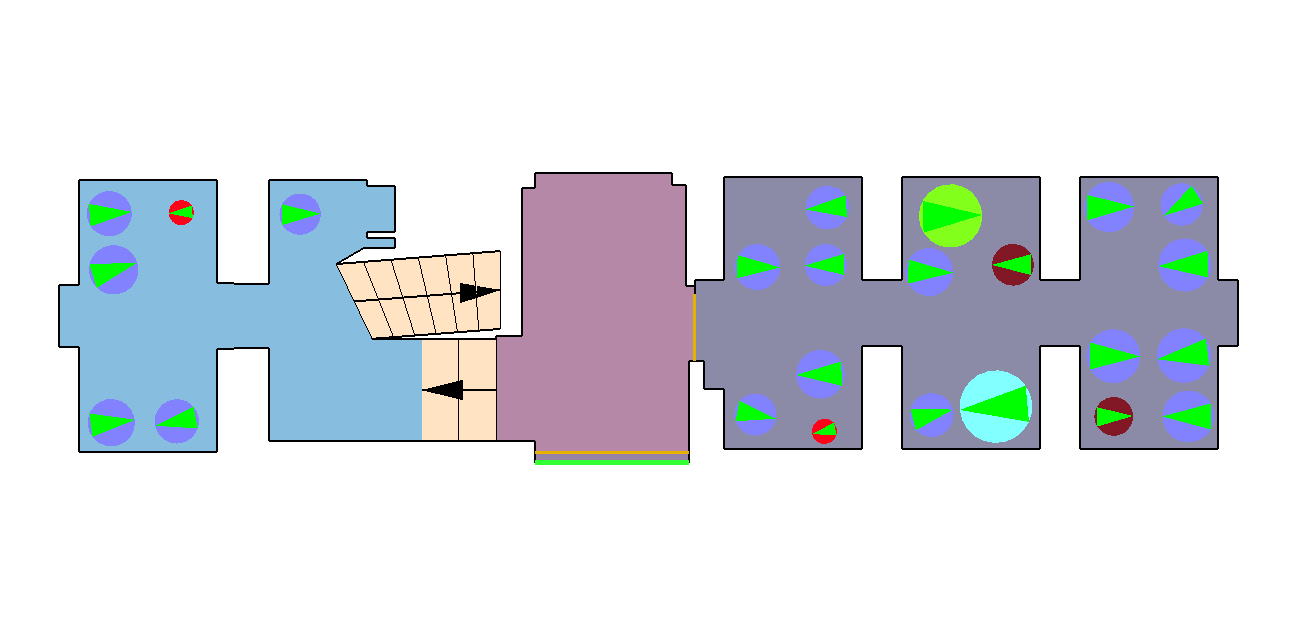} &
		\includegraphics[trim= 5pt 170pt 5pt 175pt, clip,width=\h\textwidth]{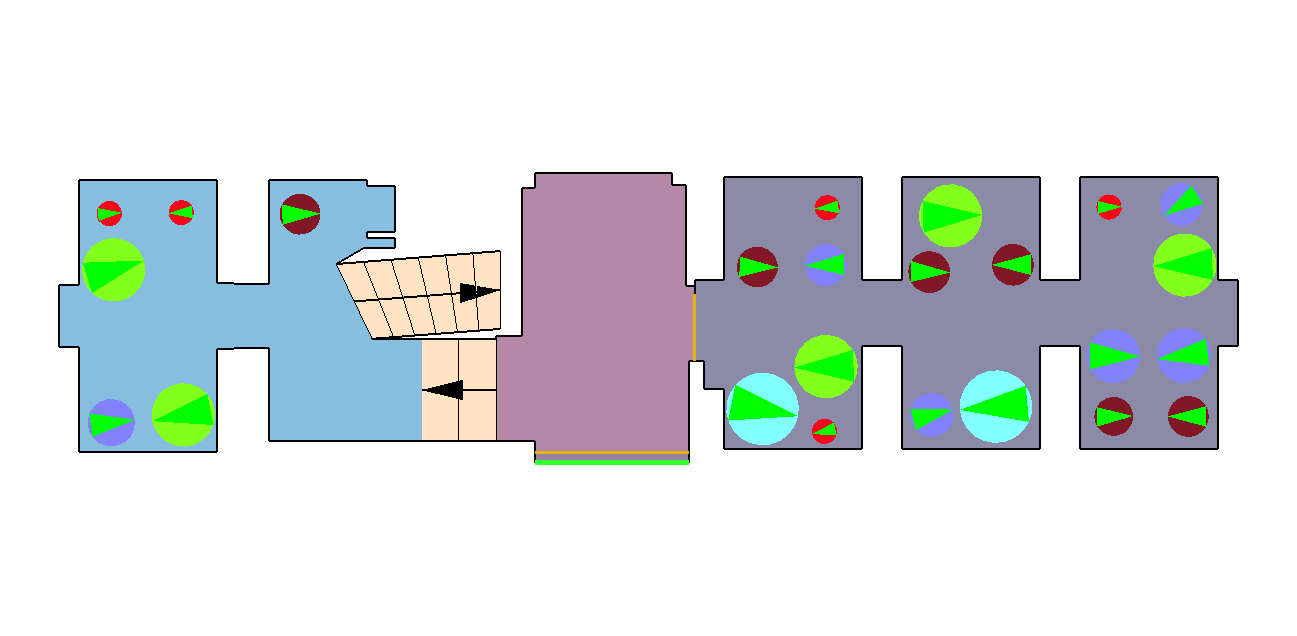} \\
	\end{tabular} 
\caption{Initial position of agents in simulations trials. The visualisation contains the information about the agent type, shoulder width, and initial orientation as well.}
\label{fig:placement}
\end{figure}

\newpage

\section{Graphs supporting the validation}
\label{app:valid}


\def\w{5.3cm}
\begin{figure}[htb!]
\begin{tabular}{rcc}
	width & HOM & HET\\
	\rotatebox{90}{\phantom{xxxxxxxxx}650~mm} & 
	 	   \includegraphics[width = \w]{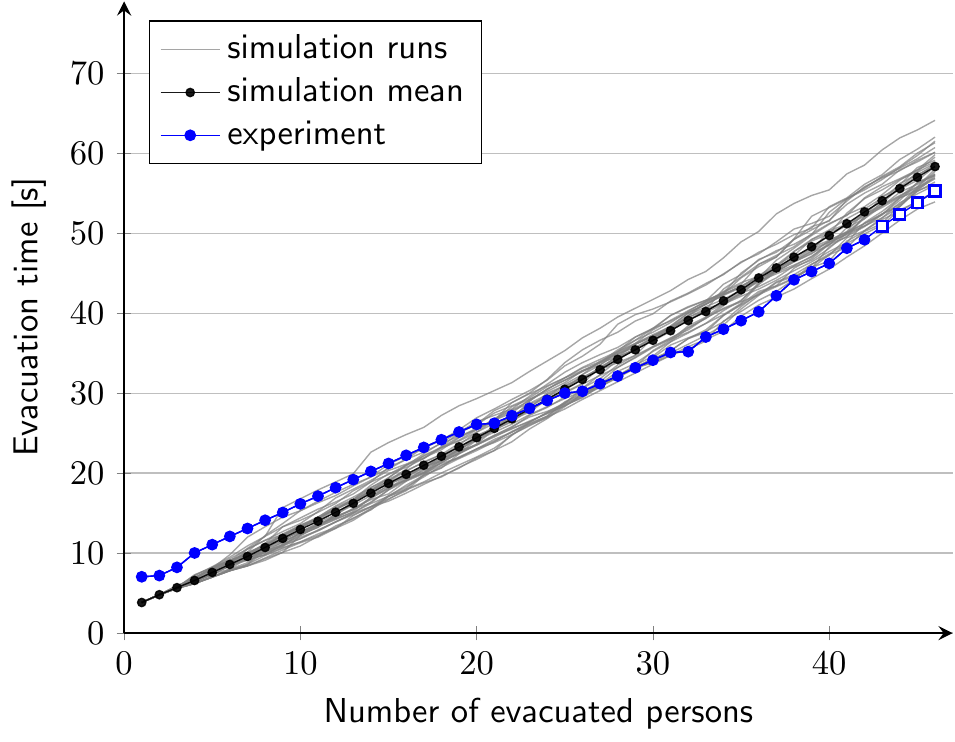}&
	 	   \includegraphics[width = \w]{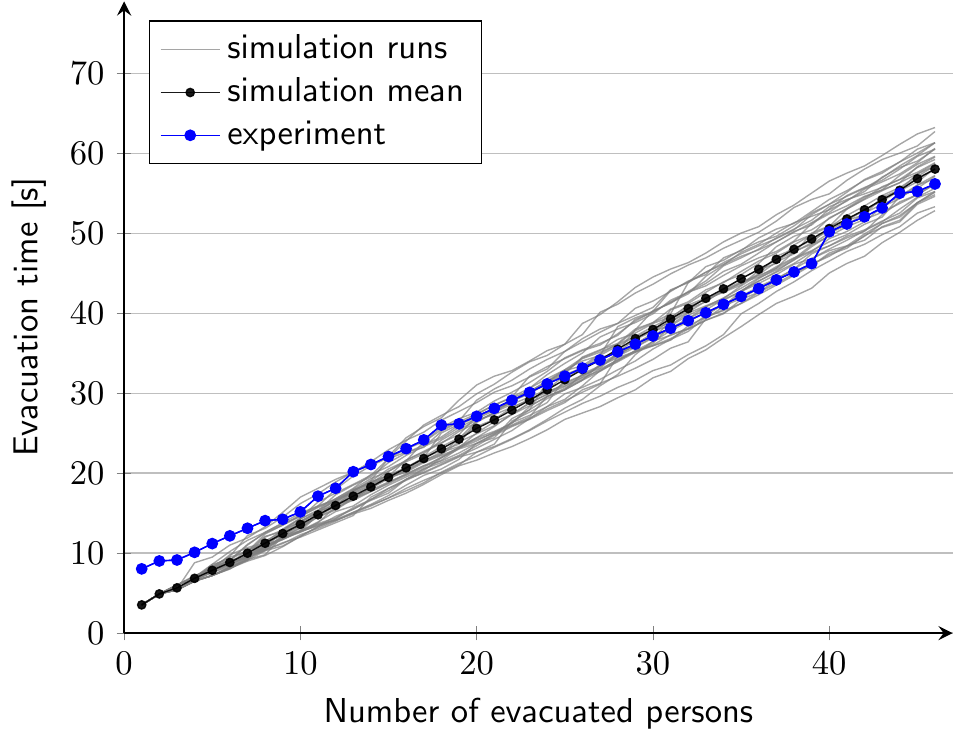}\\
	 \rotatebox{90}{\phantom{xxxxxxxxx}750~mm} & 
	 	   \includegraphics[width = \w]{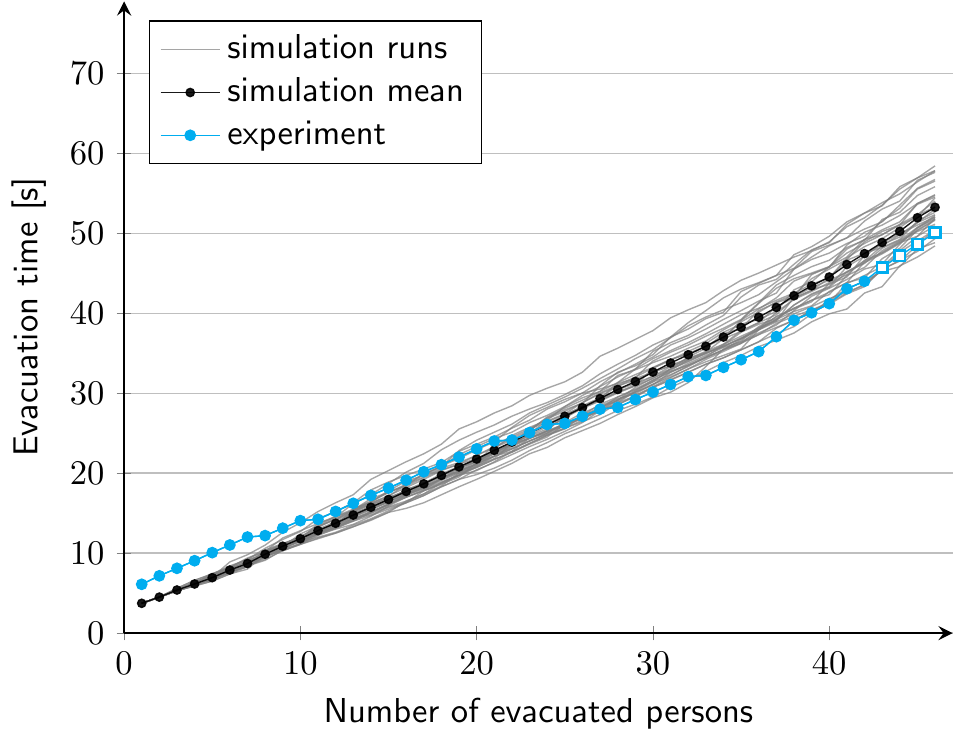}&
	 	   \includegraphics[width = \w]{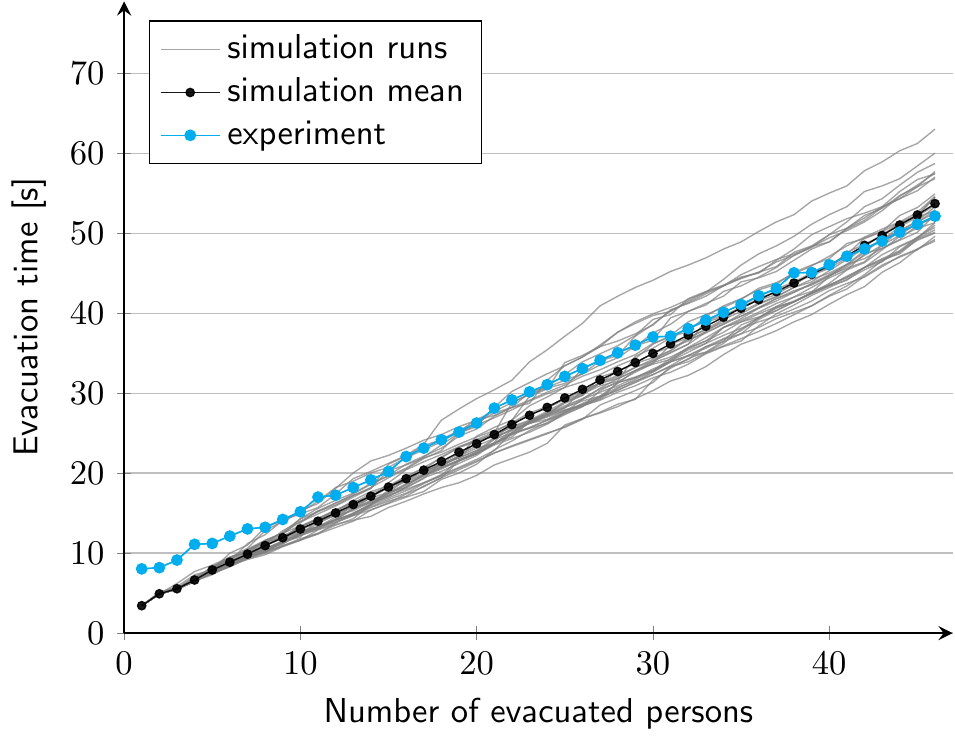}\\
	 \rotatebox{90}{\phantom{xxxxxxxxx}900~mm} & 
	 	   \includegraphics[width = \w]{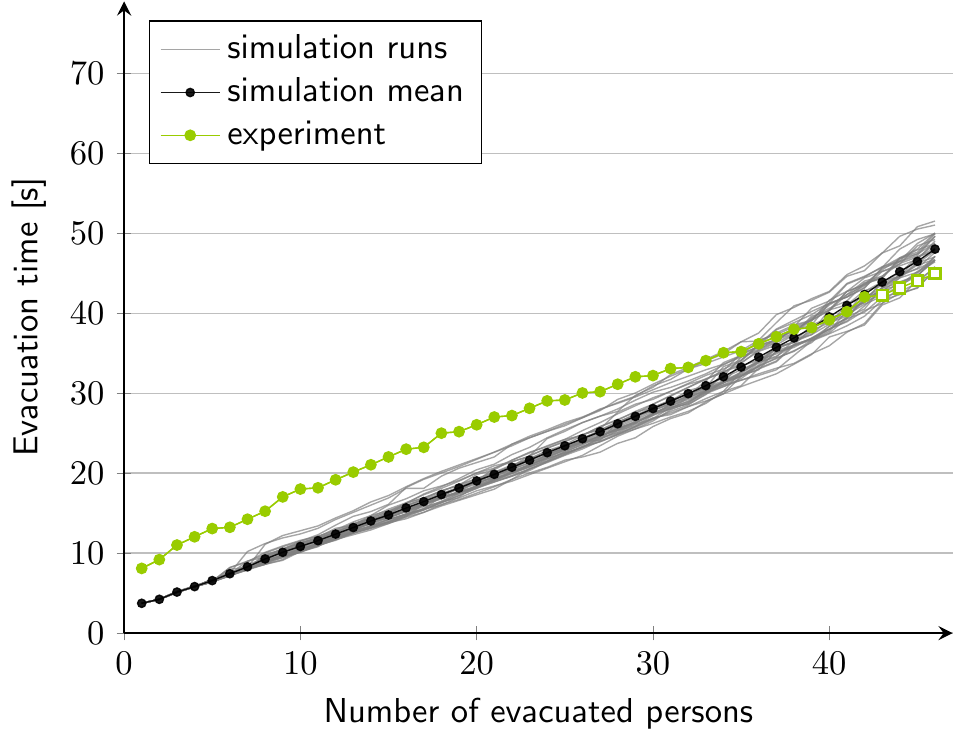}&
	 	   \includegraphics[width = \w]{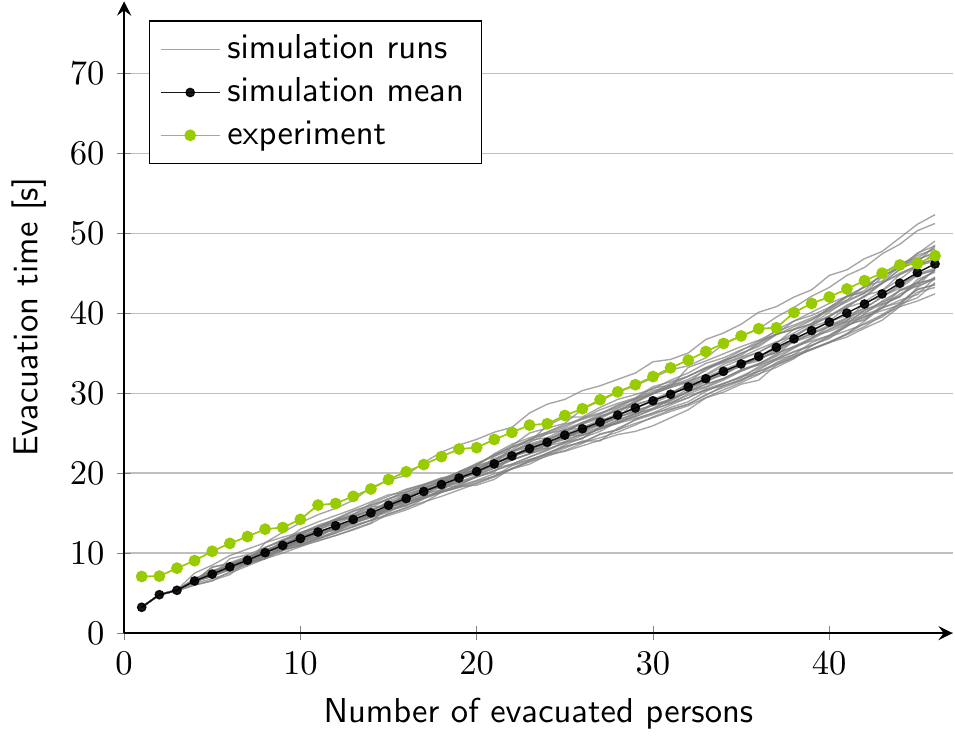}\\
	\rotatebox{90}{\phantom{xxxxxxxxx}1100~mm} & 
	 	   \includegraphics[width = \w]{figures/figure_validation_ET_platform_HOM_1100.pdf}&
	 	   \includegraphics[width = \w]{figures/figure_validation_ET_platform_HET_1100.pdf}\\
	\rotatebox{90}{\phantom{xxxxxxxxx}1340~mm} & 
	 	   \includegraphics[width = \w]{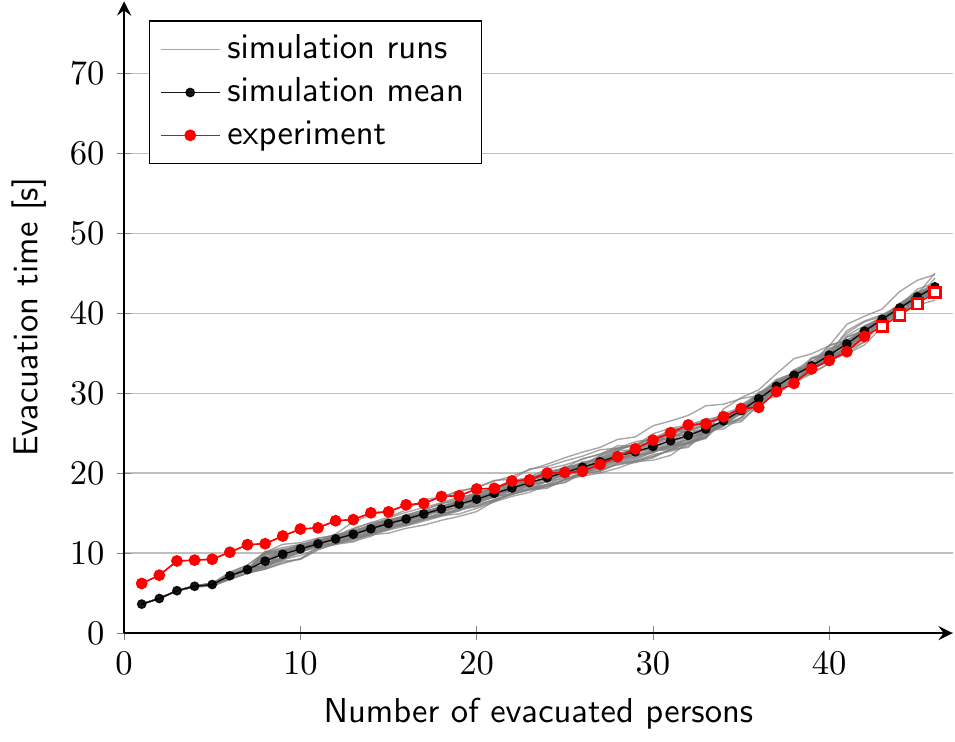}&
	 	   \includegraphics[width = \w]{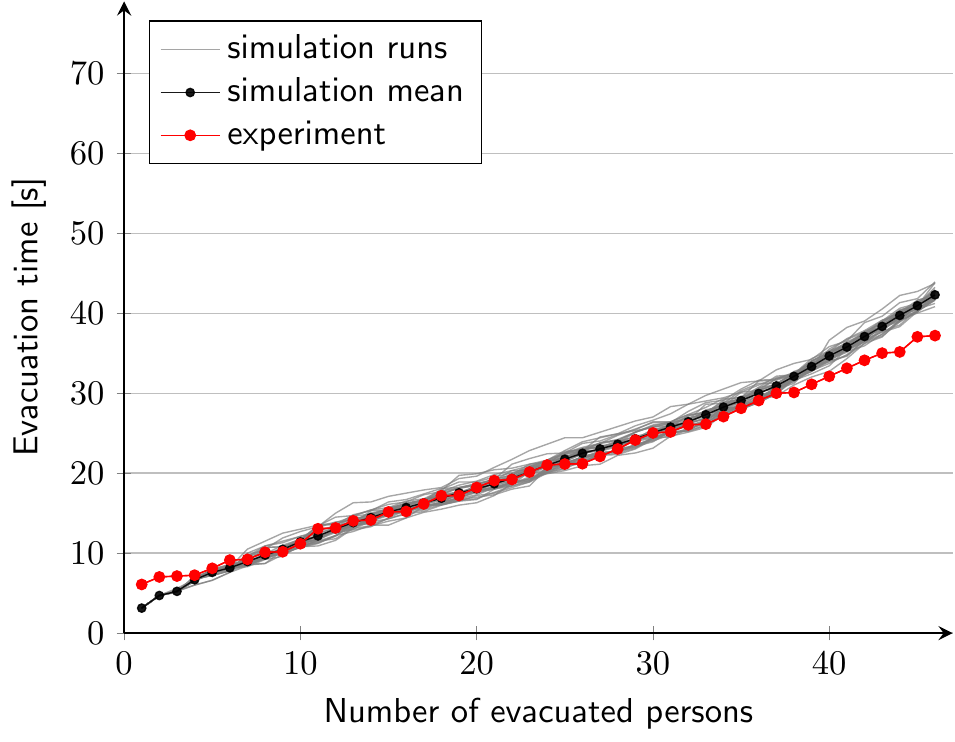}\\
\end{tabular}
\caption{Exit to high platform.}
\label{fig:valid_platform}
\end{figure}

\begin{figure}[htb!]
\begin{tabular}{rcc}
	width & HOM & HET\\
	\rotatebox{90}{\phantom{xxxxxxxxx}650~mm} & 
	 	   \includegraphics[width = \w]{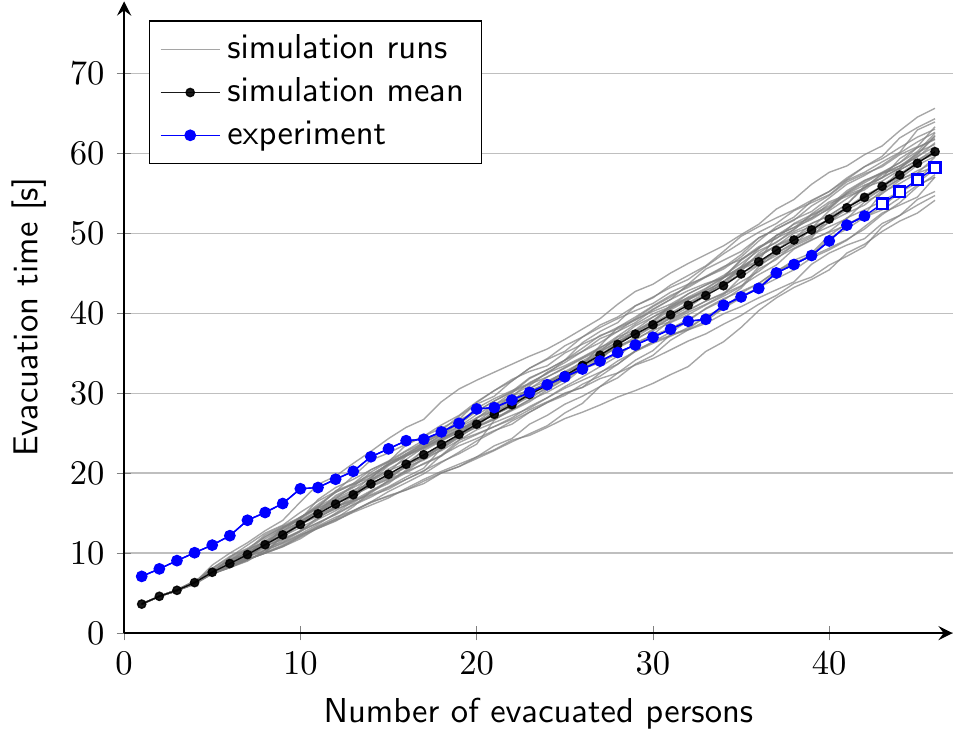}&
	 	   \includegraphics[width = \w]{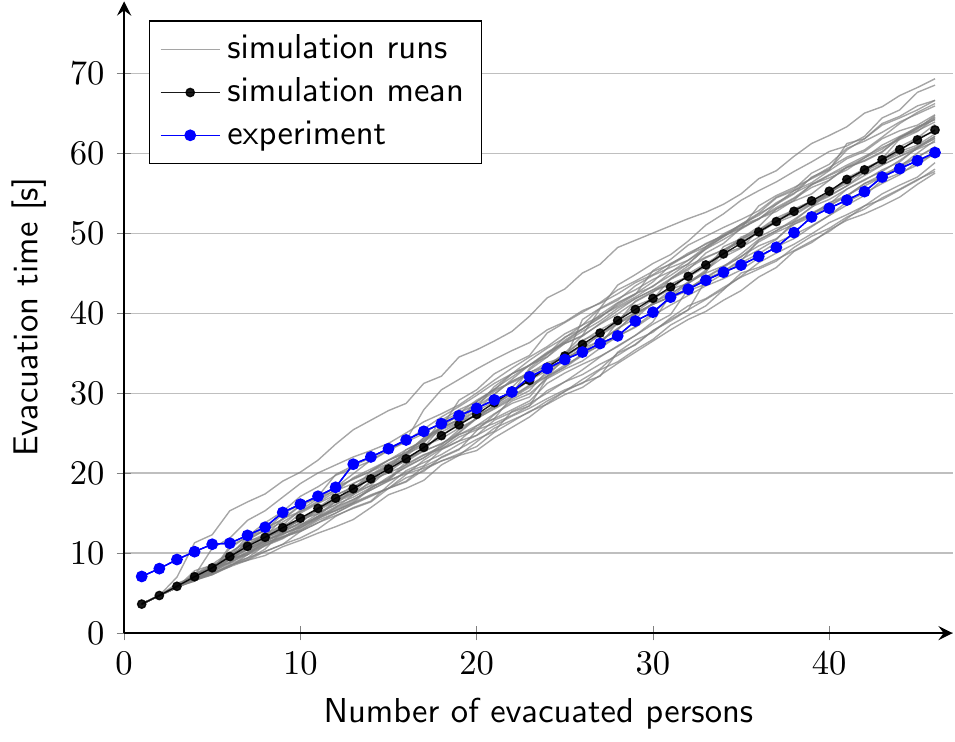}\\
	 \rotatebox{90}{\phantom{xxxxxxxxx}750~mm} & 
	 	   \includegraphics[width = \w]{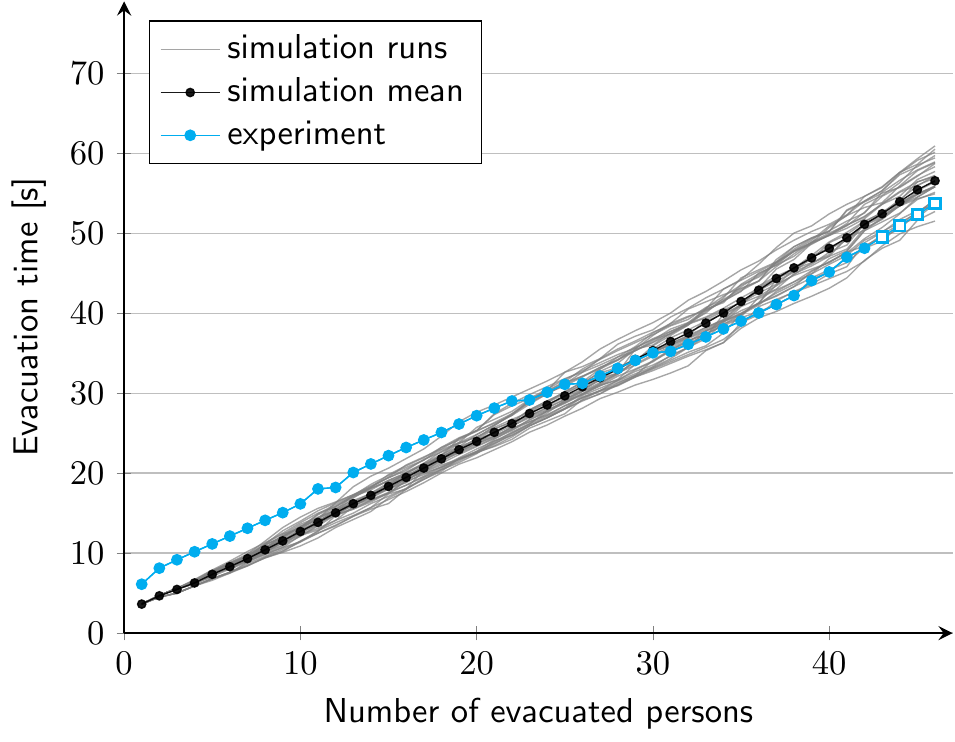}&
	 	   \includegraphics[width = \w]{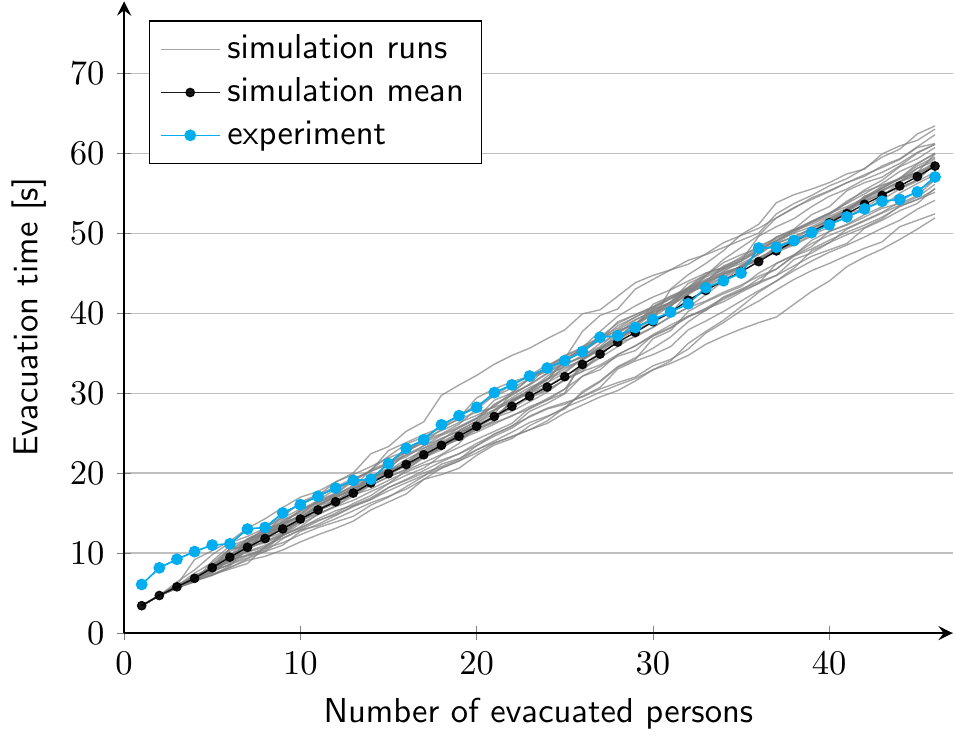}\\
	 \rotatebox{90}{\phantom{xxxxxxxxx}900~mm} & 
	 	   \includegraphics[width = \w]{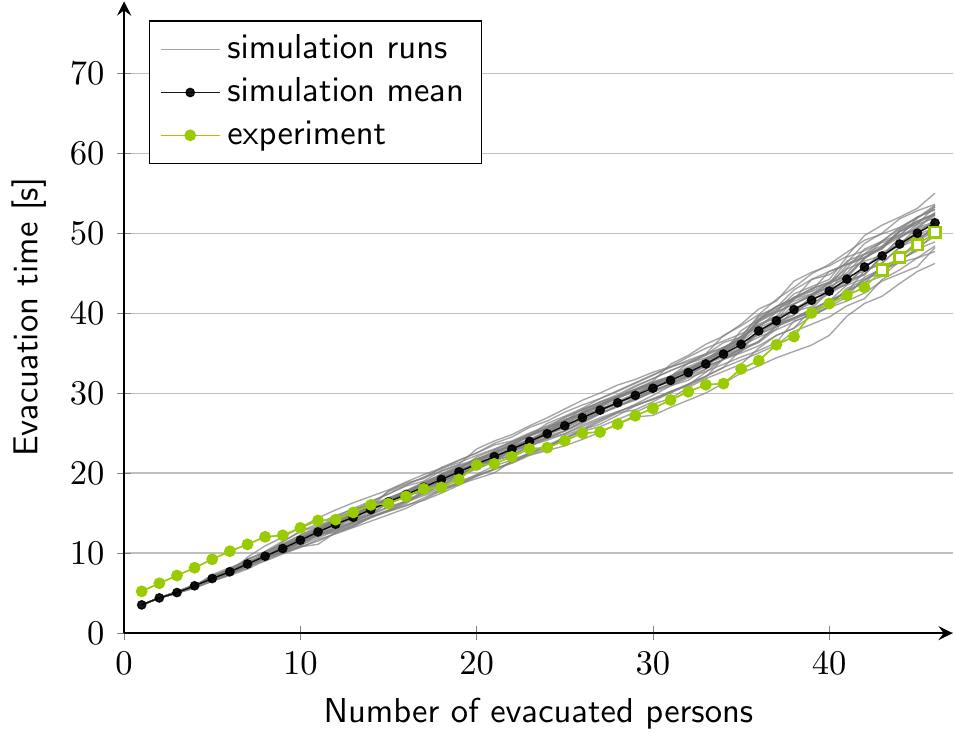}&
	 	   \includegraphics[width = \w]{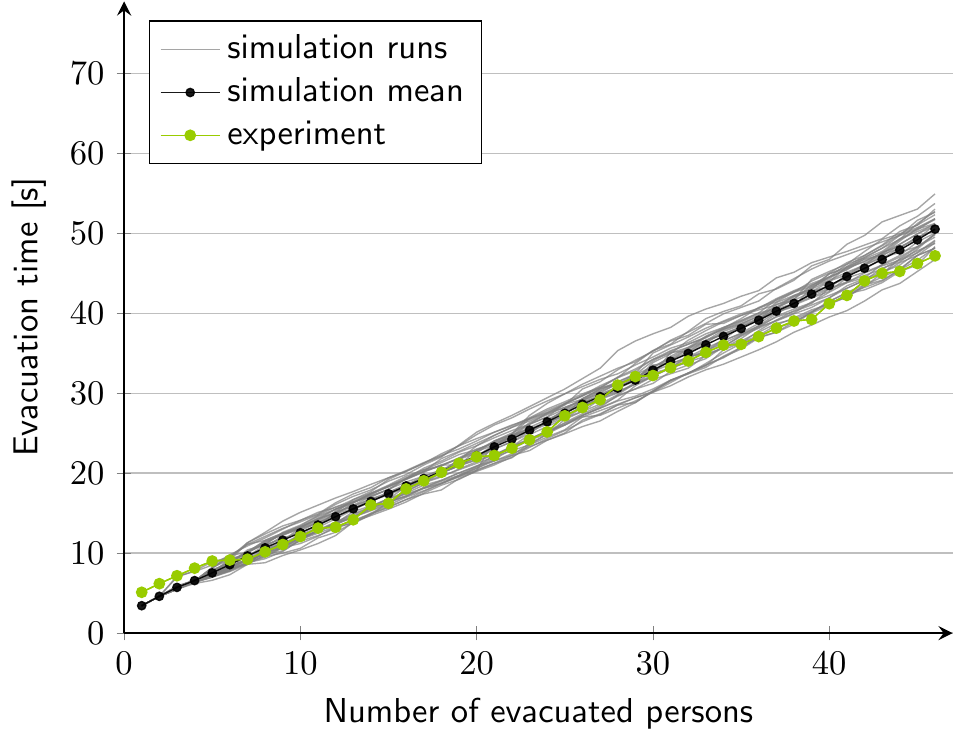}\\
	\rotatebox{90}{\phantom{xxxxxxxxx}1100~mm} & 
	 	   \includegraphics[width = \w]{figures/figure_validation_ET_stairs_HOM_1100.pdf}&
	 	   \includegraphics[width = \w]{figures/figure_validation_ET_stairs_HET_1100.pdf}\\
	\rotatebox{90}{\phantom{xxxxxxxxx}1340~mm} & 
	 	   \includegraphics[width = \w]{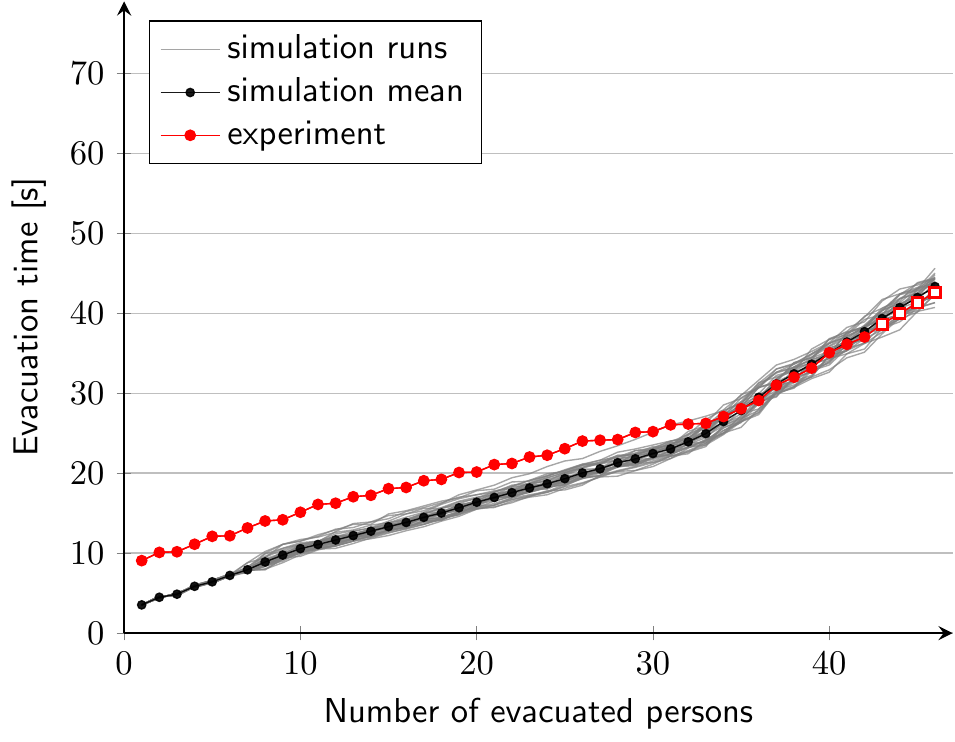}&
	 	   \includegraphics[width = \w]{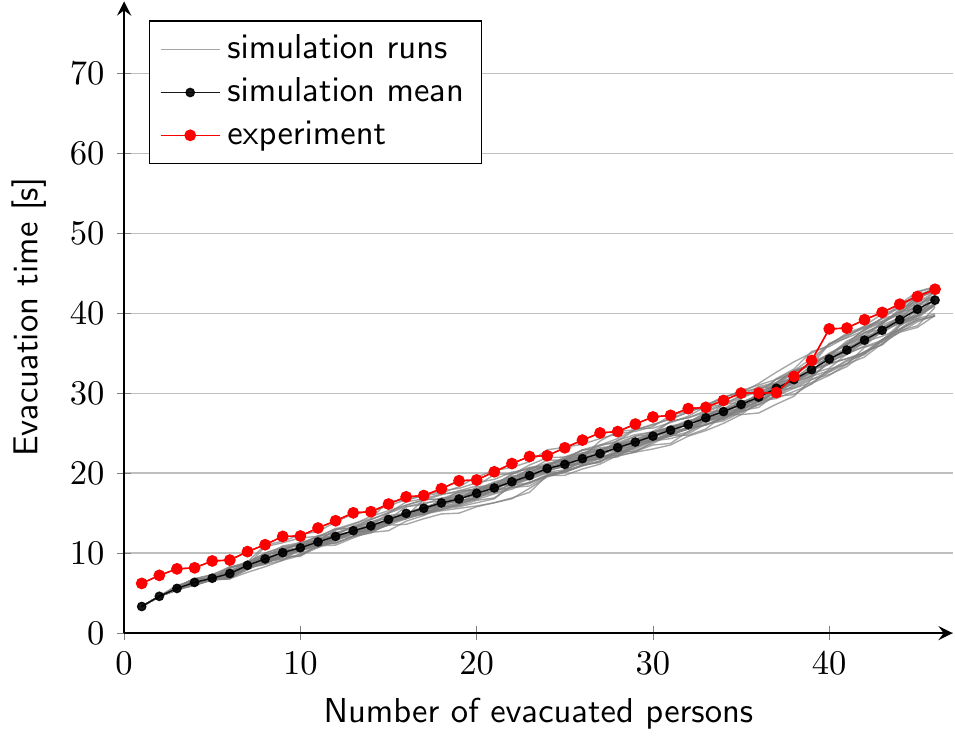}\\
\end{tabular}
\caption{Exit to an open line using stairs.}
\label{fig:valid_stairs}
\end{figure}

\begin{figure}[htb!]
\begin{tabular}{rcc}
	width & HOM & HET\\
	\rotatebox{90}{\phantom{xxxxxxxxx}650~mm} & 
	 	   \includegraphics[width = \w]{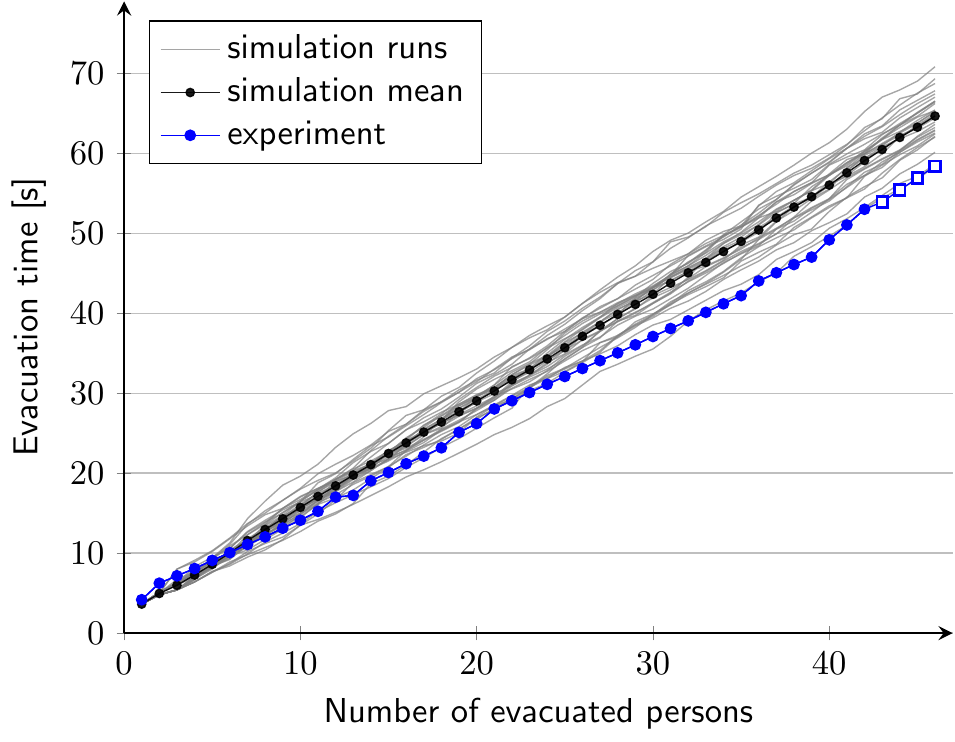}&
	 	   \includegraphics[width = \w]{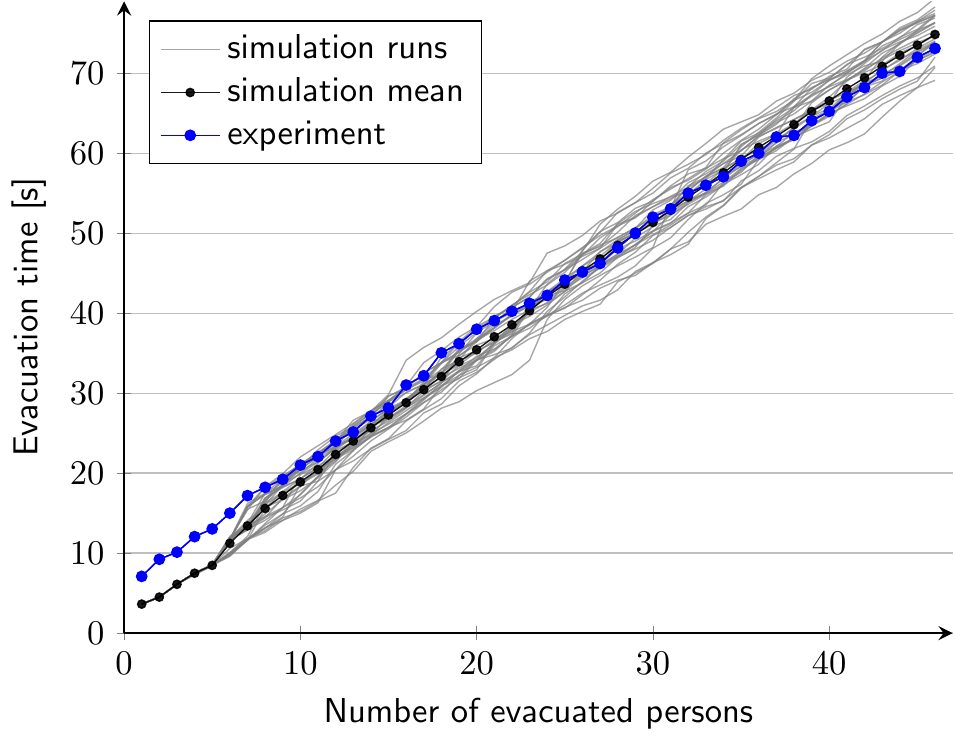}\\
	 \rotatebox{90}{\phantom{xxxxxxxxx}750~mm} & 
	 	   \includegraphics[width = \w]{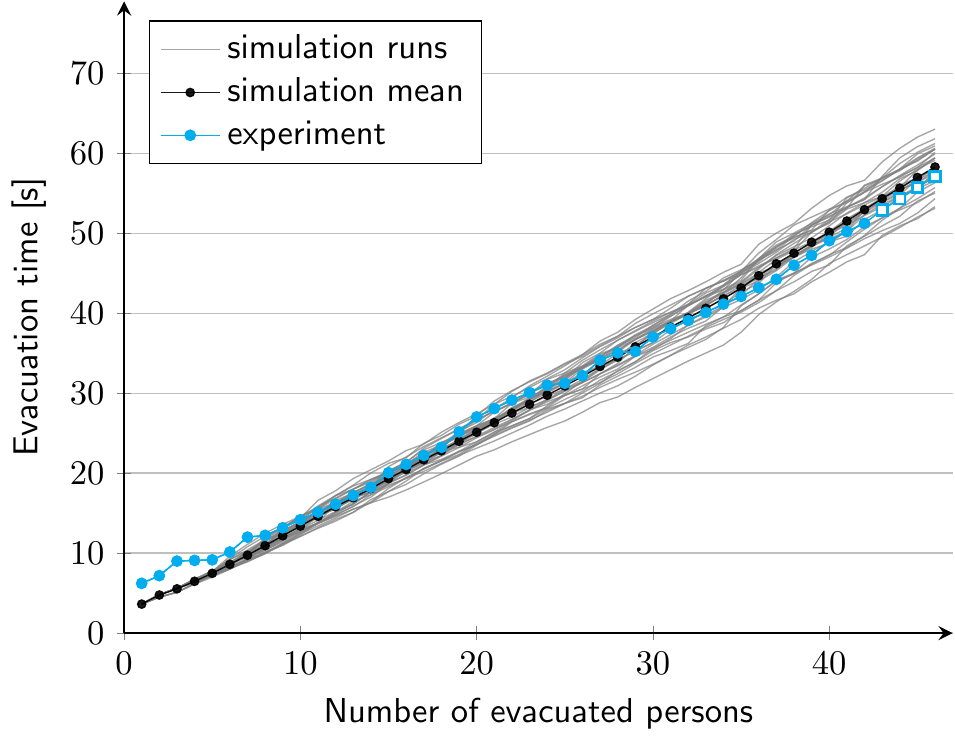}&
	 	   \includegraphics[width = \w]{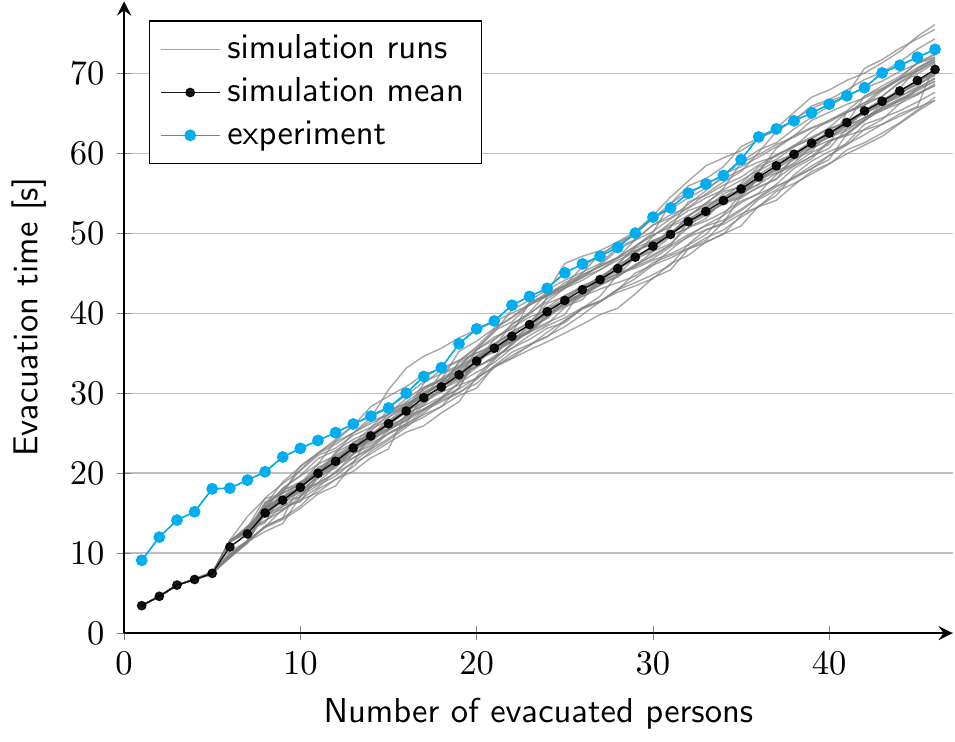}\\
	 \rotatebox{90}{\phantom{xxxxxxxxx}900~mm} & 
	 	   \includegraphics[width = \w]{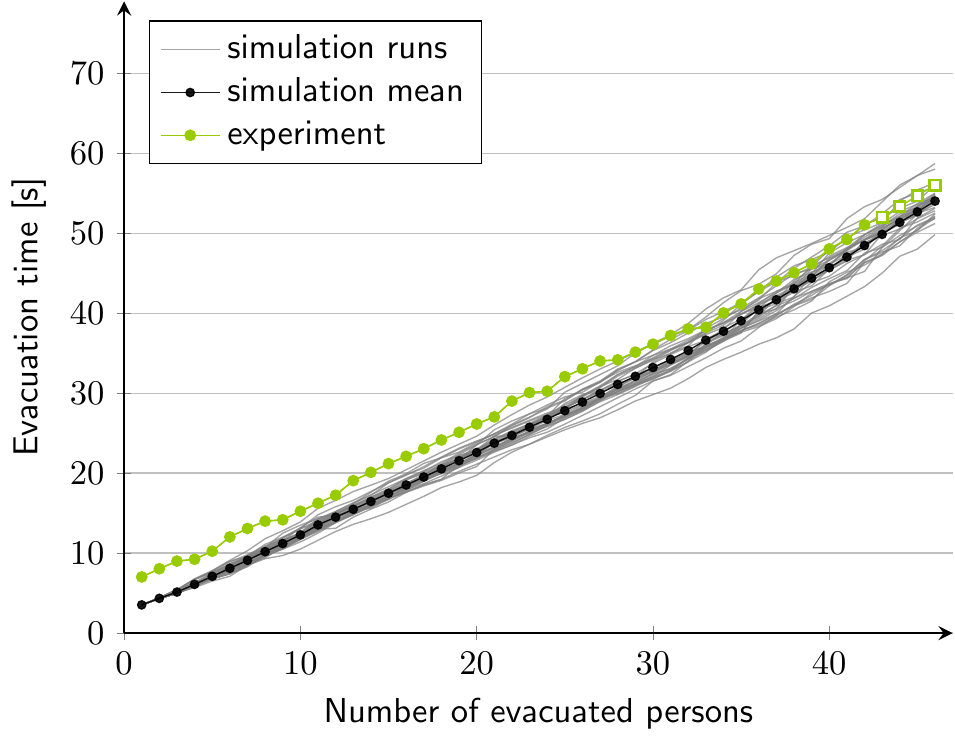}&
	 	   \includegraphics[width = \w]{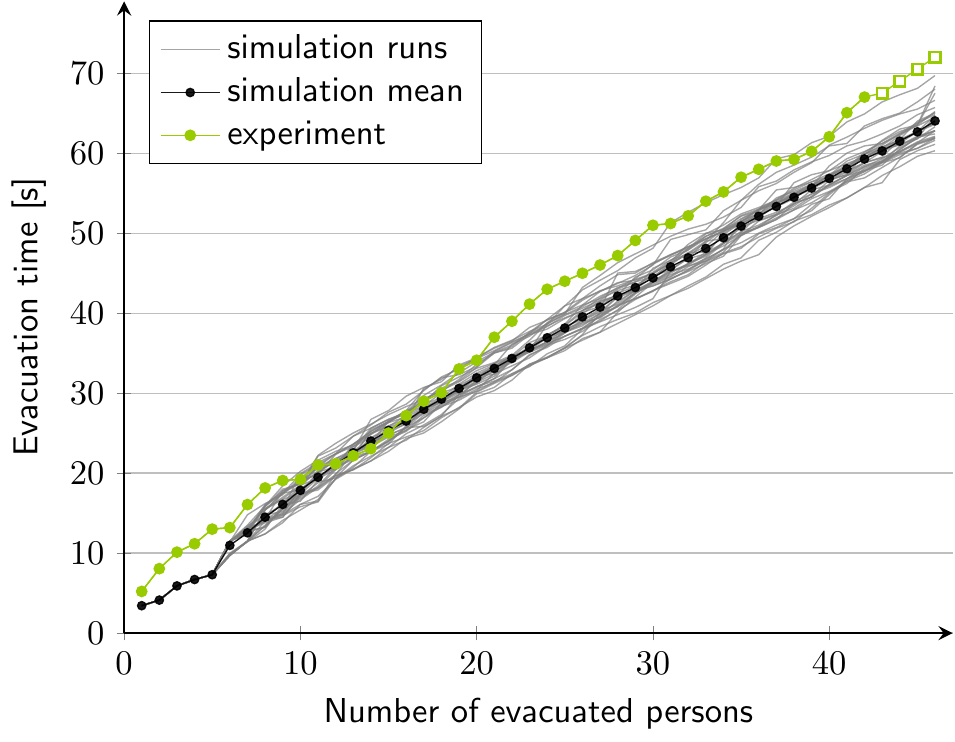}\\
	\rotatebox{90}{\phantom{xxxxxxxxx}1100~mm} & 
	 	   \includegraphics[width = \w]{figures/figure_validation_ET_ground_HOM_1100.pdf}&
	 	   \includegraphics[width = \w]{figures/figure_validation_ET_ground_HET_1100.pdf}\\
	\rotatebox{90}{\phantom{xxxxxxxxx}1340~mm} & 
	 	   \includegraphics[width = \w]{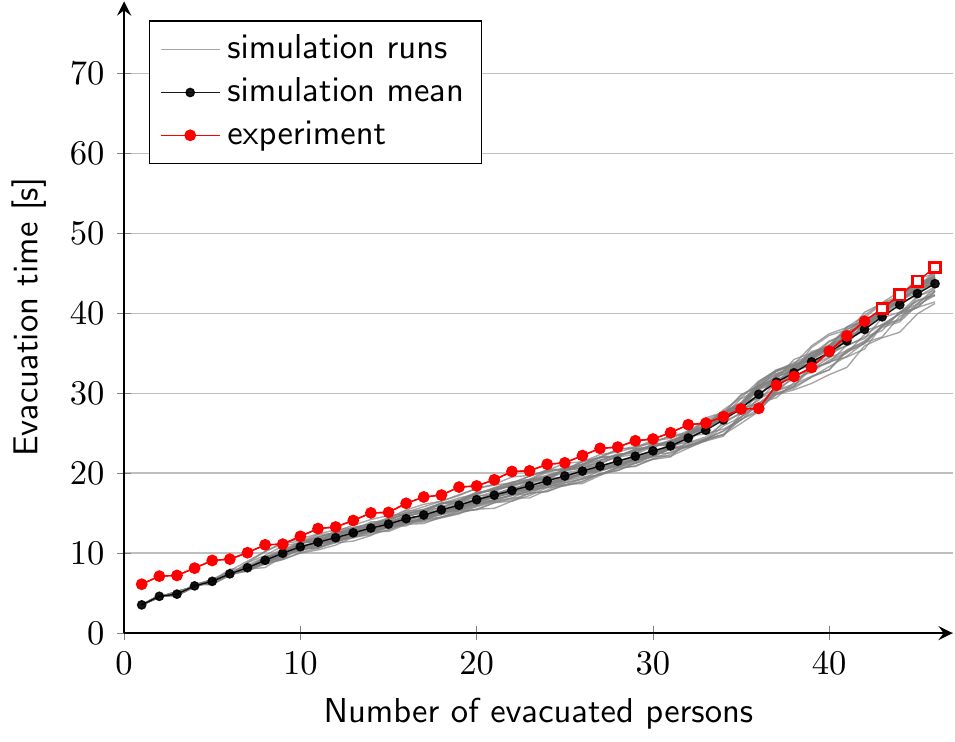}&
	 	   \includegraphics[width = \w]{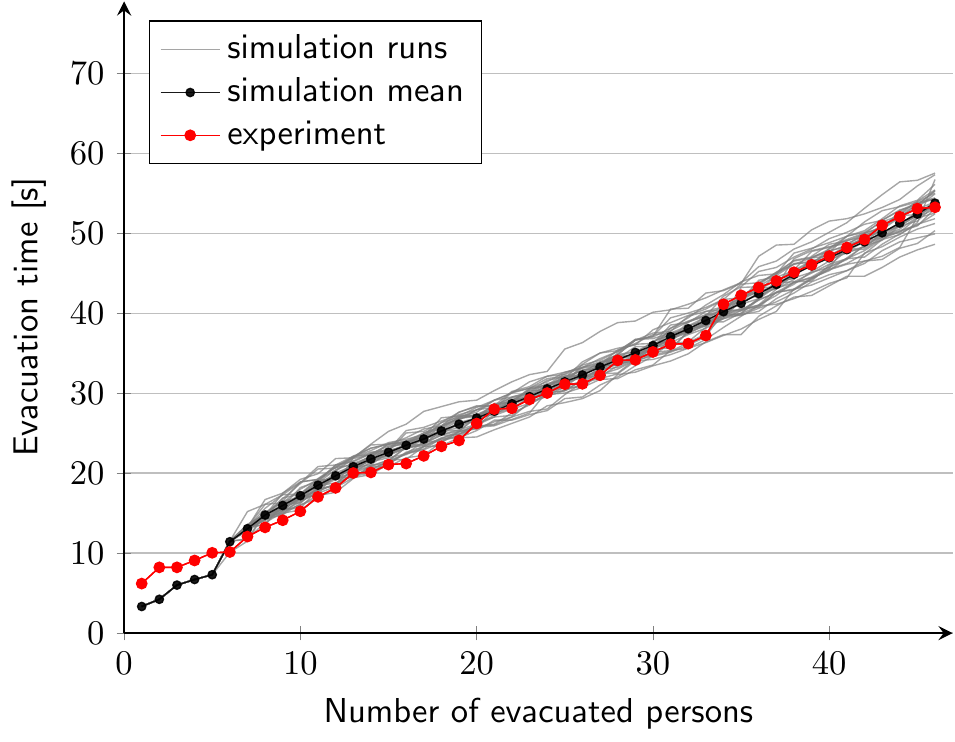}\\
\end{tabular}
\caption{Exit to an open line without any devices (750 mm jump).}
\label{fig:valid_jump}
\end{figure}

\end{document}